\documentclass[]{dukedissertation}
\usepackage{amsmath, amssymb, amsfonts, amsthm}
\usepackage{graphicx}
\usepackage{natbib}
\usepackage{color}
\usepackage{bm}
\usepackage[caption=false]{subfig}
\usepackage{subfiles}
\usepackage[export]{adjustbox}
\usepackage{bbm}
\usepackage{graphicx}
\usepackage{mathabx}
\usepackage{scalerel}

\usepackage{multirow}
\usepackage{setspace}
\usepackage{tikz}
\usetikzlibrary{decorations.shapes}
\usetikzlibrary{arrows}
\usetikzlibrary{patterns}
\graphicspath{{./Pictures/}}

\author{Emilie Huffman}
\title{Fermion Bag Approach for Hamiltonian Lattice Field Theories}
\supervisor{Shailesh Chandrasekharan}
\department{Physics} % Appears as Department of \department
\copyrighttext{ All rights reserved except the rights granted by the\\
   \href{http://creativecommons.org/licenses/by-nc/3.0/us/}
        {Creative Commons Attribution-Noncommercial Licence}
}

\member{Harold U. Baranger}
\member{Henry Greenside}
\member{Stephen W. Teitsworth}
\member{Uwe-Jens Wiese}
\usepackage[pdftex, pdfusetitle, plainpages=false, 
				letterpaper, bookmarks, bookmarksnumbered,
				colorlinks, linkcolor=black, citecolor=black,
	         filecolor=black, urlcolor=black]
				{hyperref}

\begin{document}

%-----------------------------------------------------------------------------%
% TITLE PAGE -- provides UMI abstract title page & copyright if appropriate
%-----------------------------------------------------------------------------%
\year=2018
\date{2018}
\maketitle

%-----------------------------------------------------------------------------%
% ABSTRACT -- included file should start with '\abstract'.
%-----------------------------------------------------------------------------%
\abstract

Understanding the critical behavior near quantum critical points for strongly correlated quantum many-body systems remains intractable for the vast majority of scenarios. Challenges involve determining if a quantum phase transition is first- or second-order, and finding the critical exponents for second-order phase transitions. Learning about where second-order phase transitions occur and determining their critical exponents is particularly interesting, because each new second-order phase transition defines a new quantum field theory.

Quantum Monte Carlo (QMC) methods are one class of techniques that, when applicable, offer reliable ways to extract the nonperturbative physics near strongly coupled quantum critical points. However, there are two formidable bottlenecks to the applicability of QMC: (1) the sign problem and (2) algorithmic update inefficiencies. In this thesis, I overcome both these difficulties for a class of problems by extending the fermion bag approach recently developed by Shailesh Chandrasekharan to the Hamiltonian formalism and by demonstrating progress using the example of a specific quantum system known as the $t$-$V$ model, which exhibits a transition from a semimetal to an insulator phase for a single flavor of four-component Dirac fermions.

I adapt the fermion bag approach, which was originally developed in the context of Lagrangian lattice field theories, to be applicable within the Hamiltonian formalism, and demonstrate its success in two ways: first, through solutions to new sign problems, and second, through the development of new efficient QMC algorithms. In addressing the first point, I present a solution to the sign problem for the $t$-$V$ model. While the $t$-$V$ model is the simplest Gross-Neveu model of the chiral Ising universality class, the specter of the sign problem previously prevented its simulation with QMC for 30 years, and my solution initiated the first QMC studies for this model. The solution is then extended to many other Hamiltonian models within a class that involves fermions interacting with quantum spins. Some of these models contain an interesting quantum phase transition between a massless/semimetal phase to a massive/insulator phase in the so called Gross-Neveu universality class. Thus, the new solutions to the sign problem allow for the use of the QMC method to study these universality classes.

The second point is addressed through the construction of a Hamiltonian fermion bag algorithm. The algorithm is then used to compute the critical exponents for the second-order phase transition in the $t$-$V$ model. By pushing the calculations to significantly larger lattice sizes than previous recent computations ($64^2$ sites versus $24^2$ sites), I am able to compute the critical exponents more reliably here compared to earlier work. I show that the inclusion of these larger lattices causes a significant shift in the values of the critical exponents that was not evident for the smaller lattices. This shift puts the critical exponent values in closer agreement with continuum $4-\epsilon$ expansion calculations. The largest lattice sizes of $64^2$ at a comparably low temperature are reachable due to efficiency gains from this Hamiltonian fermion bag algorithm. The two independent critical exponents I find, which completely characterize the phase transition, are $\eta=.51(3)$ and $\nu=.89(1)$, compared to previous work that had lower values for these exponents. The finite size scaling fit is excellent with a $\chi^2/DOF=.90$, showing strong evidence for a second-order critical phase transition, and hence a non-perturbative QFT can be defined at the critical point.

%-----------------------------------------------------------------------------%
% DEDICATION -- OPTIONAL.  Put the text inside the braces.
%               (Long 'dedications' probably belong in the acknowledgements)
%-----------------------------------------------------------------------------%
\dedication{\noindent For Shirley Gray Guthmann and Emilie Cobb Huffman, my grandmothers,\\ \indent and Edith Huffman Poston, my great aunt. They are all daring adventurers.}

%-----------------------------------------------------------------------------%
% FRONTMATTER -- ToC is required, LoT and LoF are required if you have any
% tables or figures, respectively. List of Abbreviations and Symbols is 
% optional.
%-----------------------------------------------------------------------------%
\small{
\tableofcontents} % Automatically generated
\listoftables	% If you have any tables, automatically generated
\listoffigures	% If you have any figures, automatically generated
\abbreviations

% You can put here what you like, but here's an example
%Note the use of starred section commands here to produce proper division
%headers without bad '0.1' numbers or entries into the Table of Contents.
%Using the {\verb \begin{symbollist} } environment ensures that entries are
%properly spaced.

\section*{Symbols}

%Put general notes about symbol usage in text here.  Notice this text is
%double-spaced, as required.

\begin{symbollist}
	\item[$\bar{\psi}_x,\psi_x,\bar{\xi}_x,\xi_x$] Grassman variable scalars.
	% Optional item argument makes the symbol/abbr
		\item[$c^\dagger_x, c_x$] Fermion creation/annihilation operators.
	\item[$\bar{\gamma}_x,\gamma_x$] Majorana fermion operators.
	\item[$S_x^1,S_x^2,S_x^3,S_x^+,S_x^-$] Quantum spin operators.
\end{symbollist}

\section*{Abbreviations}

%Long lines in the \texttt{symbollist} environment are single spaced, like in
%the other front matter tables.

\begin{symbollist}
	\item[BSS] Blankenbecler Sugar Scalapino.
  \item[CDW] Charge Density Wave. 
	\item[CT-INT] Continuous-time interaction expansion. 
	\item[RG] Renormalization group.
	\item[SSE] Stochastic series expansion.
	\item[QCP] Quantum critical point.
	\item[QMC] Quantum Monte Carlo.
\end{symbollist}
 % List of Abbreviations. Start file with '\abbreviations'

%-----------------------------------------------------------------------------%
% ACKNOWLEDGEMENTS -- included file should start with '\acknowledgements'
%-----------------------------------------------------------------------------%
\acknowledgements

I would first like to thank my advisor, Shailesh Chandrasekharan, for all his guidance and support during my time at Duke. Not only could I feel confident to ask him any question I had--no matter how basic--but he also constantly models the process of asking questions himself, stepping back often to ask just what the research is teaching us. I also want to thank Uwe-Jens Wiese for advising, mentoring, and supporting me during my time at the University of Bern and far beyond, as well as Ari Mizel for advising me for a summer at the Laboratory for Physical Sciences. My committee members Harold Baranger, Henry Greenside, and Stephen Teitsworth offered me invaluable feedback for this dissertation, and I am very grateful to them for that.

I would like to thank the other graduate students in my class, especially Lou Isabella, Anne Draelos, and Gu Zhang, for their camaraderie, friendship, and help. I am also thankful to Venkitesh Ayyar for these things, in addition to useful research discussions.

I would like to thank my friends at the University of Bern, especially Therkel Olesen, Lewis Tunstall, and Debasish Banerjee--who is also a collaborator, for useful discussion and most of all community.

I would like to thank my college friends, Caitlin Roach, Hannah Marsh, Savannah Patterson, Rebecca Harris, Kimberly Lukens, and Megan Davis, for always being one Skype call away, as well as Erica Isabella for making sure I got the most out of life in Durham. I would also like to thank Isaac Evans for putting a bounty on my Stack Overflow question and helping me skirt cruise ship wifi restrictions so I could run my jobs.

I would like to thank my grandfather, Ed Guthmann, for being invested in all of his grandchildren, and saying to me with what seemed like an annoying certainty, ``I know you're gonna be a scientist one day.''

I am very thankful for my supportive parents, Brent and Kimberly. I have always known that I have their love regardless of whether I succeed or fail. I am also thankful for my two sisters Noelle and April. They think more reasonably than anyone else I know.

Finally, I am very thankful for my husband, Benjamin. Beyond all of the concrete ways that he helps me every day, he has a way of saying ``hang in there'' that somehow strongly insinuates both (1) that he knows you will and (2) that he is very sincerely praying that you will. Both senses of the phrasing are vital.

I would like to acknowledge the support I have received from the NPSC throughout most of my time at Duke, as well as the computing time awarded by XSEDE.

%Thank anyone you like here.  It's good practice to thank every granting agency that's given you money
%since you've been ABD, any other school you visited during your research,
%and any professional society that's funded your travel.

\newpage

\small{\textit{\vspace{2cm}{}\\``In all seriousness, people think that it's the ideas that are important. Well, everyone has ideas, all the time. I tend to write mine down and remember them, but at some point you have to apply the bum to the seat and knock out about sixty five thousand words - that's how long a novel is.''\\ --Terry Pratchett}}

%==============================================================================
%-----------------------------------------------------------------------------%
%
% MAIN BODY OF PAPER
%
%
%-----------------------------------------------------------------------------%
\chapter{Introduction}
\label{chap:intro}
%Type your introduction here.  This is ``technically'' your first chapter of the dissertation/thesis.

\small{\textit{``One cannot escape the feeling that these mathematical formulae have an independent existence and an intelligence of their own, that they are wiser than we are, wiser even than their discoverers, that we get more out of them than we originally put in to them.''\\ --Heinrich Hertz}}
\section{Strongly Correlated Systems}
Strongly correlated many-body systems form an exciting area for study due in no small part to their applicability to a wide variety of physical systems, spanning many subfields of physics. Examples range from the structure of nuclear matter and the quark gluon plasma, to strongly correlated materials such as graphene and frustrated magnets, to even speculative physics beyond the standard model. Systems that arise in vastly different contexts may have an underlying unity due to similar symmetries--their phase transitions if second-order may have the same critical exponents, putting them within the same universality class \cite{Goldenfield}. The understanding of these universality classes and the models which belong to each remains an area with much need for exploration, with identifying and characterizing second-order quantum phase transitions important both for the study of quantum materials as well as the definition of continuum quantum field theories. While perturbative methods and symmetry arguments can be used to create conjectures about the behavior near quantum critical points of strongly correlated systems, a nonperturbative method is necessary to reliably study such behavior.

Because the number of  states in the Hilbert space grows exponentially with the number of quantum degrees of freedom, computing the properties of these systems nonperturbatively--especially close to a quantum phase transition--remains a formidable challenge. Techniques such as exact diagonalization methods, Quantum Monte Carlo (QMC), Density Matrix Renormalization Group (DMRG), and tensor networks have been developed to compute quantities for lattice systems in manageable times, but each method has its own limitations. Exact diagonalization is limited to small system sizes (on the order of 30 quantum spins for example) due to a calculation time that scales exponentially with the system volume \cite{doi:10.1063/1.4823192}, and these system sizes are typically too small to sufficiently understand the quantum critical behavior. DMRG and tensor networks, on the other hand, do allow for polynomial time calculations and can be used to perform calculations on systems that are orders of magnitude larger than the largest systems accessible to exact diagonalization, but have had their successes so far limited to systems with one spatial dimension \cite{RevModPhys.77.259,SCHOLLWOCK201196,ORUS2014117}. In contrast, QMC allows for polynomial time scaling even for higher dimensional systems, but this scaling is often stymied--especially for systems involving fermions--by what is known as the \textit{sign problem} \cite{PhysRevLett.94.170201}. It also can remain a challenge to design efficient algorithms even if the scaling is known to be polynomial in time: higher degree polynomials with larger prefactors will be less efficient and will leave more system sizes out of reach compared to lower degree polynomials with smaller prefactors, so part of the challenge of algorithmic development is to improve the scaling even among polynomial-time algorithms.
%We will discuss QMC and this problem in more detail later on in the introduction.

The goal of this thesis is to develop a new approach for performing QMC calculations and to subsequently uncover new critical properties of quantum systems in thermal equilibrium. Within that effort there are two subgoals: the first is to identify new many-body models that are free of sign problems, and the second is to develop polynomial time algorithms that are more efficient than those currently available. In this thesis, we find that both of these objectives can be achieved within a new class of Hamiltonian models involving interacting fermions and quantum spins. In particular, we develop an algorithm that can be used to compute the critical behavior in such systems, especially close to strongly interacting critical points with gapless fermionic excitations. We demonstrate that at least for one particular model--the $t$-$V$ model--this new approach is more efficient than the previous best QMC algorithm, known as the auxiliary field method \cite{PhysRevLett.56.2521}. We show this by studying the critical behavior near its quantum critical point, where it undergoes a semimetal-insulator phase transition, and we compute the critical exponents at the transition more reliably than previous results, due to the fact that we are able to include larger lattice sizes than those included in previous calculations. The ideas used in achieving these objectives are based on the fermion bag idea, which was originally developed in the context of the Lagrangian formalism \cite{Chandrasekharan:2013rpa}. Here we extend this idea to the Hamiltonian formalism.

The thesis is organized as follows: in the remainder of this chapter we will explain the basic ideas of the quantum Monte Carlo method, the sign problem, the Lagrangian and Hamiltonian formalisms, and the auxiliary field method. This discussion will help us to set the stage and introduce the ideas and notation we will be using throughout this work. In Chapter 2 we will discuss the fermion bag idea in the original Lagrangian context and introduce some ideas for how to extend it to the Hamiltonian picture. In Chapters 3 and 4, we will discuss new solutions to sign problems within a class of Hamiltonian models, thus allowing them to be studied by QMC. In Chapter 5, we will discuss the fermion bag algorithm suited for studying these new models, and in Chapter 6 we will discuss the results of using this algorithm to study the $t$-$V$ model. There we will present our analysis and results for the extraction of the critical exponents for its quantum critical behavior. Finally, Chapter 7 contains our conclusions and future outlook.

\section{Quantum Monte Carlo}
Quantum Monte Carlo is an excellent method for computing the equilibrium and ground state properties of quantum many body systems. Because it forms the backbone of our techniques to come, the following is a short summary of how it works. The information in this section applies equally for Monte Carlo calculations of classical and quantum systems--in fact the only characteristic that distinguishes a \textit{Quantum} Monte Carlo is that the Monte Carlo weights come from a quantum Hamiltonian or an associated Lagrangian action.

In most cases the quantity of interest, $O$, can be written as
\begin{equation}
    \left\langle O \right\rangle = \frac{1}{Z} \sum_{C} O\left(C\right) \Omega\left(C\right), \qquad Z = \sum_{C} \Omega\left(C\right),
    \label{mcavg}
\end{equation}
where $Z$ is the quantum \textit{partition function} expanded as a sum over Boltzmann \textit{weights}, $\Omega\left(C\right)$, of \textit{configurations}, $C$. It is important to note that generally this expansion is not unique, and thus the notions of configurations and their Boltzmann weights are not unique either. Later we will discuss why some choices can be more preferable for the formulation of a QMC algorithm than others. For a quantum mechanical problem, the weights $\Omega\left(C\right)$ can be written as a product of matrix elements of operators that are not necessarily Hermitian, and these products may be positive, negative, or even complex, though the full sum (the partition function) is guaranteed to be positive. In Section 1.6, we will see some more concrete examples that show how negative $\Omega\left(C\right)$ values can occur.

In the case that $\Omega\left(C\right)\geq 0$ and is computable in polynomial time, QMC can offer a way to compute $\left\langle O\right\rangle$ to a specified precision that scales in polynomial time with the system volume. The method works by allowing us to generate the configurations $C$ according to the probability distribution $\Omega\left(C\right) / Z$, and this distribution is typically sharply spiked with significant contributions coming from only an exponentially small fraction ($\propto \: e^{- {\rm f Vol.}}$ where $\rm Vol.$ is the volume of the system) of $C$ configurations in the full configuration space, which in the case of a quantum problem is typically exponentially large in the number of degrees of freedom. These configurations $C$ are generated by movement through the configuration space:
\begin{equation}
C^{\rm initial} \rightarrow C^{(1)} \rightarrow C^{(2)} \rightarrow ... \rightarrow C^{(k)}\rightarrow ... \: .
\end{equation}
These moves are called updates, and (as we will explain later) so long as the updates satisfy (1) detailed balance and (2) ergodicity, the set of configurations $\left\{C^{(k)}\right\}$ as $k\rightarrow\infty$ will be distributed according to the probability $\Omega\left(C\right) / Z$, regardless of $C^{\rm initial}$ \cite{Gubernatis}.

In practice, we do not have to wait for an infinite number of updates to occur to produce properly distributed $C^{(k)}$ configurations. The minimum number of updates, $a$, which separate two configurations, $C^{(i)}$ and $C^{(i+a)}$, such that making a measurement for a particular observable at one configuration, $O(C^{(i)})$, yields a result that is completely uncorrelated from another measurement of the same observable at the second configuration, $O(C^{(i+a)})$, is known as the \textit{autocorrelation time}, and the maximum autocorrelation time for all observables of interest is known as the \textit{equilibration time} \cite{Gubernatis}. We can assume that once a sufficient number of updates to achieve equilibration have been performed, any new configurations obtained by further updates, $\left\{C^*\right\}$, are distributed according to the probability $P\left(C\right) = \Omega\left(C\right) / Z$ (as long as we continue to generate new configurations in the chain without restarting). It is then possible to compute observables from $M$ of these equilibrated configurations using
\begin{equation}
    \left\langle O \right\rangle \approx \frac{1}{M} \sum_{C^*} O\left(C^*\right).
    \label{avocalc}
\end{equation}
If statistically independent configurations can be obtained quickly enough during the update process, $\left\langle O \right\rangle$ can be determined with the required precision by increasing $M$. If all the $M$ configurations in (\ref{avocalc}) can be assumed to be statistically independent, then the error in the observable is given by
\begin{equation}
    \Delta O = \sqrt{\frac{\left\langle O^2 \right\rangle - \left\langle O\right\rangle^2}{M}}.
    \label{errorex}
\end{equation}
Again, the time to obtain two statistically independent configurations is called the autocorrelation time, with efficient algorithms having small autocorrelation times. Autocorrelation time in general also depends on the observable in question.

Why do detailed balance and ergodicity guarantee the correct distribution of $C^{(k)}$ as $k\rightarrow \infty$? \textit{Detailed balance} or \textit{local balance} was formulated as a concept important to equilibrium physics studies in the early twentieth century \cite{Fowler400}. It is a key component in Metropolis QMC algorithms \cite{met} and is the statement that
\begin{equation}
    \Omega\left(C_1\right) P_{C_1\rightarrow C_2} = \Omega\left(C_2\right) P_{C_2\rightarrow C_1},
    \label{detbal}
\end{equation}
where $P_{C_1\rightarrow C_2}$ is the probability that when configuration $C_1$ is updated, the new configuration is a particular configuration $C_2$, and ergodicity is the guarantee that any configuration $C''$ can be reached from any other configuration $C'$ through a finite series of updates.

If for simplicity we assume a finite number of configurations and define a vector $\vec{\Omega}$ with entries $\Omega_1 = \Omega\left(C_1\right),...,\Omega_{N_c}=\Omega\left(C_{N_c}\right)$, where $N_c$ is the total number of configurations, and a transition probability matrix $P$, where $P_{ij} = P_{C_i\rightarrow C_j}$, then the detailed balance condition (\ref{detbal}) can be written as
\begin{equation}
    \Omega_i P_{ij} = \Omega_j P_{ji},
\end{equation}
where there is no implied summation over repeated indices. Summing over the index $i$ in particular yields
\begin{equation}
\begin{aligned}
    \sum_i \Omega_{i} P_{ij} &= \sum_i \Omega_j P_{ji}= \Omega_j,
    \end{aligned}
    \label{eigenid}
\end{equation}
where the second equality comes from the fact that the sum of the probabilities for $C_i$ transitioning to any $C_j$ must be unity. The relation (\ref{eigenid}) is called the \textit{balance} or \textit{global balance} relation and implies that $\vec{\Omega}$ is a left eigenvector $\vec{\Omega}^\lambda$ of the probability matrix $P$ with eigenvalue $\lambda=1$ (i.e. $\vec{\Omega} = \vec{\Omega}^{\lambda =1}$). In fact, balance alone combined with ergodicity is sufficient to achieve the correct equilibration.

If the matrix $P$ is ergodic, then it can be shown that all other eigenvalues of $P$ have a magnitude less than 1, i.e. $\left|\lambda\right|<1$ for all other eigenvalues $\lambda$ \cite{Gubernatis}. Therefore if we start with an initial set of configuration weights $\vec{\Omega}_0$, which can be written as a linear combination of eigenvectors $\vec{\Omega}^\lambda$ of the matrix $P$ (so $\vec{\Omega}_0 = \sum_\lambda a_\lambda \vec{\Omega}^\lambda$), we see that upon applying $P$ an infinite number of times,
\begin{equation}
    \lim_{k\rightarrow \infty} \sum_i \left(\vec{\Omega}_0\right)_i \left(P^k\right)_{ij} %= \vec{W}^T + \lim_{k\rightarrow \infty} \sum_i {\lambda_i}^k \vec{W_i}^T
    = a_{\lambda=1}\left(\vec{\Omega}\right)_j.
    \label{equilkey}
\end{equation}
Here $\vec{\Omega}$ is the correctly distributed set of weights we wanted, and $a_{\lambda=1}$ is a constant. Because $\vec{\Omega}$ was the only eigenvector with an eigenvalue of 1, it is the only contribution that remains as $k\rightarrow \infty$. The next largest eigenvalue then has the largest effect on how quickly we can obtain an equilibrated configuration. The effect of equilibration is clear from (\ref{equilkey}): no matter which sets of weights the starting configurations are originally distributed according to in ($\vec{\Omega}_0$), after a sufficient number of transitions according to the probabilities $P$, the configurations will eventually be distributed according to $\vec{\Omega}$, the desired weights.

Over time many efficient kinds of algorithms based on the above ideas have been developed for both quantum spin systems and fermionic systems such as worm algorithms \cite{PROKOFEV1998253}, loop cluster algorithms \cite{PhysRevLett.58.86,WIESE1993235}, and hybrid Monte Carlo algorithms \cite{DUANE1987216}. All methods however have to reckon with the fact that in many cases $\Omega\left(C\right)$ is not guaranteed to be positive, which leads to the \textit{sign problem}. As we next explain, a naive way to deal with this problem leads to exponential scaling of the time required to compute observables with a given error.

\section{The Sign Problem}
When $\Omega\left(C\right)$ in (\ref{mcavg}) is not positive for all $C$, which is often the case in quantum mechanical problems and especially those containing fermions, there is no natural probability distribution for constructing the QMC. As we explain below, a naive choice often leads to severe cancellations and the number of QMC samples necessary to obtain a reliable answer scales exponentially with the system size \cite{PhysRevLett.94.170201}. %Fermionic systems often have severe sign problems which stem from the Pauli exclusion principle, which mathematically leads to the anticommutation relations:
%\begin{equation}
 %   \left\{c_i^\dagger, c_j\right\} = \delta_{ij}, \qquad \left\{c^\dagger_i,c^\dagger_j\right\} = \left\{c_i,c_j\right\} = 0.
%\end{equation}
%The following is a proof for the exponentially scaling, as shown in \cite{PhysRevLett.94.170201}. As described previously, the calculation for the observable is

It is clear that if $\Omega\left(C\right)$ in (\ref{mcavg}) is both positive and negative, then $\Omega\left(C\right)/Z$ cannot be a probability distribution. The naive way to remedy this problem consists of introducing a new \textit{sign} observable $s\left(C\right)= {\rm sgn} \left(\Omega\left(C\right)\right)$ and a new quantity $Z'$, defined by
\begin{equation}
Z' = \sum_{C} \left|\Omega\left(C\right)\right|.
\end{equation}
This new quantity can be viewed as a partition function of some other (perhaps unphysical) problem with positive Boltzmann weights. We can then re-express (\ref{mcavg}) in the following way:
\begin{equation}
    \left\langle O \right\rangle=\frac{\sum_C O\left(C\right)s\left(C\right)\left|\Omega\left(C\right)\right|/Z'}{\sum_C s\left(C\right)\left|\Omega\left(C\right)\right|/Z'} = \frac{\left\langle O s \right\rangle'}{\left\langle s\right\rangle '}.
\end{equation}
Here $\left\langle \quad \right\rangle '$ denotes an average using the weights $\left |\Omega\left(C\right)\right|$, which are guaranteed to be positive. Thus we can imagine designing an algorithm that is based on $\left|\Omega(C)\right|/Z'$ as the probability distribution. Using this algorithm $\left\langle O \right\rangle$ is then obtained as a ratio of the quantities $\left\langle O s \right\rangle'$ and $\left\langle  s \right\rangle'$.

While this re-expression is a simple way to always allow for Monte Carlo calculations, it leads to a method that requires computations that scale exponentially with the system size. To see why, recall that a partition function, $Z$, is related to a system's free energy density, $f$, by
\begin{equation}
    Z = e^{-\beta N f},
\end{equation}
where $N$ is the number of spatial sites in the system \cite{pathria}. Thus we can also expect $Z' = e^{-\beta N f'}$ with $f' < f$. This means $\left\langle s \right\rangle' = Z / Z' = e^{-\beta N \Delta f}$, where $\Delta f = f-f'$. Therefore $\left\langle O \right\rangle$ is a ratio of two exponentially small quantities, and can be very noisy. For example, the relative error in the calculation of $\left\langle s\right\rangle'$ using $M$ measurements, (referencing the expression for the absolute error in (\ref{errorex})), is given by
\begin{equation}
    \frac{\Delta s}{\left\langle s \right \rangle'} = \frac{\sqrt{\left(\left\langle s^2 \right\rangle' -\left\langle s \right\rangle'^2\right)/M}}{\left\langle s\right\rangle'} = \frac{\sqrt{1-\left\langle s \right\rangle'^2}}{\sqrt{M}\left\langle s \right\rangle'}\sim \frac{e^{\beta N \Delta f}}{\sqrt{M}}.
    \label{errscale}
\end{equation}
Because $s$ measures the sign of each $\Omega\left(C\right)$, it is clear that $\left\langle s^2 \right\rangle' = 1$. Thus (\ref{errscale}) shows us that the number of measurements necessary for a desired error scales exponentially with the system size. This is the origin of the sign problem.

This way of reweighting then is not the way to simulate a system that has a sign problem. Instead it is necessary to find a way to re-express the original partition function $Z$ as a sum of positive terms only, either by finding a basis where all terms are positive, or by doing some kind of analytic sum first and proving that the new terms are all positive. If $\left\langle O\right \rangle$ can be computed to a given precision in polynomial time, then the sign problem is defined to be solved.

\section{The Lagrangian and Hamiltonian Pictures}
Traditionally there are two major formalisms used for defining and exploring quantum many-body systems: the Lagrangian picture and the Hamiltonian picture. In the following discussion we explain these pictures briefly in order to bring out their main differences and describe the typical computational techniques used within each formalism, focusing on purely fermionic systems on a lattice since these are the systems we will primarily consider in our work. The discussions in the next couple of sections will help us explain what we mean by a ``Hamiltonian lattice field theory,'' as referenced in the title of this thesis.

In the Lagrangian picture, the partition function of a lattice theory is written in terms of an action $S\left(\bar{\psi},\psi\right)$, and a path integral over Grassmann fields $\bar{\psi}_x,\psi_x$, where $x$ labels the spacetime lattice sites. It is given by:
\begin{equation}
    Z = \int \left[ d\bar{\psi} d\psi \right] e^{-S\left(\bar{\psi},\psi\right)}.
    \label{lapart}
\end{equation}
Assuming spacetime is a finite lattice with $V$ sites, the measure of integration is given by $\left[d\bar{\psi} d\psi\right] = d\bar{\psi}_{x_1} d\psi_{x_1} ... d\bar{\psi}_{x_V} d\psi_{x_V}$. The action $S$ is a function of the Grassmann fields and since the time dimension is on equal footing with the space dimension (Euclidean time), the formalism is well-suited for relativistic models in particle physics, although this approach is equally applicable for non-relativistic models as well. More details on the properties of Grassmann numbers and their integrals are given in Appendix A.

By contrast, in the Hamiltonian picture the partition function is written in terms of a Hamiltonian operator $H\left(c^\dagger,c\right)$, which is a function of fermionic creation and annihilation operators $c_i^\dagger$ and $c_i$, and a trace over all states  $\left| n \right\rangle$ for some basis in the corresponding Fock space of the system \cite{negele1988quantum}:
\begin{equation}
    Z = {\rm Tr}\left(e^{-\beta H\left(c^\dagger,c\right)}\right) = \sum_{n} \left\langle n \right| e^{-\beta H\left(c^\dagger,c\right)} \left| n \right\rangle .
    \label{hapart}
\end{equation}
The $i$ subscripts on the creation and annihilation operators label spatial lattice sites only. The constant $\beta$ is the extent in imaginary time and is equal to inverse temperature. In this formalism space is treated differently from time, and while there is a lattice in space, time may be taken to be continuous. Thus the Hamiltonian formalism is often natural for modeling nonrelativistic condensed matter materials that consist of strongly correlated fermions, though it is important to note that even in nonrelativistic models, relativistic physics can emerge at low energies near critical points.

\begin{center}
\begin{figure}[t]
\begin{center}
\includegraphics[width=4.5in]{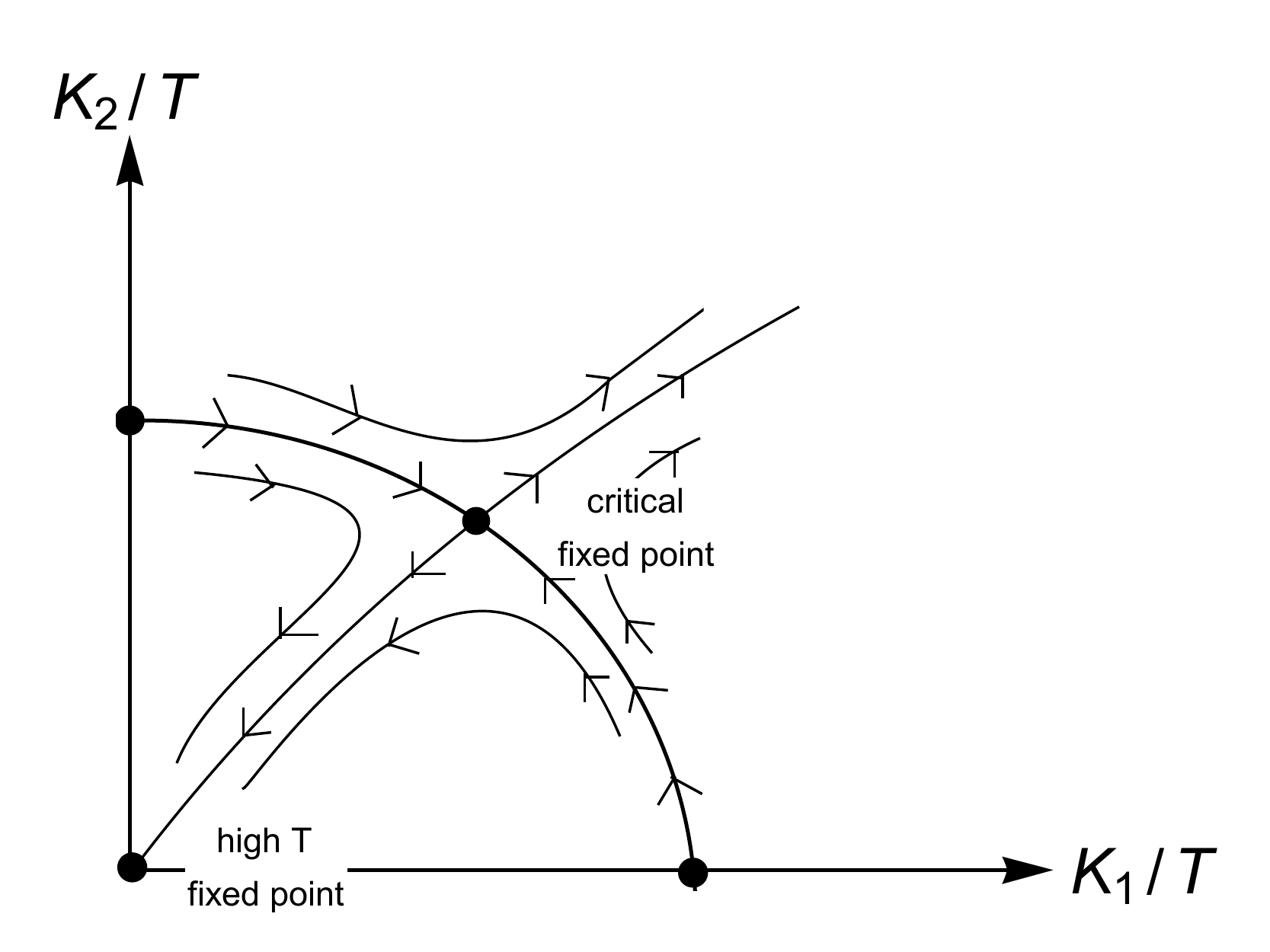}
\end{center}
\caption[The RG flow for a two-dimensional Ising model.]{The RG flow for a two-dimensional Ising model in the $h=0$ plane \cite{Goldenfield}. This is a simplified picture of the RG flow to illustrate how universality works. }
\label{rgdiagram}
\end{figure}
\end{center}

While in principle we can map a problem from the Lagrangian picture into the Hamiltonian picture and vice versa, in practice and especially for theories involving relativistic fermions, the mapping is not easy. Thus the physics community typically studies these two pictures separately and uses what is known as \textit{universality} to relate the two. Universality is a concept whose origin can be traced to the study of renormalization group (RG) flow techniques (see Figure \ref{rgdiagram}). Such techniques, which are based on an insight by L.P. Kadanoff \cite{Kadanoff:1966wm} that was subsequently developed rigorously by K.G. Wilson  \cite{PhysRevB.4.3174,WILSON197475}, show that when a correlation length diverges, especially close to second order critical points, different microscopic theories behave identically at long distances \cite{Goldenfield}. Wilson developed a quantitative understanding of this phenomenon through the notion of RG group flows. He argued that under RG transformations many models will flow to the same fixed point. It is common to refer to RG fixed points in terms of universality classes, which usually depend on the dimension of spacetime, the symmetries, and the field content. 

A simplified picture of an RG flow is given in Figure \ref{rgdiagram}. It is for a two-dimensional Ising model with a nearest neighbor coupling $K_1$ and next-to-nearest neighbor coupling $K_2$. It is placed in an external field $h$.
\begin{equation}
H = K_1 \sum_{\left\langle ij \right\rangle} S_i S_j + K_2 \sum_{\left\langle\left\langle ij \right\rangle\right\rangle} S_i S_j + h\sum_i S_i.\label{rgmodelex}
\end{equation}
The notation $\left\langle ij\right\rangle$ denotes sites $i$ and $j$ that are nearest neighbors, and $\left\langle \left\langle ij\right\rangle \right\rangle$ denotes next to nearest neighbor sites. The flow diagram is in the $h=0$ plane, with all combinations of $K_1$ and $K_2$ couplings. The arrows in the diagram are the flow and represent the results of multiple RG transformations, where degrees of freedom are grouped into a smaller set of degrees of freedom with each transformation. The flow toward the critical fixed point in the center shows that a system described by a next-to-nearest neighbor Hamiltonian ($K_1=0$) is in the same universality class as one described by a nearest neighbor Hamiltonian ($K_2 = 0$). These will both undergo second-order finite-temperature phase transitions with the same critical exponents, which correspond to the critical fixed point. There are a couple of unstable directions: one to a high temperature fixed point and the other out of the $h=0$ plane.

Based on the concept of universality, both the Lagrangian and Hamiltonian formalisms offer complementary ways to explore second order quantum critical behavior. Quantum critical points belonging to models in different formalisms will end up in the same universality class because of the RG flow, despite appearing superficially very different \cite{PhysRevLett.50.1153}. However, when the microscopic symmetries are different the reverse is also true: sometimes similar-looking models will end up in different universality classes. This is an important area of research that is not completely well understood, as we will explain below.

\begin{figure}
    \centering
    \includegraphics[width=12cm]{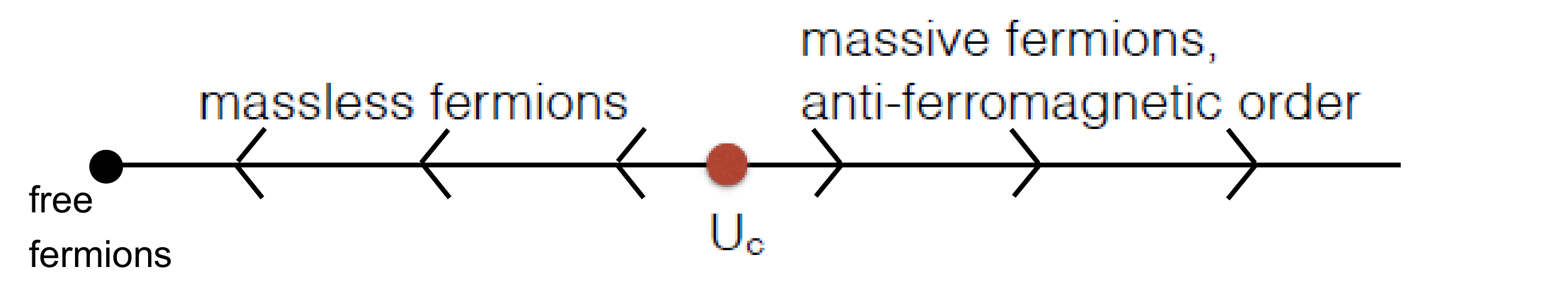}
    \caption[Transition from massless to massive fermions with $SU(2)$ symmetry breaking.]{Transition from massless to massive fermions accompanied by an $SU(2)$ symmetry breaking. The RG flow arrows are given as well.}
    \label{hontran}
\end{figure}

For example, consider the well-known Hubbard model on a honeycomb lattice, whose Hamiltonian is given by
\begin{equation}
    H = -t \sum_{\left\langle ij\right\rangle,\sigma} \left(c_{i,\sigma}^\dagger c_{j,\sigma} + c_{j,\sigma}^\dagger c_{i,\sigma}\right) + U \sum_{i} \left(n_{i,\uparrow} - \frac{1}{2}\right)\left(n_{i,\downarrow} - \frac{1}{2}\right),
    \label{hubbmod}
\end{equation}
where $\left\langle ij\right\rangle$ represents neighboring site pairs as in (\ref{rgmodelex}), but on the two-dimensional honeycomb spatial lattice, and $\sigma=\uparrow,\downarrow$ represents a spin degree of freedom for the fermions. Among other symmetries, the model has an $SU(2)$ spin symmetry that is spontaneously broken for couplings larger than a critical coupling ($U> U_c$). The free part of the Hamiltonian, which is the first term, causes the theory to have two flavors of four-component massless Dirac fermions at small couplings. This can be seen in the low energy limit, where there are two Dirac cones for each spin value with each cone consisting of two zero modes, giving a total of eight zero modes in the Brillouin zone. Thus the model describes an interesting quantum phase transition between a semi-metal with massless fermions and an anti-ferromagnet with massive fermions but massless Goldstone bosons. Close to the critical coupling we expect interesting universal physics that belongs to what is known as the Gross-Neveu chiral Heisenberg universality class (see Figure \ref{hontran}). The critical exponents for this transition have been calculated recently to be $\nu=1.02(1)$ and $\eta=0.49(2)$ using studies on large lattices \cite{sorella}.

Interestingly, it is not easy to study the universal critical properties of the Gross-Neveu chiral Heisenberg universality class in the Lagrangian formulation. Attempts to study it using what is referred to as the staggered fermion approach have appeared in the literature \cite{PhysRevB.79.241405,Armour:2009vj}. For example, one can begin with the following action,
\begin{equation}
    S = \sum_{xy} \bar{\psi}_x M_{xy} \psi_y - U \sum_{\left\langle xy\right\rangle} \bar{\psi}_x \psi_x \bar{\psi}_y \psi_y ,
    \label{xyuniv}
\end{equation}
on a cubic lattice. The $M_{xy}$ matrix is the well-known staggered fermion matrix \cite{Montvay:1994cy} that connects nearest neighbors and ensures the free part of this action produces two flavors of four-component massless Dirac fermions, just as in the Hamiltonian model. The four-fermion coupling is chosen such that when it is larger than $U_c$, again there is a phase transition to a spontaneously broken phase with massive fermions. However, unlike the Hamiltonian model, the Lagrangian model only breaks a $U\left(1\right)$ symmetry rather than an $SU(2)$ symmetry. Thus, the transition most likely belongs to the Gross-Neveu chiral XY universality class. Initial studies had ignored this difference \cite{PhysRevB.79.241405,Armour:2009vj}. Recently the critical exponents at the transition were found to be $\nu = .85(1), \eta=.65(1)$ \cite{PhysRevLett.108.140404}, significantly different from those for the Hamiltonian model. As far as we know, a Lagrangian version of the Hubbard model has still not yet been studied.

%Because some models are defined more naturally in the Hamiltonian formalism than the Lagrangian formalism, and vice versa, it would serve us well to develop efficient ways of studying critical behavior for both formalisms. In this thesis we have made progress specifically in the Hamiltonian formalism: both by expanding the class of models of this type that can be studied, and by contributing to the algorithmic toolbox for studying Hamiltonian systems.
\section{Mapping a Hamiltonian into a Lagrangian}
While the Hamiltonian and Lagrangian formalisms are usually constructed and studied independently, we can in principle begin with the Hamiltonian formalism and construct the Lagrangian approach in a systematic way. However, such an approach breaks the symmetry between space and time, which is important for relativistic applications. Also, the approach is not guaranteed to be free of sign problems, especially in the presence of interactions. For these two reasons this systematic mapping has not traditionally been pursued in lattice field theory. We summarize below how the process works.
%\begin{equation}
 %   H = \sum_{i,n=1,2} \left( c^\dagger_{i} c_{i+\hat{e}_n} + c^\dagger_{i} c_{i-\hat{e}_n}\right),
%\end{equation}

The key for moving into the Lagrangian picture is that the fermionic trace in (\ref{hapart}) can be translated into a Grassmann integral using coherent states \cite{Shankar} (alternatively one can also move in the other direction--from the Lagrangian to the Hamiltonian formalism--using  transfer matrix constructions \cite{PhysRevD.15.1128,lüscher1977}, but we focus on only one direction for simplicity here). We will begin with a generic Hamiltonian $H\left(c^\dagger,c\right)$, whose creation and annihilation operators are in normal-ordered form. That is, all annihilation operators are to the right of all creation operators.

We next discretize time by introducing an $\epsilon$ such that $N\epsilon = \beta$, write the exponentiated Hamiltonian as 
\begin{equation}
e^{-\beta H\left(c^\dagger, c\right)} \approx \prod_{i=1}^N e^{-\epsilon H\left(c^\dagger, c\right)},
\end{equation}
and introduce coherent states, defined as
\begin{equation}
     \left|\psi_t\right\rangle = \prod_{\bigotimes i} \left|\psi_{x=\left\{i,t\right\}}\right\rangle_i , \qquad   \left|\psi_x\right\rangle_i = \left|0\right\rangle_i - \psi_x \left|1\right\rangle_i .
     \label{coherentmultidim}
\end{equation}
This expression for the states is the multi-particle form of the fermionic coherent state introduced in Appendix A. In (\ref{coherentmultidim}), $x$ is a spacetime point, with $i$ representing the spatial coordinate of $x$. The key property of the kets that make up these states is that $c_i \left|\psi_x\right\rangle = \psi_x \left|\psi_x\right\rangle$. We can thus re-express the trace as a Grassmann integral (as in (\ref{gstrace})):
\begin{equation}
   Z= {\rm Tr}\left(e^{-\beta H}\right) = \int \left[d\bar{\psi}_{N_t-1} d\psi_0\right] e^{-\bar{\psi}_{N_t-1} \psi_0} \left\langle -\bar{\psi}_{N_t-1} \right| \prod_{t=1}^{N_t} e^{-\epsilon H\left(c^\dagger,c\right)} \left|\psi_0\right\rangle .
    \label{grasint}
\end{equation} %Introducing a state notation
%\begin{equation}
%    \left|\psi_t\right\rangle = \prod_{\bigotimes i} \left|\psi_{x=\left\{i,t\right\}}\right\rangle_i ,
%\end{equation}
%which is the Kronecker product over the spatial volume of all coherent states for a particular time $t$, 
Here $\bar{\psi}_{N_t-1} \psi_0 = \sum_i \bar{\psi}_{(i,N_t-1)} \psi_{(i,0)}$, the measure $\left[d\bar{\psi}_{N_t -1} d\psi_0\right] = \prod_i d\bar{\psi}_{(i,N_t-1) }d\psi_{(i,0)}$, and $N_t \epsilon = \beta$. The next step is to turn (\ref{grasint}) into a Feynman path integral by inserting the identity $I = \int \left[d\bar{\psi}_{t} d\psi_{t+1}\right] e^{-\bar{\psi}_{t}\psi_{t+1}} \left|\psi_{t+1}\right\rangle \left\langle \bar{\psi}_t\right|$ at each timeslice $t$. Upon simplification we obtain
\begin{equation}
    Z = \int \left[d\bar{\psi} d\psi\right] e^{\sum_{t=0}^{N_t -1} \left(\bar{\psi}_t \psi_t - \bar{\psi}_t \psi_{t+1} -\epsilon H\left(\bar{\psi}_t,\psi_t\right)\right)} = \int \left[d\bar{\psi}d\psi\right] e^{-S}.
    \label{partsh}
\end{equation}

The expression (\ref{partsh}) holds for a general normal-ordered Hamiltonian, $H\left(c^\dagger,c\right)$. To see concretely how this mapping works, we specifically consider the tight-binding Hamiltonian for free fermions hopping on a two-dimensional square lattice, given by
\begin{equation}
     H = \sum_{ij} c_i^\dagger M_{ij} c_j,
     \label{freeham}
\end{equation}
where $M_{ij}$ is a generic Hermitian matrix. Later in this work we will introduce the so-called $\pi$-flux (or staggered fermion) Hamiltonian, given by
\begin{equation}
  \qquad M_{ij} = -\sum_{n=1,2}t\eta_{i,n}\left(\delta_{i+\hat{e}_n,j} + \delta_{i-\hat{e}_n, j}\right).
    \label{hammatrix}
\end{equation}
The $\hat{e}_1,\hat{e}_2$ vectors are orthogonal unit vectors on the square lattice. The factors $\eta_{i,n}$ are defined by $\eta_{i,1}=1$ and $\eta_{i,2} = \left(-1\right)^{i_x+i_y}$, where $i_x$ and $i_y$ are the $x$ and $y$ components of $i$. These factors introduce a $\pi$-flux in each plaquette of the square lattice, corresponding to a constant magnetic field. In the low energy limit, free fermions hopping on a $\pi$-flux lattice produce a single flavor of relativistic four-component Dirac fermions.

Using the specific Hamiltonian in (\ref{hammatrix}) in the expression (\ref{partsh}) we obtain the following expression for the action, $S$:
\begin{equation}
\begin{aligned}
    S &= \sum_{\left(i,t\right),\left(j,t'\right)} \bar{\psi}_{\left(i,t\right)} \tilde{M}_{\left(i,t\right),\left(j,t'\right)} \psi_{\left(j,t'\right)},\\ \tilde{M}_{\left(i,t\right),\left(j,t'\right)} &= \left(\delta_{t+1,t'} - \delta_{t,t'}\right)\delta_{ij} + \epsilon \delta_{t,t'}\sum_{n=1,2}t\eta_{i,n}\left(\delta_{i+\hat{e}_n,j} + \delta_{i-\hat{e}_n,j}\right).
    \end{aligned}
\end{equation}
While $M$ is an $N\times N$ matrix, where $N$ is the number of sites in the spatial lattice, $\tilde{M}$ is a $V\times V$ matrix, where $V$ is the number of spacetime lattice sites. Note that the time coordinate in $\tilde{M}$ is treated differently from the spatial coordinates, even though the theory is expected to be relativistic. As we explain in the next section, the above mapping can also be implemented in the presence of interactions. Thus we see that any Hamiltonian theory on a lattice can be mapped exactly into a Lagrangian lattice field theory in a straightforward way. We refer to the approach where we study Hamiltonian models in a spacetime setting as \textit{Hamiltonian lattice field theory}.

%Thus this type of model with the asymmetry between space and time is still called a Hamiltonian model, as it is defined more naturally in the Hamiltonian picture. When we refer to Lagrangian models, we will mean models whose actions treat the space and time dimensions equivalently.
\section{Auxiliary Field Techniques}
In order to apply QMC techniques to lattice field theories containing fermions, it is important to understand how to deal with Grassmann variables in the path integral expression (\ref{lapart}). When the action or Hamiltonian is ``quadratic,'' the Grassmann path integral or Fock space trace can be computed exactly, as explained in Appendix A. In problems containing four-fermion interactions, such as the ones we will consider in this thesis, the traditional method is to use auxiliary field techniques and convert an interacting problem into a free problem. Some of the largest lattice calculations of critical behavior involving gapless fermions with strong four-fermion couplings have been accomplished using these methods. For example, in \cite{sorella}, Sorella et. al. are able to perform calculations for the Hubbard model (defined in (\ref{hubbmod})) up to 2592 sites.

Both the Lagrangian and Hamiltonian pictures use auxiliary field techniques tailored for the formalisms, and the key in both pictures is to rewrite quartic terms as quadratic terms by inserting an auxiliary field. In the Lagrangian picture for example, if we assume the action is of the form
\begin{equation}
S = \sum_{xy} \bar{\psi}_x M_{xy} \psi_y + \sum_{xy}U_{xy}\bar{\psi}_x \psi_y \bar{\psi}_y \psi_x,
\label{quadaction}
\end{equation}
where $M_{xy}$ is a generic matrix, the quartic terms in the action can be rewritten as
\begin{equation}
    e^{-U_{xy}\bar{\psi}_x \psi_y \bar{\psi}_y \psi_x} = \frac{1}{2} \sum_{\sigma_{xy} = \pm 1} e^{\sigma_{xy} \sqrt{U_{xy}/2}  \left( \bar{\psi}_x \psi_y - \bar{\psi}_y \psi_x\right)},
    \label{coeffi}
\end{equation}
where $\sigma_{xy}$ is an auxiliary field that lives on the bond connecting the sites $x$ and $y$. The partition function in (\ref{lapart}) can then be written as
\begin{equation}
\begin{aligned}
    Z &= \int \left[d\bar{\psi} d\psi\right] e^{-\sum_{xy} \bar{\psi}_x M_{xy} \psi_y -\sum_{xy} U_{xy}\bar{\psi}_x \psi_y \bar{\psi}_y \psi_x}\\
    &=C\int \left[d\bar{\psi} d\psi\right] \sum_{\left\{\sigma\right\}}e^{-\sum_{xy} \bar{\psi}_x M_{xy} \psi_y -\sum_{xy}\sigma_{xy}\sqrt{U_{xy}/2}\left( \bar{\psi}_x \psi_y - \bar{\psi}_y \psi_x\right)}\\
    &=C\sum_{\left\{\sigma\right\}} \int \left[d\bar{\psi} d\psi\right] e^{-\sum_{xy} \bar{\psi}_x {M'}_{xy}\left[\left\{\sigma\right\}\right] \psi_y},
    \end{aligned}
    \label{auxla}
\end{equation}
where each $\left\{\sigma\right\}$ is a combination of $+1$ and $-1$ values, with a value for every $xy$ bond, and
\begin{equation}
    M'\left[\left\{\sigma\right\}\right]_{xy} = M_{xy} + \sigma_{xy}\sqrt{U_{xy}/2} -\sigma_{yx}\sqrt{U_{yx}/2}.
\end{equation}
The constant $C$ factor in the second two lines of (\ref{auxla}) comes from the $1/2$ coefficients that are found in front of the sum in (\ref{coeffi}). Grassmann integrals of quadratic actions lead to determinants as explained in (\ref{gsxpints}) of Appendix A. Thus
\begin{equation}
    Z = C\sum_{\left\{\sigma\right\}} \det M'\left[\left\{\sigma\right\}\right].
    \label{auxaction}
\end{equation}
If $\det\left(M'\left[\left\{\sigma\right\}\right]\right)$ is positive for all $\left\{\sigma\right\}$ configurations, the sign problem is solved and it is possible to generate a QMC algorithm to generate them. However, there is nothing to guarantee the terms in (\ref{auxaction}) will always be positive and here we see a potential manifestation of the sign problem. Each model must be looked at individually to see if there is one. Traditionally, a method known as the Hybrid Monte Carlo (HMC) algorithm has been used for efficent updates, but for massless fermions this algorithm is known to suffer from critical slowing down and is usually not applied. On the other hand the HMC algorithm has been very successful for studies of Quantum Chromodynamics where the ``auxiliary gauge fields'' are known to be smoother.
%\begin{equation}
 %   H = H_0 + H_{\rm int}, \quad H_0 = \sum_{ij} c_i^\dagger M_{ij} c_j, \quad H_{\rm int} =  \sum_{ijkl} V_{ijkl} c^\dagger_i c^\dagger_j c_k c_l + O\left(c^\dagger c\right) .
%\end{equation}

In the Hamiltonian formalism, auxiliary fields can similarly be introduced to convert quartic terms into quadratic terms involving the creation and annihilation operators. For example, consider $H=H_0 + H_{\rm int}$, where $H_0$ is the generic free part given in (\ref{freeham}) and $H_{\rm int}$ is quartic. Again discretizing $\beta$ into $N_t$ steps of size $\epsilon$, we can write (\ref{hapart}) as
\begin{equation}
    Z = {\rm Tr}\left[ \left( e^{-\epsilon H_0}e^{-\epsilon H_{\rm int}} \right)^{N_t}\right]={\rm Tr}\left[ \left( e^{-\epsilon H_0}\textstyle{\prod_{b=1}^{N_V}}e^{-\epsilon  H_{\rm int}\left(b\right)} \right)^{N_t}\right],
    \label{discpart}
\end{equation}
%Again we define $\epsilon$ and $N_t$ so that $N_t \epsilon = \beta$. The second term with the product over the $m$ variable comes from the fact that it will be helpful for the auxiliary field step to separate the interaction piece into local contributions. We assume $N_V$ contributions total.
where we have assumed that the quartic interaction is written as a sum of $N_V$ local terms $H_{\rm int}\left(b\right)$ and the exponent of the sum of these terms is approximated as a product. For example, in the Hubbard model (\ref{hubbmod}) we could define $H_{\rm int}\left(i\right) = U \left(n_{i\uparrow}-1/2\right)\left(n_{i\downarrow}-1/2\right)$. In this case $N_V$ would be the number of sites in the spatial lattice, and the $b$ labels would simply correspond to the sites $i$.

The expression in (\ref{discpart}) would be simple to compute if the trace over the fermionic Hilbert space involved only exponentiated operators that were quadratic, similar to the free part. The quartic terms in the interaction pieces thwart that scheme, however each quartic interaction contribution in the trace can be written in terms of exponentiated quadratic operators if we again introduce an auxiliary field. This type of transformation is known as a Hubbard-Stratonovich transformation. There are many ways to accomplish this, and one example that works for the Hubbard model is \cite{PhysRevB.28.4059}
\begin{equation} 
\begin{aligned}
e^{-\epsilon U \left(n_{i\uparrow}-1/2\right)\left( n_{i\downarrow}-1/2\right)} &\approx  e^{-\epsilon U/4} \sum_{s_i \pm 1} e^{A_{1,\rm int} \left(i,s_i\right)} e^{A_{2, \rm int} \left(i,s_i\right)}\\
A_{1,\rm int}\left(i,s_i\right)  = \left(i s_{i} \sqrt{\epsilon U}+\epsilon U/2\right)n_{i\uparrow}, & \qquad A_{2,\rm int}= \left(i s_{i} \sqrt{\epsilon U} +\epsilon U /2\right)n_{i\downarrow}.
\end{aligned}
\label{hstrans}
\end{equation}

%that if the $H_{\rm int}$ can be rewritten in terms of $H_{\rm int}\left(m\right)$ terms that are each squares of operators $\hat{A}$, the powers of these operators may be reduced in the following way \cite{Bercx:2017pit,doi:10.1143/JPSJ.66.1872}
%\begin{equation}
 %   e^{\epsilon \hat{A}^2} = \sum_{s=\pm1,\pm2} \gamma\left(s\right) e^{\sqrt{\epsilon} \eta\left(s\right) \hat{A}} + O\left(\epsilon^4\right).
%\end{equation}
%The $\gamma$ and $\eta$ constants take on the values
%\begin{equation}
 %   \begin{aligned}
 %   \gamma\left(\pm1\right) &= 1+\sqrt{6}/3, \qquad \eta\left(\pm 1\right) = \pm\sqrt{2(3-\sqrt{6})}\\
 %       \gamma\left(\pm2\right) &= 1-\sqrt{6}/3, \qquad \eta\left(\pm 2\right) = \pm\sqrt{2(3+\sqrt{6})}.
 %   \end{aligned}
%\end{equation}
In general then, the partition function $Z$ may be written as a sum over exponentiated quadratic operators, which are referenced as configurations of the auxiliary field $\left\{s\right\}$ that lives on spacetime lattice sites:
\begin{equation}
    Z = C\sum_{\left\{s\right\}}{\rm Tr}\left[ \left( e^{-\epsilon H_0}\textstyle{\prod_{m=1}^{N_V}}e^{  A_{1,\rm int}\left(m,s\right)} e^{  A_{2,\rm int}\left(m,s\right)} \right)^{N_t}\right],
    \label{sumauxfield}
\end{equation}
where now every factor in the trace is an exponentiated quadratic term. The constant $C$ will vary from one transformation to another. For the example in (\ref{hstrans}) it comes from the factors of $e^{-\epsilon U/4}$. If the operators $A_{n,\rm int}$ are real and two identical operator layers can be identified (as will be explained further in Chapter 3), all terms in the expansion (\ref{sumauxfield}) will be positive. However operators that involve imaginary portions, even if two identical layers are involved, do not have to lead to expansions with only positive terms, and again here we can see the potential manifestations of the sign problem. Indeed the particular operators given in (\ref{hstrans}) are complex, though as we will see in Chapter 3, there is no sign problem for the half-filled Hubbard model.

There are two ways to proceed in the computation of (\ref{sumauxfield}). We could introduce fermionic coherent states and write a Grassmann path integral for the fermionic trace in terms of quadratic action $S\left(\bar{\psi},\psi\right)$,
\begin{equation}
    Z = \sum_{\left\{s\right\}} \int \left[d\bar{\psi} d\psi\right] e^{-\sum_{\left(i,t\right),\left(j,t'\right)} \bar{\psi}_{\left(i,t\right)} \tilde{M}_{\left(i,t\right),\left(j,t'\right)}\left[\left\{s\right\}\right] \psi_{\left(j,t'\right)}},
    \label{partitionzz}
\end{equation}
where $\tilde{M}\left[\left\{s\right\}\right]$ is a $V\times V$ matrix (i.e. spacetime $\times$ spacetime matrix). However, we could also proceed differently as first pointed out by Blankenbecler, Sugar, and Scalapino. If we write $A_{n,\rm int}\left(m,s\right) = \sum_{ij} c^\dagger_i {a_{n,\rm int}\left(m,s\right)}_{ij} c_j$, where $h_0$ and $a_{n,\rm int}\left(m,s\right)$ are $N\times N$ matrices, with $N$ being the spatial volume of the system, it is possible to perform the trace over the Hilbert space using the BSS (Blankenbecler, Sugar, Scalapino) formula \cite{PhysRevD.24.2278}, which we will discuss in more detail in Chapter 4 (it is also possible to reach this formula from the Lagrangian picture). We can then write $Z$ as a sum of $N\times N$ determinants:
\begin{equation}
    Z = C\sum_{\left\{s\right\}}\det\left[\mathbbm{1} +  \left( e^{-\epsilon h_0}\textstyle{\prod_{m=1}^{N_V}}e^{  a_{1,\rm int}\left(m,s\right)} e^{  a_{2,\rm int}\left(m,s\right)} \right)^{N_t}\right].
    \label{auxsum}
\end{equation}
If the determinants for all $\left\{s\right\}$ configurations are positive, then there is no sign problem and again we can develop a QMC algorithm that generates these $\left\{s\right\}$ configurations. The computational time for each update in such an algorithm usually scales as $O\left(\beta N^3\right)$. For QMC studies of critical phenomena in strongly coupled theories containing gapless fermions, the BSS approach seems to be better than the HMC approach using (\ref{partitionzz}).

\section{Recent Fermion Algorithms}
It is important to reiterate that the fermion determinants in (\ref{auxla}), (\ref{partitionzz}), and (\ref{auxsum}) are not guaranteed to be positive. When they are negative or complex, we encounter a sign problem. Even in cases where the determinants are positive, some types of fermion algorithms may be much more efficient than others. Thus, formulating an efficient QMC approach to solve fermion problems remains an exciting area of research.

Over the years, new ways to think about QMC for fermions have emerged. For example in \cite{Chandrasekharan:1999ys} a novel meron cluster solution to the fermion sign problem led to a new class of algorithms. More recently, this idea was adapted into the fermion bag
approach \cite{Chandrasekharan:2013rpa}. There have also been developments of diagrammatic determinantal Monte Carlo methods \cite{Wang:2015rga} based on the CT-INT expansion. In this thesis we will focus on the fermion bag approach since it unifies all these developments. It is based on regrouping degrees of freedom more thoughtfully so that we can solve sign problems while simultaneously making algorithms efficient.

While previous developments of the fermion bag approach were restricted to the Lagrangian picture, in this thesis we extend this approach to the Hamiltonian picture. We will combine insight from the fermion bag ideas of the Lagrangian picture with techniques from the BSS algorithms of the Hamiltonian picture.

%As we will see in the next section, the fermion bag idea can improve the efficiency of algorithms by taking advantage of locality to compute QMC weights more quickly. Its usefulness has been demonstrated in the Lagrangian picture, but its application in the Hamiltonian context had not been developed before this work.

%Our goal is then is to see how fermion bag ideas apply to the Hamiltonian picture and develop them there, just as auxiliary field techniques have been developed for the Hamiltonian picture in addition to the Lagrangian picture. The Hamiltonian picture has the advantage of a better scaling of $O\left(\beta N^3\right)$ instead of $O\left(\beta^3 N^3\right)$, which gives added incentive for developing new methods in this picture. Locality will be key for us as it was a key factor in making efficient fermion bag algorithms in the Lagrangian picture. For example, the matrices of the type $e^{-\epsilon h}$ currently used in the auxiliary field expansion  (\ref{auxsum}) are very nonlocal. One goal will be to use fermion bag techniques to make these matrices local and thus increase the efficiency of the BSS approach. The idea of fermion bags implores us to be careful to avoid matrices that connect the whole spatial volume.
\chapter{The Fermion Bag Idea}
\label{chap:fb}
\small{\textit{ ``You should be happy I'm not dealing with bosons.'' \\
``Yeah, I have to deal with those sorts of people every day at work.''
\\ --Private Communication}}

\section{Lagrangian Approach}

The Fermion Bag idea was originally developed by Shailesh Chandrasekharan in the context of Lagrangian four-fermion models \cite{Chandrasekharan:2009wc}. Key aspects involve writing the partition function in a new way, and grouping the degrees of freedom thoughtfully so that it is possible to sum over them analytically while simultaneously achieving two objectives: (1) avoiding sign problems and (2) developing more efficient algorithms.

These ideas offer an alternative to auxiliary field methods, mentioned in Chapter 1. For example, consider the model (\ref{quadaction}) and its partition function (\ref{auxaction}) that we encountered in the Lagrangian approach,
\begin{equation}
\begin{aligned}
    Z &=C\sum_{\left\{\sigma\right\}} \det \left(M'\left[\left\{\sigma\right\}\right]\right).
    \end{aligned}
\end{equation}
%In the Lagrangian picture, a key identity for the auxiliary fields used in four-fermion models is
%\begin{equation}
 %   e^{-U\bar{\psi}_x \psi_y \bar{\psi}_y \psi_x} = \frac{1}{2} \sum_{\sigma_{xy} = \pm 1} e^{\sigma_{xy} \sqrt{U/2}  \left( \bar{\psi}_x \psi_y - \bar{\psi}_y \psi_x\right)}.
%\end{equation}
%Here, as in Chapter 1, $\bar{\psi}_x$ and $\psi_x$ are Grassmann number scalars, with $x$ labeling a point in spacetime. The properties of Grassmann numbers are given in Appendix A.
%The partition function can then be written as
%\begin{equation}
%\begin{aligned}
 %   Z &= \int \left[d\bar{\psi} d\psi\right] e^{-S_0\left(\bar{\psi},\psi\right)-U\sum_{xy} \bar{\psi}_x \psi_y \bar{\psi}_y \psi_x}\\
  %  &=\int \left[d\bar{\psi} d\psi\right] \sum_{xy}e^{-S_0\left(\bar{\psi},\psi\right)-\sqrt{U/2}\sigma_{xy}\left( \bar{\psi}_x \psi_y - \bar{\psi}_y \psi_x\right)}\\
   % &=\sum_{\sigma} \det \left(M\left[\sigma\right]\right),
%    \end{aligned}
%\end{equation}
Here the configurations for a QMC are defined as combinations of the auxiliary field variables $\left\{\sigma\right\}$. As mentioned previously, there is no guarantee that the terms in this sum will be positive. Even if they are positive, these determinants--which define the Boltzmann weights--are nonlocal despite the fact that interactions are usually local, and this can hinder the ability to create fast updates from one configuration to another.

%the fermion bag technique looks for alternative ways to sum over the degrees of freedom in a system, both to solve sign problems, and to take advantage of locality in order to create more efficient algorithms. For example, the method enabled more efficient calculations for the four-fermion Thirring model \cite{Chandrasekharan:2009wc}, with an expansion given by
Let us explore how fermion bag ideas try to remedy these problems for the action (\ref{quadaction}). For concreteness we take
\begin{equation}
   S = \sum_{xy} \bar{\psi}_x {M}_{xy} \psi_y + U \sum_{\left\langle xy\right\rangle} \bar{\psi}_x \psi_y \bar{\psi}_y \psi_x ,
\end{equation}
where $M$ is the well-known free staggered fermion matrix which is known to be antisymmetric and connects only even sites with odd sites, and $U_{xy} = U > 0$ but connects only nearest neighbors. This is in fact the action for the model referenced in (\ref{xyuniv}). The partition function is then
\begin{equation}
\begin{aligned}
    Z  &= \int \left[d\bar{\psi} d\psi\right] e^{-\sum_{xy}\bar{\psi}_x{M}_{xy}\psi_y-U\sum_{\left\langle xy\right \rangle} \bar{\psi}_x \psi_y \bar{\psi}_y \psi_x}.
    \end{aligned}
    \end{equation}

Now instead of using the auxiliary field method, we expand the interaction,
\begin{equation}
    \begin{aligned}
    Z&= \int \left[d\bar{\psi} d\psi\right]e^{-\sum_{xy}\bar{\psi}_x{M}_{xy}\psi_y} \prod_{\left\langle xy\right\rangle} e^{-U\bar{\psi}_x \psi_y \bar{\psi}_y \psi_x} \\
    &= \int \left[d\bar{\psi} d\psi\right]e^{-\sum_{xy}\bar{\psi}_x{M}_{xy}\psi_y} \prod_{\left\langle xy\right\rangle} \left(1-U\bar{\psi}_x \psi_y \bar{\psi}_y \psi_x\right) \\
     &= \sum_{\left\{n\right\}}\int \left[d\bar{\psi} d\psi\right]e^{-\sum_{xy}\bar{\psi}_x{M}_{xy}\psi_y} \prod_{\left\langle xy\right\rangle} \left(U\bar{\psi}_x \psi_x \bar{\psi}_y \psi_y\right)^{n_{xy}}.
     \label{grassbagpart}
    \end{aligned}
\end{equation}
Here we introduce the notion of dimer configurations $\left\{n\right\}$, given by a set of $n_{xy}\in\left\{0,1\right\}$ for every nearest neighbor combination of $x$ and $y$. Note that due to the Grassmann nature of $\bar{\psi}_x$ and $\psi_x$, the exponentials in the product of the first line have only two terms in their expansion, and so everything in (\ref{grassbagpart}) is exact.

\begin{figure}
    \centering
    \includegraphics[width=6cm]{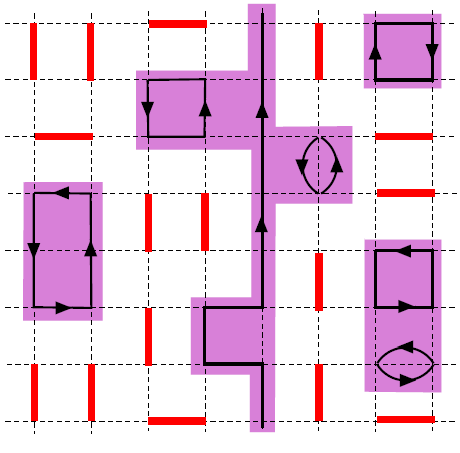}
    \includegraphics[width=6cm]{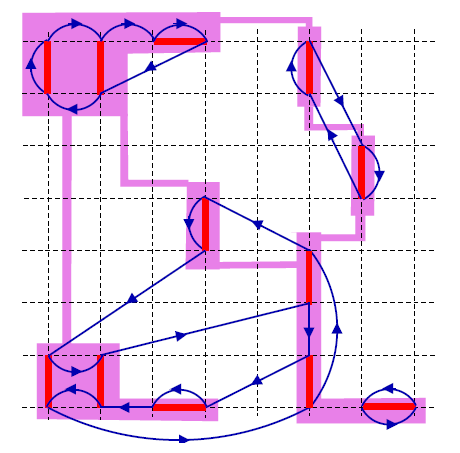}
    \caption[Fermion bags for strong and weak couplings.]{The image to the left illustrates the fermion bag technique for strong coupling, and the image to the right illustrates it for weak coupling. The red lines are the dimers and the purple-shaded regions are fermion bags. The arrows in the diagram to the left are fermionic worldlines and illustrate some of the paths fermions can take from the $S_0\left(\bar{\psi},\psi\right)$ portion of the action.}
    \label{strwk}
\end{figure}

We can now perform the Grassmann integrals over the sites $x,y$ where $n_{xy}=1$ (in other words, sites touched by dimers), each of which simply gives a factor of unity. The remaining Grassmann integrals involve terms coming from $\sum_{xy}\bar{\psi}_x{M}_{xy}\psi_y$, and we can drop all terms corresponding to the sites we have already integrated over. We can thus use a smaller matrix $W$, which is $M$ without the rows and columns corresponding to the sites touched by dimers, to write this modified action with fewer terms: $\sum_{xy}\bar{\psi}_x{W}_{xy}\psi_y$. Now recalling the Grassmann form for a determinant, which is given in (\ref{gsxpints}) of Appendix A,
%\begin{equation}
 %   \int \left[d\bar{\psi}d\psi\right] e^{-\bar{\psi} M \psi} = \det M,
%\end{equation}
we can write the partition function in (\ref{grassbagpart}) as
\begin{equation}
    Z = \sum_{\left\{n\right\}} \det\left(W\left[\left\{n\right\}\right]\right).
\end{equation}
As mentioned before, the $W\left[\left\{n\right\}\right]$ matrices are submatrices of $M$, obtained by dropping the rows and columns corresponding to sites touched by dimers. The left image in Figure (\ref{strwk}) illustrates a dimer configuration. The red lines connecting nearest neighbors are the dimers which correspond to values where $n_{xy}=1$. The remaining sites correspond to the rows and columns that remain in the reduced $W\left[\left\{n\right\}\right]$ matrix, and free fermions can be imagined to be hopping inside these regions, as illustrated by the arrows.
%Because there are regions isolated from each other, we do in fact have several fermion bags, indicated in purple, and so the determinant of $W\left[\left\{n\right\}\right]$ can be rewritten as a product of smaller determinants, one for each bag.

When the coupling $U$ is strong and there are many dimers in the equilibriated configurations, the sites that are in the reduced $W$ matrices may become isolated from other remaining sites, due to the red dimer lines separating them into regions that do not touch, as illustrated by the purple highlighting in the left side of Figure \ref{strwk}. These highlighted regions are the free fermion bags, and in fact the determinant of $W$ can be rewritten as a product of smaller determinants--one for each of the free fermion bags. In this way we have identified some locality in the fermion dynamics which helps to make faster updates in the QMC.

An important point to remember is that the above
identification of fermion bags is not unique.
This is helpful and can be exploited. For example, the above identification of fermion bags is not ideal at weak couplings. When the coupling is weak, the density of dimers is small and the matrix $W$ can become as large as the full spacetime volume. The purple regions illustrated on the left side of Figure \ref{strwk} would grow to connect all sites that were not touched by the red dimer lines. In that case it is better to use Wick's theorem to write:
\begin{equation}
     \int \left[d\bar{\psi}d\psi\right] e^{-\bar{\psi} M_0 \psi} \left(\bar{\psi}_{x_1} \psi_{x_1} \bar{\psi}_{y_1} \psi_{y_1}\right)... \left(\bar{\psi}_{x_k} \psi_{x_k} \bar{\psi}_{y_k} \psi_{y_k}\right)=\det M_0 \det G.
    \label{wkcp}
\end{equation}
Here $G$ is a $2k\times 2k$ propagator matrix with rows and columns corresponding to the sites in the $\bar{\psi}_{x} \psi_{y} \bar{\psi}_{y} \psi_{x}$ terms. The image on the right in Figure \ref{strwk} illustrates this scenario, with the purple highlighting defining a new concept of a \textit{dual} fermion bag in this context. Now fermions ``propagate'' through the $G$ matrix across the spacetime lattice as shown by the arrows in the right side of Figure \ref{strwk}. For weak coupling there are fewer dimers and thus the size of $G$ is small. Ultimately it is the weight ratios that matter for the QMC algorithm, and so the $\det M_0$ factors will be irrelevant and the determinants to calculate will be proportional to the size of the fermion bag. This weak coupling approach is nothing but the determinantal diagrammatic Monte Carlo. The fermion bag ideas help us unify strong and weak couplings into a single umbrella through this notion of dual fermion bags \cite{Chandrasekharan:2009wc}.

At an intermediate coupling, updates can be made faster using a combination of the two pictures in Figure \ref{strwk}. Using a notion of a background configuration with weight $\Omega_B$, subsequent configuration weights $\Omega_1$ and $\Omega_2$ can be obtained from small fluctuations over this background configuration. Ratios such as $\Omega_1/\Omega_B$ and $\Omega_2/\Omega_B$ will be determinants with a size that is proportionate to the number of dimer fluctuations from the background configuration. The update ratio $\Omega_1/\Omega_2$ can then be calculated as
\begin{equation}
    \frac{\Omega_1}{\Omega_2} = {\frac{\Omega_1}{\Omega_B}}/{\frac{\Omega_2}{\Omega_B}}.
    \label{wtbgrat}
\end{equation}
This allows for more efficient updates regardless of the update size.

Since its development, there have been clear successes for the fermion bag technique in the Lagrangian picture. It has been used to so far to study a variety of critical behavior in fermionic systems with exactly massless fermions \cite{PhysRevLett.108.140404,Chandrasekharan:2009wc,PhysRevD.85.091502,PhysRevD.86.021701,PhysRevD.91.065035,PhysRevD.93.081701,PhysRevD.96.114506,Ayyar:2017xmi}, which are very difficult to study with traditional methods. It has also been used to solve a severe sign problem in lattice Yukawa models involving interacting fermions and bosons \cite{PhysRevD.86.021701}. Additionally, by using lattices up to $60^3$, it has been employed to provide the strongest evidence yet that an $SU(4)$ symmetric model involving two flavors of staggered fermions with an onsite four-fermi interaction exhibits an exotic second-order phase transition from massless fermions to massive fermions that is not accompanied by symmetry breaking \cite{PhysRevD.91.065035,PhysRevD.93.081701}.

\section{Hamiltonian Approach}
\begin{figure}
    \centering
    \includegraphics[width=6cm]{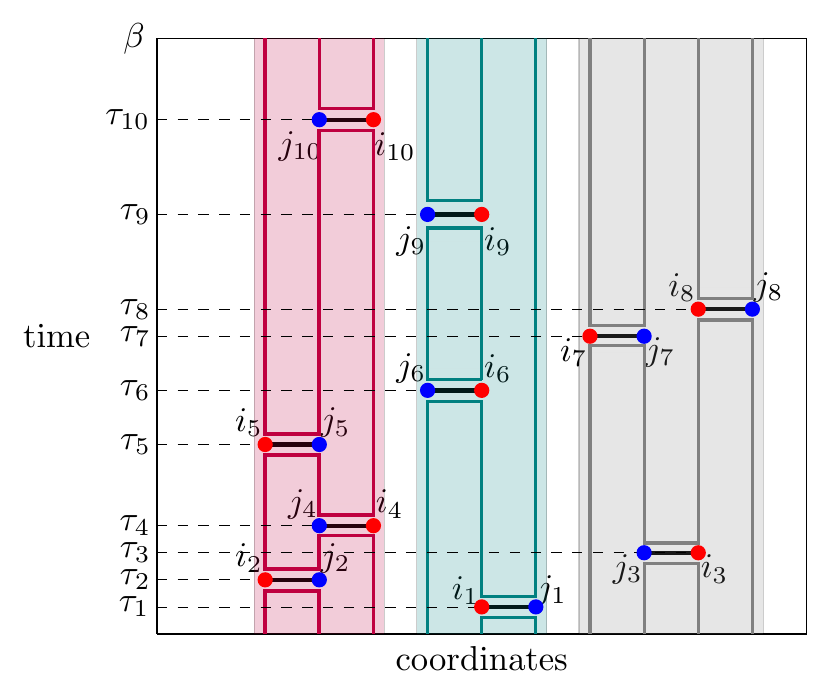}
    \includegraphics[width=6cm]{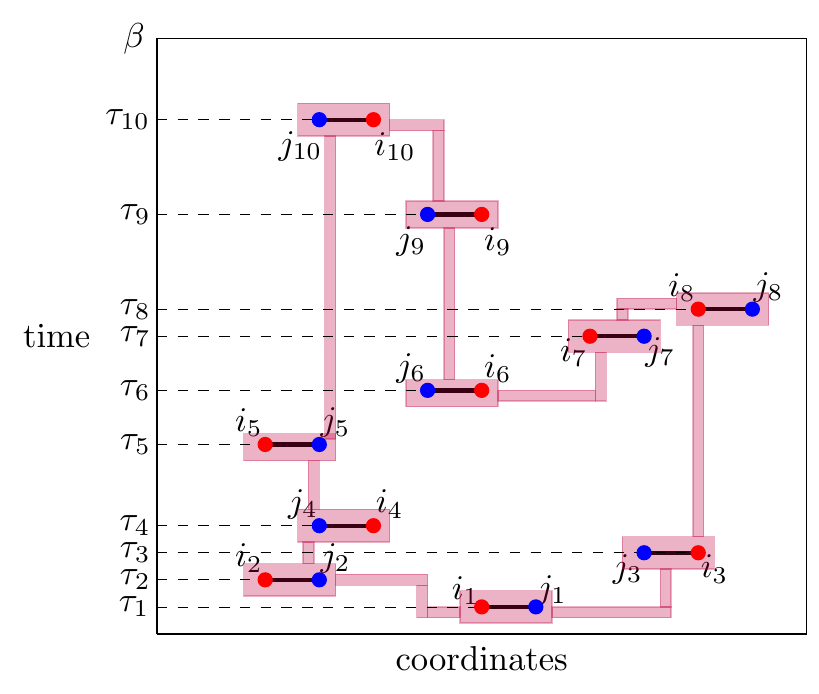}
    \caption[An illustration of fermion bags in the Hamiltonian approach.]{An illustration of fermion bags in the Hamiltonian approach. The left image shows an example where temperature as a parameter is used to give us multiple fermion bags, and the right image shows an example of the analog to the weak coupling expansion in the Lagrangian approach, where all bonds are in the same bag.}
    \label{fig:fbillus}
\end{figure}

%Now that fermion bag ideas in the Lagrangian picture have been described in some detail, we turn our attention to the extension of these ideas to the Hamiltonian picture.
A central goal in this work is to explore if fermion bag ideas discussed above can be useful for studying models in the Hamiltonian formulation. To recall from Chapter \ref{chap:intro}, the partition function (\ref{hapart}) in the Hamiltonian picture is given by
\begin{equation}
    Z = {\rm Tr}\left(e^{-\beta H\left(c^\dagger,c\right)}\right).
\end{equation}
Just as we did in the Lagrangian picture, we can explore alternative ways to expand the partition function compared to the usual auxiliary field approach detailed in Chapter \ref{chap:intro}. As before, we first split $H\left(c^\dagger, c\right)$ into a ``free'' part $H_0$ and an ``interaction'' part $H_{\rm int}$,
\begin{equation}
    H = H_0 + H_{\rm int},\qquad H_0 = \sum_{ij} c^\dagger_i M_{ij} c_j,
\end{equation}
where for the moment the only condition is that $H_0$ should be quadratic in $c^\dagger_i$ and $c_i$. As in the weak coupling Lagrangian picture, we can in principle expand the partition function in the interaction term using the CT-INT expansion \cite{PhysRevLett.81.2514,PhysRevB.72.035122,PhysRevE.74.036701,PhysRevLett.101.090402,PhysRevA.82.053621,RevModPhys.83.349},
\begin{equation}
    Z = \sum_k\int \left[d\tau\right]\left(-1\right)^k{\rm Tr}\left(e^{-\left(\beta-\tau_k\right)H_0} H_{\rm int} ...H_{\rm int} e^{-\left(\tau_2-\tau_1\right) H_0} H_{\rm int} e^{-\tau_1 H_0}\right).
    \label{partctint}
\end{equation}
Here $k$ is the number of $H_{\rm int}$ insertions in each term, and $\int\left[d\tau\right]$ is the time-ordered integral $\int_{\tau_{k-1}}^\beta d\tau_k ... \int_{\tau_2}^{\tau_3} d\tau_2 \int_0^{\tau_2} d\tau_1$. Again as before, the Hamiltonian portion $H_{\rm int}$ can then be broken down further into a sum over smaller local terms $H_{\rm int} = \sum_b H_{\rm int}\left(b\right)$. As we will see in Chapter \ref{chap:example}, such an expansion helped us solve an unsolved sign problem within a class of models. In particular we derived a formula in \cite{PhysRevB.89.111101} that is somewhat analogous to the weak coupling formula (\ref{wkcp}) from the Lagrangian picture. For example, if the interaction was a nearest neighbor interaction (like in the Lagrangian example) given by $H_{\rm int}=\sum_{\left\langle ij \right\rangle} c^\dagger_i c_i c^\dagger_j c_j$, then upon substituting these expressions in for $H_{\rm int}$ in (\ref{partctint}) and expanding, we get that the partition function is a sum of terms that are of the form 
\begin{equation}
     {\rm Tr}\left(e^{-\left(\beta - \tau_k\right)H_0}c^\dagger_{i_k} c_{i_k} c^\dagger_{j_k} c_{j_k} ...e^{-\left(\tau_2 - \tau_1\right)H_0}c^\dagger_{i_1} c_{i_1} c^\dagger_{j_1} c_{j_1}e^{- \tau_1 H_0}\right) ={\rm Tr}\left(e^{-\beta H_0}\right) \times \det G,
    \label{proptr}
\end{equation}
where again $G$ is a $2k\times 2k$ matrix that grows with the number of $c^\dagger c$ pairs. Within a certain class of problems, as we discuss in the next chapter, we can show $\det G \geq 0$. The right side of Figure \ref{fig:fbillus} illustrates how we can view the formula (\ref{proptr}) as a weak coupling fermion bag formulation similar to (\ref{wkcp}), as illustrated in the right side of Figure \ref{strwk}. Formula (\ref{proptr}) can be a useful form for proving properties of the configuration weights, which we will see in Chapter 3, as well as for weak coupling calculations--just as (\ref{wkcp}) was. However, just as (\ref{wkcp}) connects every site belonging to the dimers with every other site belonging to the dimers, so does this expression connect every site belonging to the bonds with every other site belonging to the bonds. No matter how isolated spatially two of the bonds are, they will still remain connected.

For strong couplings we know that the $G$ matrix in (\ref{proptr}) will get very large, so it would be valuable to have some kind of expression in the Hamiltonian picture that works in a way more akin to the left side of Figure \ref{strwk}, where we are able to isolate regions of the diagram with no correlations between these distinct regions. In this way we can have multiple fermion bags and reduce the calculation time if we are only making local changes, because a local change would only affect one fermion bag. In fact the meron cluster approach discovered by Chandrasekharan and Wiese \cite{Chandrasekharan:1999ys} is one such a strong coupling approach and works within a class of models. Unfortunately, it seems quite restrictive and is not yet applicable widely.

The approach we take here is based on the discovery that we can use temperature instead of the coupling as a parameter to help isolate spatial regions into fermion bags in the Hamiltonian approach. To understand this, let us look at a slightly more specific expansion within the (\ref{partctint}) class of CT-INT expansions. In particular if we set $H_0 =0$ and define $H_{\rm int} =H$, we get
\begin{equation}
\begin{aligned}
    Z &= \sum_k\int \left[d\tau\right]\left(-1\right)^k{\rm Tr}\left( H ...H  H\right)\\
    &= \sum_{k,\left\{b\right\}}\int \left[d\tau\right]\left(-1\right)^k{\rm Tr}\left( H_{b_k} ...H_{b_2}  H_{b_1}\right).
    \end{aligned}
    \label{sseeq}
\end{equation}
This expansion in terms of the full Hamiltonian $H$ is related to the stochastic series expansion (SSE), or high temperature approach \cite{PhysRevB.43.5950,PhysRevB.93.155117}. The second equation in (\ref{sseeq}) assumes the Hamiltonian can be rewritten as $H = \sum_b H_b$, where each term involves only local degrees of freedom. Configurations are now defined as $\left[b,\tau\right]$, depending on the locations of the various $H_b$ terms in space-time.

By using (\ref{sseeq}) we can take advantage of the locality of $H_b$ to define fermion bags in this picture. For example we know that $\left[H_{b_1},H_{b_2}\right] = 0$ so long as $H_{b_1}$ and $H_{b_2}$ share no degrees of freedom in common. Thus for the example of $b=\left\langle ij \right\rangle$, where $i$ and $j$ are nearest neighbor sites, we can define a fermion bag as including a group of spatial sites that are connected to each other by the $H_b$ insertions in (\ref{sseeq}), but are not connected to any of the other sites by these insertions. The left side of Figure \ref{fig:fbillus} gives an example of fermion bags for a configuration consisting of ten insertions of local ``bond'' terms, $H_{b_1},...,H_{b_{10}}$, at imaginary times $\tau_1,...,\tau_{10}$. For this example configuration, there are three fermion bags, indicated by the purple, teal, and gray shaded regions. These regions indicate how we have three isolated groups of spatial points that are not connected to the other groups by $H_b$ insertions. Similar to the strong coupling portion of the Lagrangian picture example, it is possible to calculate new weights using determinants that are proportionate in size to the number of sites included in a fermion bag, thus allowing for more efficient updates. Note that at high temperatures on large lattices, it is more likely that we will have many fermion bags. More details on how this could be used in a fermion bag algorithm will be given in Chapters 5 and 6. In Chapters 3 and 4 we will explain why expressions of the type (\ref{proptr}) also lead to positive weights in a class of models.

\chapter{Solutions to Sign Problems in the Hamiltonian Picture}
\label{chap:example}
{\small\textit{``What distinguishes the good theorists from the bad ones[?] The good ones always make an even number of sign errors, and the bad ones always make an odd number.'' \\
--Anthony Zee}}
\newline
\newline

%This chapter is an example of how to format normal material in the
%dissertation style.  Most of this information is standard to \LaTeX.

\section{The $t$-$V$ Model}

We now show how fermion bag ideas for the Hamiltonian approach--as discussed in the previous chapter--led us to a solution to the sign-problem for the $t$-$V$ model, whose Hamiltonian is given by
%While later a simpler way of proving positivity was discovered \cite{Li:2014tla}, this solution as given in \cite{PhysRevB.89.111101} was the first solution for this model, and led to a flurry of Quantum Monte Carlo calculations for its universality class. \cite{Wang:2015vha,Li:2016gte,1367-2630-16-10-103008,1367-2630-17-8-085003,PhysRevB.95.035108,Hesselmann:2016tvh}
\begin{equation}
    H= \sum_{\left\langle ij \right\rangle} -t_{ij}\left(c^\dagger_i c_j + c_j^\dagger c_i\right) + V\sum_{\left\langle ij\right\rangle} \left(n_i-\frac{1}{2}\right)\left(n_j-\frac{1}{2}\right).
    \label{model}
\end{equation}
The lattice is assumed to be bipartite and nearest neighbor notation $\left\langle ij\right\rangle$ is used here, as introduced in (\ref{rgmodelex}). All sites $i$ are assumed to be on one sublattice, and all sites $j$ on the other. This model with $t_{ij}=t$ was considered on small square lattices in the eighties \cite{PhysRevB.29.5253,PhysRevB.32.103}, however the sign problem was ignored. Shailesh Chandrasekharan and Uwe-Jens Wiese later discovered a meron cluster solution to the sign problem for $V\geq 2t$ \cite{Chandrasekharan:1999ys}. By choosing a honeycomb lattice or by choosing $t_{ij} = t \eta_{ij}$ on a square lattice, where $\eta_{ij}$ are phases that introduce a $\pi$-flux (as in (\ref{hammatrix})) on each plaquette, the above t-V model becomes an interesting model of interacting massless Dirac fermions, with the version involving $\pi$-fluxes in particular also referred to as lattice Hamiltonian staggered fermions. The model then undergoes a second-order quantum phase transition at a critical coupling $V=V_c$ between a semimetal phase for small $V$ and an insulator phase for large $V$ with a charge density wave ordering, with the field theory near the critical point expected to be described by a Gross-Neveu model with a single flavor of four-component Dirac fermions \cite{HANDS199329}. Again this can be seen in the low energy limit (as in (\ref{hubbmod})), where for this model there is one Dirac cone in the Brillouin zone that has four zero modes. Due to sign problems, the properties of this critical point had never been studied using Quantum Monte Carlo methods--either in the Lagrangian or the Hamiltonian method. We will give further details on the physics and the studies of this model in Chapter 6.
 %Similar difficulties also arise in models containing an odd number of fermion species that are of interest in the theory of strong interactions.
\section{The Sign Problem in the $t$-$V$ Model}
It is important to note that although the $t$-$V$ model (\ref{model}) is at half-filling, no obvious solution to the sign problem  exists either for $V\leq 0$ or $V\geq 0$. This is in contrast to the half-filled Hubbard model given in (\ref{hubbmod}), which has no sign problem for all values of $U$ on a bipartite lattice. To see this easily, note that for the Hubbard model with $U\leq 0$, the Hubbard-Stratonovich transformation in (\ref{hstrans}) can be implemented in this case and we get
\begin{equation}
\begin{aligned}
    e^{-\epsilon U \left(n_{i\uparrow}-1/2\right) \left(n_{i\downarrow}-1/2\right)} &\approx
    e^{-\epsilon U/4} \sum_{s_i \pm 1} e^{A_{\uparrow, \rm int}\left(i,s_i\right)} e^{A_{\downarrow, \rm int}\left(i,s_i\right)}\\
    A_{\uparrow,\rm int}\left(i,s_i\right) = \left(s_i \sqrt{\epsilon \left|U\right|}+\epsilon U/2\right) n_{i\uparrow} &\qquad A_{\downarrow,\rm int}\left(i,s_i\right) =  \left(s_i \sqrt{\epsilon \left|U\right|} + \epsilon U/2\right) n_{i\downarrow} ,
    \end{aligned}
\end{equation}
where we have written the square root terms using $\left|U\right|$ to make it clear that the action is in fact real when $U\leq 0$. We use the $\uparrow,\downarrow$ subscripts in this case because it makes the separation of the spin layers more obvious. The equation that replaces (\ref{sumauxfield}) is given by
\begin{equation}
\begin{aligned}
    Z &= C\sum_{\left\{s\right\}} {\rm Tr}\left[\left(e^{-\epsilon \left(H_{0,\uparrow} + H_{0,\downarrow}\right)} \textstyle{\prod_{m=1}^{N_V}} e^{A_{\uparrow,\rm int}\left(m,s\right)} e^{A_{\downarrow,\rm int}\left(m,s\right)}\right)^{N_t}\right] \\
    &= C\sum_{\left\{s\right\}} {\rm Tr_\uparrow}\left[\left(e^{-\epsilon H_{0,\uparrow}} \textstyle{\prod_{m=1}^{N_V}} e^{A_{\uparrow,\rm int}\left(m,s\right)} \right)^{N_t}\right]^2,
    \label{spinsquare}
    \end{aligned}
\end{equation}
where the ${\rm Tr}_\uparrow$ signifies a trace over only states in the spin $\uparrow$ subspace. Since the operators in both the spin spaces appear identically, we obtain the square of the trace in one of the spaces. Thus we have a sum of squares of real numbers, and the sign problem is absent. In the case where $U\geq 0$, we can perform a particle-hole transformation on one spin component but not the other:
\begin{equation}
   c_{i\uparrow}^\dagger \rightarrow \sigma_i c_{i\uparrow}, \qquad c_{i\uparrow} \rightarrow \sigma_i c_{i\uparrow}^\dagger \qquad c_{i\downarrow}^\dagger \rightarrow c_{i\downarrow}^\dagger \qquad c_{i\downarrow} \rightarrow c_{i\downarrow}.
   \label{phtran}
\end{equation}
Here $\sigma_i=\pm 1$ is the parity of site $i$ and is defined as $+1$ on one sublattice and $-1$ on the other, so note that we need a bipartite lattice to make this work. This maps $U\rightarrow -U$ and so it is clear that there is no sign problem.

On the other hand, in the $t$-$V$ model there is no analogous particle-hole transformation that would map $V > 0$ into $V < 0$. Moreover, since the interactions involve a single spin component, it is not clear how to get two identical copies of subspace traces in the equation similar to (\ref{sumauxfield}) that lead to a square of a real number. Hence for the $t$-$V$-model no sign problem solution has been found by traditional methods for generic values of $V$. Unfortunately even today the problem remains unsolved by any method for $V\leq 0$. We will discuss below the solution we discovered for $V\geq 0$ using the fermion bag ideas we discussed in Chapter 2.
%In this paper we solve the sign problem for all values of $V > 0$. While most of our discussion will be focused on model (\ref{model}) for concreteness, many of the ideas behind the solution are general and easily extendable to other models within a class. We will mention some of these extensions towards the end.

\section{Solution to the Sign Problem in the $t$-$V$ Model}
We take as a starting point the CT-INT expansion of the partition function, as given in (\ref{partctint}),
\begin{equation}
    Z = \sum_k\int \left[d\tau\right]\left(-1\right)^k{\rm Tr}\left(e^{-\left(\beta-\tau_k\right)H_0} H_{\rm int} ...H_{\rm int} e^{-\left(\tau_2-\tau_1\right) H_0} H_{\rm int} e^{-\tau_1 H_0}\right).
    \label{ctintagain}
\end{equation}
This means we need to divide the $t$-$V$ Hamiltonian (\ref{model}) into a ``free'' part $H_0$ and an ``interaction'' part $H_{\rm int}$. We set those operators for (\ref{model}) in the following way,
\begin{equation}
\begin{aligned}
    H_0 &= -t_{ij}\sum_{\left\langle ij\right\rangle} \left(c_i^\dagger c_j + c_j^\dagger c_i\right), \qquad H_{\rm int} = V\sum_{\left\langle ij \right\rangle} \left(n_i-1/2\right) \left(n_j -1/2\right).
    \end{aligned}
\end{equation}

While there is no particle-hole transformation that maps $V$ to $-V$ for the $t$-$V$ model (\ref{model}), the particle-hole (p-h) symmetry can be observed through the following transformation:
\begin{equation}
    c^\dagger_i \rightarrow \sigma_i c_i, \qquad c_i \rightarrow \sigma_i c^\dagger_i.
    \label{phsymm}
\end{equation}
Here again $\sigma_i$ is the site parity factor defined in (\ref{phtran}). We can write $H$ in a form that makes the symmetry more explicit by writing
\begin{equation}
\begin{aligned}
H_0 &= \sum_{i,j} c^\dagger_i M_{ij} c_j, \\ H_{\rm int} &= \sum_{\left\langle ij\right\rangle}\frac{V}{4} \big(n^+_i - n^-_i) (n^+_j - n^-_j) = \sum_{\left\langle ij\right\rangle}\sum_{s,s'=\pm1}\frac{V}{4} \left(s n_i^{s}\right) \left(s' n_j^{s'}\right) .
\label{model1}
\end{aligned}
\end{equation}
Here $M_{ij}$ is a matrix that connects the nearest neighbors, the operator $n^+_i = n_i = c^\dagger_i c_i$ represents the particle number operator, and $n^-_i = (1-n_i) = c_i c^\dagger_i$ is the hole number operator at site $i$. The p-h symmetry (\ref{phsymm}) for the quadratic part of the Hamiltonian is related to the matrix relation
\begin{equation}
    M^T = -DMD,
    \label{phmatrix}
\end{equation}
where $D_{ij} = \sigma_i \delta_{ij}$ is an $N\times N$ diagonal matrix with the rows/columns corresponding to the spatial lattice sites.

 %We call the free term $H_0$ and the interaction term
%\begin{equation}
%H_{\rm int} = \frac{V}{4}\ \sum_{b,s_i,s_j} (s_i n^{s_i}_i) \ (s_j n^{s_j}_j).
%\label{intchoice}
%\end{equation}
%Here $b = \langle ij\rangle$ labels the bond connecting the nearest neighbor sites $i$ and $j$\textit{,} and $\{s_i,s_j\} = \pm 1$ label the presence of either $n^+$ or $-n^-$ at the sites $i$ and $j$. The free matrix $M$ has the special property that it is real and only connects sites on opposite sublattices. This implies that
%\begin{equation}
%M^T = -D M D
%\label{symm}
%\end{equation}
%where $D_{ij} = \sigma_i \delta_{ij}$ is a %diagonal matrix with elements $\sigma_i = +1$ if $i$ belongs to the even sublattice and $\sigma_i = -1$ if $i$ belongs to the odd sublattice. This property of $M$ will play an important role in the solution to the sign problem.

\begin{figure}
    \centering
    \includegraphics[width=8cm]{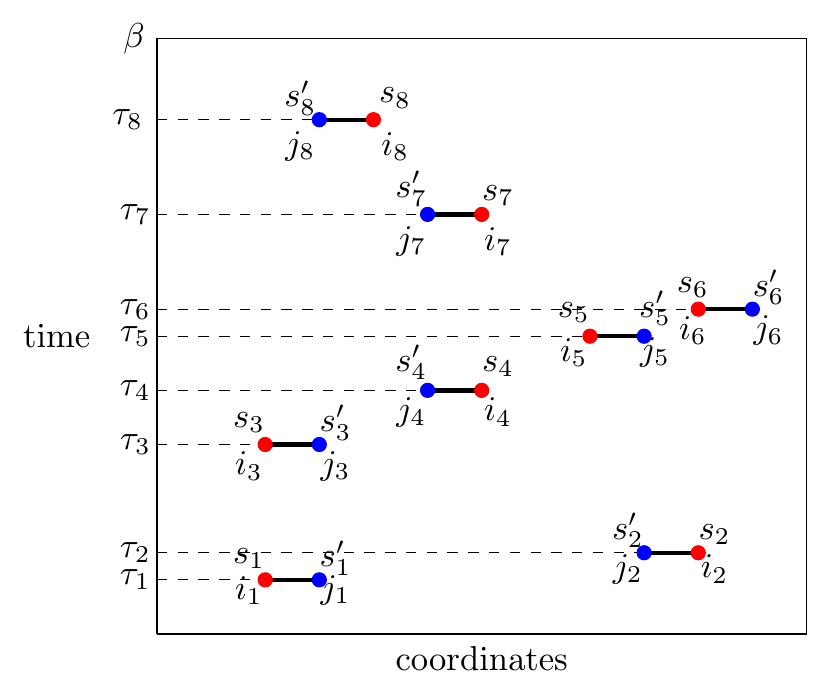}
    \caption[One configuration in the $CT$-$INT$ expansion of the $t$-$V$ model.]{An illustration of a $[b,s,s',\tau]$ configuration. The vertical axis represents continuous imaginary time and the horizontal axis represents the spatial lattice. Each bond has a coordinate $i$ on one sublattice and a coordinate $j$ on the other sublattice, and $s$ and $s'$ that label particle/hole operator insertions.}
    \label{fig:ctintdiag}
\end{figure}

%Instead of the traditional auxiliary field method\textit{,} we use the well known series expansion of the partition function that %are
%\textit{is} used often these days in continuous time Monte Carlo methods \cite{PhysRevLett.81.2514,PhysRevB.72.035122,PhysRevE.74.036701,PhysRevLett.101.090402,PhysRevA.82.053621,RevModPhys.83.349}. 
%The expansion is in powers of interaction vertices and in our model we obtain
Substituting the interactions in the form given by (\ref{model1}), we can now write the CT-INT expansion of (\ref{ctintagain}) as
\begin{eqnarray}
Z \ &=& \ \sum_k \sum_{\left\{b\right\},\left\{s\right\},\left\{s'\right\}}\ \int [d\tau] \big(-V/4\big)^k \nonumber\\
&&\times \mathrm{Tr}\Big(\mathrm{e}^{-(\beta-\tau_k) H_0} (s_k n^{s_k}_{i_k}) ({s_{k}}' n^{{s_{k}}'}_{j_{k}}) ...\mathrm{e}^{-(\tau_{2}-\tau_{1}) H_0} (s_{1} n^{{s_1}}_{i_{1}}) ({s_{1}}' n^{{s_{1}}'}_{j_{1}}) \mathrm{e}^{-\tau_1 H_0} \Big).
\label{ctctint}
\end{eqnarray}
Here $b=\left\langle ij\right\rangle$ is a bond variable introduced to label the nearest neighbor spatial site pairs more easily, $[d\tau]$ represents the $k$ time-ordered integrations from $0$ to $\beta$ over the locations of the interaction bonds, and the notation $[b,s,s',\tau]$ will be used to define a configuration of $k$ interaction bonds located at times $\tau_1 \leq \tau_2 \leq \tau_3,\leq...\leq \tau_{k}$. Each of the $k$ bond interactions contain two insertions of particle/hole operators. Figure \ref{fig:ctintdiag} gives an illustration of what one of these configurations looks like.

%We label these $2k$ interaction vertices with the index $q=1,2...,2k$ such that $i_q$ labels the spatial site of the vertex, $t_q$ labels the temporal location of the vertex\textit{,} and $s_q$ labels the particle-hole operator that is inserted at the vertex. Further since the interactions are bonds, we will naturally have $t_1 = t_2 \geq t_3 = t_4 \geq ...\geq t_{2k-1}= t_{2k}$. Thus%, the integration $[dt]$ involves only $k$ integrations as explained above. 

The task at hand is then to develop an expression that allows us to compute the trace terms in (\ref{ctctint}). We begin with the well-known two-point correlation function for a free particle, given by,
\begin{subequations}
\begin{eqnarray}
\left\langle c_i \left(\tau\right) c^\dagger_j\left(0\right) \right\rangle =\mathrm{Tr}\left(\mathrm{e}^{-(\beta-\tau)H_0} c_{i}  \mathrm{e}^{-\tau H_0} c^\dagger_{j}\right) / {\rm Tr}\left(e^{-\beta H_0}\right) &=& \left(\frac{\mathrm{e}^{-\tau M}}{\mathbbm{1} + \mathrm{e}^{-\beta M}}\right)_{ij} \\
\left\langle c_i^\dagger \left(\tau\right) c_j\left(0\right) \right\rangle =\mathrm{Tr}\left(\mathrm{e}^{-(\beta-\tau)H_0} c^\dagger_{i}  \mathrm{e}^{-\tau H_0} c_{j}\right) /{\rm Tr}\left(e^{-\beta H_0}\right) &=& \left(\frac{\mathrm{e}^{\tau M^T}}{\mathbbm{1} + \mathrm{e}^{\beta M^T}}\right)_{ij}
\end{eqnarray}
\label{correlators}
\end{subequations}
and build up an expression for the full trace using these expressions. For example, suppose we wish to compute the follow object,
\begin{equation}
\begin{aligned}
  \left\langle n^+_i\left(\tau\right) n^-_j \left(0\right) \right\rangle
   & = {\rm Tr}\left(e^{-(\beta-\tau)H_0} n_i^+ e^{-\tau H_0} n_j^- \right) / {\rm Tr}\left(e^{-\beta H_0}\right) \\
   & = {\rm Tr}\left(e^{-(\beta-\tau)H_0} c_i^\dagger c_i  e^{-\tau H_0} c_j c_j^\dagger\right) / {\rm Tr}\left(e^{-\beta H_0}\right).
   \end{aligned}
\end{equation}
Using the formula $e^{\alpha H_0} c_i^\dagger = \sum_q c^\dagger_q e^{\alpha H_0} \left(e^{\alpha M}\right)_{qi}$ as well as the fermionic anticommutation relations, we can commute the $c^\dagger_i$ operator around the trace once and obtain
\begin{equation}
\begin{aligned}
   \left\langle n^+_i\left(\tau\right) n^-_j \left(0\right) \right\rangle &= -\sum_q \left\langle c_q^\dagger c_i\left(\tau\right) n^-_j \left(0\right) \right\rangle \left(e^{-\beta M}\right)_{qi}\\
    &+ {\rm Tr}\left(e^{-(\beta -\tau)H_0} c_i e^{-\tau H_0} c_j^\dagger\right) \left(e^{-(\beta-\tau)M}\right)_{ji}/{\rm Tr}\left(e^{-\beta H_0}\right)\\
    &- {\rm Tr}\left(e^{-\beta H_0}c_j c_j^\dagger\right) \left(e^{-\beta M}\right)_{ii} /{\rm Tr}\left(e^{-\beta H_0}\right).
    \label{inprogress}
    \end{aligned}
\end{equation}
Now by substituting the expressions from (\ref{correlators}) into (\ref{inprogress}) and solving for $\left\langle n^+_i\left(\tau\right) n^-_j \left(0\right) \right\rangle$, we obtain
\begin{equation}
   \left\langle n^+_i\left(\tau\right) n^-_j \left(0\right) \right\rangle =  \Big(\frac{\mathrm{e}^{-\tau M}}{\mathbbm{1} + \mathrm{e}^{-\beta M}}\Big)_{ij} \Big(\frac{\mathrm{e}^{-(\beta -\tau)M}}{\mathbbm{1} + \mathrm{e}^{-\beta M}}\Big)_{ji} -\Big(\frac{\mathbbm{1}}{\mathbbm{1} + \mathrm{e}^{-\beta M}}\Big)_{jj} \Big(\frac{\mathrm{e}^{-\beta M}}{\mathbbm{1} + \mathrm{e}^{-\beta M}}\Big)_{ii}.
   \label{twotwo}
\end{equation}

We can see that this expression for the trace with two bilinear operators resembles the determinant of a $2\times 2$ matrix. Using similar manipulations that we used in the derivation of (\ref{twotwo}), we can derive a general result for a correlator with $k$ insertions of $s n_i^{s} {s}' n_j^{s'}$ (in the bond configuration $\left[b,s,s'\right]$) at arbitrary times $\tau_1,...,\tau_k$ (that label the time configuration $\left[\tau\right]$). This turns out to be a determinant of a $2k \times 2k$ propagator matrix $G([b,s,s',t])$:
\begin{equation}
\begin{aligned}
    &\left\langle s_k n_{i_k}^{s_k}  s_k' n_{j_k}^{s_{k}'}\left(\tau_k\right)...s_1 n_{i_1}^{s_{1}} s_1' n_{j_1}^{s_{1}'}\left(\tau_1\right)\right\rangle\\
   &={\rm Tr}\left(e^{-\left(\beta-\tau_k\right)H_0}s_k n_{i_k}^{s_k} s_k' n_{j_k}^{s_k'} ... e^{-\left(\tau_2-\tau_1\right)H_0}s_k n_{i_1}^{s_k} s_k' n_{j_1}^{s_1'}e^{-\tau_1 H_0}\right)/{\rm Tr}\left(e^{-\beta H_0}\right) \\
   &= \det G\left(\left[b,s,s',\tau\right]\right).
   \label{keyform}
  \end{aligned}
\end{equation}

Every insertion of $n^{s}_i n^{s'}_j$ adds two rows and two columns to the matrix $G\left(\left[b,s,s',\tau\right]\right)$. If we arrange the rows and columns based on the time-ordered list of these insertions, then the off-diagonal elements of $G\left(\left[b,s,s',\tau\right]\right)$ in the upper triangle ($q' > q$), are given by
\small{\begin{equation}
\begin{aligned}
 G_{2q-1,2q'-1}  &= \left(\frac{\mathrm{e}^{-(\tau_q-\tau_{q'}) M}}{\mathbbm{1} + \mathrm{e}^{-\beta M}}\right)_{i_q,i_{q'}}, G_{2q,2q'}  = \left(\frac{\mathrm{e}^{-(\tau_q-\tau_{q'}) M}}{\mathbbm{1} + \mathrm{e}^{-\beta M}}\right)_{j_q,j_{q'}} 
\\
%G_{q'q}([b,t]) \ &=&\ -\ \sigma_{i_q} \ \sigma_{i_{q'}}\ G_{qq'}([b,s,t]).
G_{2q-1,2q'} &= \left(\frac{\mathrm{e}^{-(\tau_q-\tau_{q'}) M}}{\mathbbm{1} + \mathrm{e}^{-\beta M}}\right)_{i_q,j_{q'}}, G_{2q,2q'-1} = \left(\frac{\mathrm{e}^{-(\tau_q-\tau_{q'}) M}}{\mathbbm{1} + \mathrm{e}^{-\beta M}}\right)_{j_q,i_{q'}} .
\end{aligned}
\label{offdiag}
\end{equation}}
Note that $1\leq q\leq k$ and $\tau_q \geq \tau_{q'}$ in the expressions above.

The elements in the lower triangle are very similar in form to those in the upper triangle. The difference is that $\tau_q-\tau_{q'}\rightarrow \beta - \left(\tau_q - \tau_{q'}\right)$, and the elements have an additional overall negative sign:
\begin{equation}
\begin{aligned}
 G_{2q'-1,2q-1}  &= -\left(\frac{\mathrm{e}^{-(\beta-(\tau_{q'}-\tau_{q})) M}}{\mathbbm{1} + \mathrm{e}^{-\beta M}}\right)_{i_{q'},i_{q}}, G_{2q',2q}  = -\left(\frac{\mathrm{e}^{-(\beta-(\tau_{q'}-\tau_{q})) M}}{\mathbbm{1} + \mathrm{e}^{-\beta M}}\right)_{j_{q'},j_{q}} 
\\
%G_{q'q}([b,t]) \ &=&\ -\ \sigma_{i_q} \ \sigma_{i_{q'}}\ G_{qq'}([b,s,t]).
G_{2q'-1,2q} &= -\left(\frac{\mathrm{e}^{-(\beta-(\tau_{q'}-\tau_{q})) M}}{\mathbbm{1} + \mathrm{e}^{-\beta M}}\right)_{i_{q'},j_{q}}, G_{2q',2q-1} =- \left(\frac{\mathrm{e}^{-(\beta-(\tau_{q'}-\tau_{q})) M}}{\mathbbm{1} + \mathrm{e}^{-\beta M}}\right)_{j_{q'},i_{q}} .
\label{offdiag2}
\end{aligned}
\end{equation}
The negative signs in (\ref{offdiag2}) are due to the usual anti-periodic boundary conditions in time that must be introduced when the trace is written as a determinant.

It is important to note that the elements in (\ref{offdiag}) and (\ref{offdiag2}) do not depend on $[s]$ at all, because both $n^+$ and $n^-$ can be treated as the same operator for these off-diagonal matrix elements. Only the diagonal elements distinguish between these two operators. They are given by
\begin{equation}
    G_{2q-1,2q-1}= \frac{s_q}{2}, \qquad  G_{2q,2q} = \frac{s_q'}{2}, 
    \label{gdiag}
\end{equation}
and are only dependent on the $s$ and $s'$ fields and not on their location in time.

The identity in (\ref{phmatrix}) gives us some additional important information about the relationships between the $G$-matrix elements. Observe that
\begin{equation}
\begin{aligned}
    G_{2q-1,2q'-1} &= \left(\frac{\mathrm{e}^{-(\tau_q-\tau_{q'}) M}}{\mathbbm{1} + \mathrm{e}^{-\beta M}}\right)_{i_q,i_{q'}} = \left(\frac{\mathrm{e}^{(\tau_q-\tau_{q'}) DMD}}{\mathbbm{1} + \mathrm{e}^{\beta DMD}}\right)_{i_{q'},i_{q}}\\
    &= \sigma_{i_q} \sigma_{i_{q'}} \left(\frac{\mathrm{e}^{(\tau_q-\tau_{q'}) M}}{\mathbbm{1} + \mathrm{e}^{\beta M}}\right)_{i_{q'},i_{q}}= -\sigma_{i_q} \sigma_{i_{q'}} G_{2q'-1,2q-1},
    \end{aligned}
    \label{grelation}
\end{equation}
where the equality in the first line comes from first replacing $M$ with $M^T$ and flipping the $i_q,i_{q'}$ indices and then replacing $M^T$ with $-DMD$, and the equalities in the second line come from moving the relevant diagonal elements from the $D$ matrices outside the expression and then multiplying the numerator and denominator of the expression by $e^{-\beta M}$.

\begin{figure*}[t]
\begin{center}
\hbox{
\includegraphics[width=0.5\textwidth]{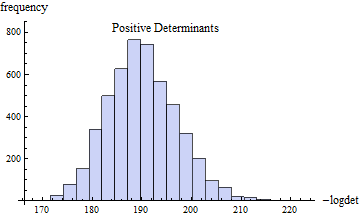}
\hskip0.2in
\includegraphics[width=0.5\textwidth]{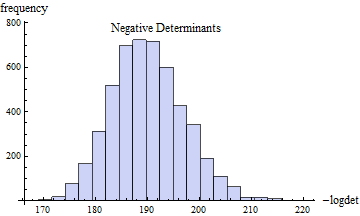}
}
\caption[Positive and negative determinant histograms for $t$-$V$ model.]{ In a sample of 10000 randomly generated configurations $[b,s,s',\tau]$ on an $8 \times 8$ square lattice at $\beta=10$ involving the matrices of size  $250 \times 250$, we obtained $4972$ positive determinants and $5028$ negative determinants. Histograms of positive determinants (left) and negative determinants (right) in this sample are plotted above. The similarity of the two distributions suggests the existence of a severe sign problem.
}
\label{fig1}
\end{center}
\end{figure*}

The partition function is then given as a sum of these $G$ determinants, each dependent on a $\left[b,s,s',\tau\right]$ configuration:
%change t to \tau
%keep t_ij general
%put G alternative expansion in third chapter
%section 1 CT-INT expansion
%alpha and delta need to become omega and delta omega=omega_{ij}
%section 2 alternative expansion with majoranas (we put our expansion in chapter 3)
%section 3 derive the BSS formula for the trace--motivate it
%section 4 derive a formula for the pfaffian--we check both the BSS and the pfaffian

\begin{equation}
Z \ = \ Z_0 \sum_k \sum_{\left\{b\right\},\left\{s\right\},\left\{s'
\right\}}\ \int\ [d\tau]\ (-V/4)^k \  \det G([b,s,s',\tau]).
\label{sse1}
\end{equation}
Here $Z_0 = {\rm Tr}\left(e^{-\beta H_0}\right)$. Unfortunately, this expansion still suffers from a sign problem. We can see this by studying the behavior of $\det G([b,s,s',\tau])$ numerically. On a two dimensional periodic square lattice of length $L=8$ at $\beta = 8$, we have generated $10000$ random $[b,s,s',\tau]$ configurations containing $k=125$ interaction bonds (or equivalently $250$ interaction vertices). We found $4972$ configurations with positive determinants and $5028$ configurations with negative determinants. Figure \ref{fig1} shows the distribution of positive and negative determinants. As expected, the similarity of the two distributions suggests a severe sign problem.

While the representation (\ref{sse1}) of the partition function in terms of configurations $\left[b,s,s',\tau\right]$ has a severe sign problem, there is still a summation that is simple to do analytically. Note that we can write $G([b,s,s',\tau]) = D_0([s,s']) + A([b,\tau])$,  where $D_0([s,s'])$ is the diagonal part defined in (\ref{gdiag}) and $A([b,\tau])$ is the off-diagonal part defined in (\ref{offdiag}) and (\ref{offdiag2}).  Using the Grassmann integral representation for determinants we can then write
\begin{equation}
\begin{aligned}
\sum_{\left\{s\right\},\left\{s'\right\}} \det G[b,s,s',\tau] &= \sum_{\left\{s\right\},\left\{s'\right\}} \int [d\bar{\psi} d\psi] \mathrm{e}^{-\bar{\psi} (D_0([s,s']) + A([b,\tau]))\psi} . 
\end{aligned}
\label{detda}
\end{equation}
The variables $[s,s']$ appear only through diagonal terms, and an analytical sum over all possible $[s,s']$ configurations is possible by expanding in these diagonal terms. Each term in such an expansion can be viewed as having a fermion bag for every $\bar{\psi}_q \psi_q$ pair coming from the $D_0$ matrix portion of the expansion. Viewing the diagonal elements as fermion bags was in fact explored earlier in \cite{Chandrasekharan:2013rpa}.

\begin{figure}
    \centering
    \includegraphics[width=8cm]{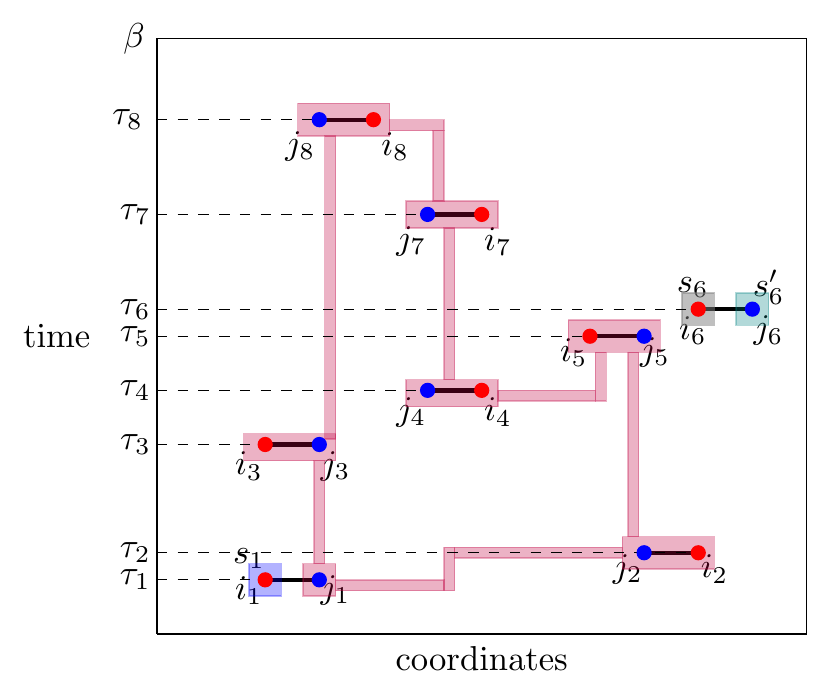}
    \caption[An illustration of fermion bags for the $CT$-$INT$ expansion of the $t$-$V$ model.]{A diagram representing one of the terms in (\ref{bagexpansion}), with the bags highlighted. The three sites belonging to the diagonal portion for this particular combination of $n,n'$ are each in their own bag, as shown by the individual blue, gray, and green square patches. The remaining sites are still connected into a single bag by the $A\left(\left[b,\tau\right]\right)$ matrix. Each diagonal site has an $s$ or $s'$ variable attached to it, which will be summed over.}
    \label{fig:ctintbag}
\end{figure}

To make the above description of these fermion bags more concrete, let us look at (\ref{detda}) expanded in the diagonal terms. Substituting $D_0([s,s'])$ from (\ref{gdiag}), we get
\begin{equation}
\begin{aligned}
    \sum_{\left\{s\right\},\left\{s'\right\}} &\det G[b,s,s',\tau] \\
    &= \sum_{\left\{s\right\},\left\{s'\right\},\left\{n\right\},\left\{n'\right\}} \int [d\bar{\psi} d\psi] \mathrm{e}^{-\bar{\psi} A([b,\tau])\psi} \prod_q \left(\frac{s_{q}}{2} \bar{\psi}_{2q-1} \psi_{2q-1}\right)^{n_q} \left(\frac{s_{q}'}{2} \bar{\psi}_{2q} \psi_{2q}\right)^{n_q'}.
    \end{aligned}
    \label{bagexpansion}
\end{equation}
Here every term in the sum, characterized by $\left[b,s,s',\tau,n,n'\right]$, has two types of fermion bags: the bags coming from the diagonal terms which are attached to an $s$ or $s'$ field, and one remaining bag that connects the other sites to each other through the $A\left(\left[b,\tau\right]\right)$ matrix. Figure \ref{fig:ctintbag} illustrates one of these terms and shows the two types of bags.

Because we want to sum over the $s,s'$ variables, we will rewrite the portion of (\ref{bagexpansion}) that corresponds to the sum over these variables in a form that is easier to sum analytically. Doing so yields
\begin{equation}
\begin{aligned}
\sum_{\left\{s\right\},\left\{s'\right\}}&  e^{-\bar{\psi} D_0([s,s'])\psi}\\
&= \prod_q \sum_{s_{2q-1}=1,-1} \left(1 - \frac{s_{2q-1}}{2} \bar{\psi}_{2q-1} \psi_{2q-1}\right) \sum_{s_{2q'}=1,-1} \left(1 - \frac{s_{2q}'}{2} \bar{\psi}_{2q} \psi_{2q}\right)  = 4^k .
\label{diagsum}
\end{aligned}
\end{equation}
In the current problem, since the variables $s$ and $s'$ multiply the diagonal terms, a sum over $s,s'$ for these bags will lead to a zero weight. This means all terms in the sum (\ref{bagexpansion}) that contain diagonal fermion bags will not contribute to the partition function, as (\ref{diagsum}) illustrates.

Thus $\sum_{\left\{s\right\},\left\{s'\right\}} \det(G[b,s,s',\tau]) = 4^k \det (A([b,\tau]))$. Substituting this result into (\ref{sse1}) we obtain
\begin{equation}
Z = Z_0\sum_{\left\{b\right\}} \int [d\tau] (-V)^k  \det (A([b,\tau])),
\label{sse2}
\end{equation}
where we have now performed the sum over all $[s,s']$ configurations.

The matrix $A([b,\tau])$, which consists of zeroes on the diagonal, and terms defined by (\ref{offdiag}) and (\ref{offdiag2}) on the off-diagonal, has a very nice property that comes from (\ref{grelation}). Because $M^T = -DMD$ (see (\ref{phmatrix})), we can in fact write that
\begin{equation}
    A([b,\tau]) = -\tilde{D} A([b,\tau]) \tilde{D},
\end{equation}
where $\tilde{D}$ is a $2k\times 2k$ diagonal matrix obtained from $D$ by using its entries that correspond to the $2k$ interaction sites rather simply one of each of the $N$ spatial lattice sites, as $D$ contains. The above relation implies that $A([b,\tau])\tilde{D}$ is a real anti-symmetric matrix whose determinant must be positive. But $\det(\tilde{D}) = (-1)^k$ since $k$ sites belong to the even sublattice and $k$ sites belong to the odd sublattice. Thus,
\begin{equation}
\begin{aligned}
 (-1)^k\det(A([b,\tau])) = \det(A([b,\tau])\tilde{D})  \geq 0 .
\label{adtil}
\end{aligned}
\end{equation}
Hence by combining (\ref{sse2}) with (\ref{adtil}) we finally obtain
\begin{equation}
Z  = Z_0\sum_{\left\{b\right\}} \int [d\tau] V^k  \det (A([b,\tau])\tilde{D}),
\label{sgnexpansion}
\end{equation}
which is a sum over positive terms for $V > 0$. Thus, the expansion (\ref{sgnexpansion}) is a solution to the sign problem for the $t$-$V$ model for all $V> 0$.

\section{Alternative $t$-$V$ Model Expansion Solution}
The expansion (\ref{ctctint}) that led to the solution (\ref{sgnexpansion}) for the sign problem contains both local operators such as $H_{\rm int, b=\left\langle ij \right\rangle}$, and nonlocal operators such as $e^{-t H_{0}}$. In discussing the fermion bag ideas in Chapter \ref{chap:fb} however, we saw how locality was an important concept in creating efficient algorithms. As we will show now, it is possible to express the $t$-$V$ model partition function  in terms of only local terms if we set $H_0=0$ and $H_{\rm int} = H$ as discussed in (\ref{sseeq}). For the $t$-$V$ model we set $H=\sum_b H_b$ and obtain
\begin{equation}
\begin{aligned}
    Z 
    &= \sum_{k,\left\{b\right\}} \int \left[d\tau\right] \left(-1\right)^k {\rm Tr}\left(H_{b_k} H_{b_{k-1}} ... H_{b_1}\right).
    \end{aligned}
    \label{ssep}
\end{equation}
Every operator $H_b$ in the expansion (\ref{ssep}) is an operator that depends on sites $\left\langle ij\right\rangle$ that are connected by the bond. Hence additions or subtractions of $H_b$ operators only implement local changes. This expansion of the partition function is related to the Stochastic Series Expansion (SSE) approach, as explored in \cite{PhysRevB.93.155117}.

It turns out that if we define the $t$-$V$ model Hamiltonian (\ref{model}) with an extra additive constant and choose
\begin{equation}
H_{b}= -t_{ij}(c_{i}^\dagger c_{j} + c_{j}^\dagger c_{i}) + V\left(n_{i} - \frac{1}{2}\right)\left(n_{j}-\frac{1}{2}\right) - \frac{{t_{ij}}^2}{V},
\label{hamsquare}
\end{equation}
there is no sign problem in the expansion (\ref{ssep}). The additive constant has no effect on the physics, but key is the fact that (\ref{hamsquare}) allows us to write
\begin{equation}
    H_{b} = -\sum_{\left\langle ij \right\rangle} \omega_{ij} e^{2\alpha_{ij}\left(c_i^\dagger c_j + c_j^\dagger c_i\right)},
    \label{expham}
\end{equation}
where $\omega_{ij}\geq 0$, and $\alpha_{ij}$ are related to $t_{ij}$ and $V$ by
\begin{equation}
    \begin{aligned}
     \omega_{ij} & = \frac{t_{ij}^2}{V}\frac{1}{1-\left(V/2t_{ij}\right)^2}\\
     \sinh2\alpha_{ij} &=\frac{V}{t_{ij}}\frac{1}{1-\left(V/2t_{ij}\right)^2}\\
     \cosh2\alpha_{ij} &=\frac{1+\left(V/2t_{ij}\right)^2}{1-\left(V/2t_{ij}\right)^2}.
    \end{aligned}
    \label{oatv}
\end{equation}
Let us introduce the operator
\begin{equation}
    h_b = 2\alpha_{ij} \left( c_i^\dagger c_j + c_j^\dagger c_i \right) = c^\dagger m_{b} c,
\end{equation}
where $c^\dagger$ and $c$ are vectors of the operators $c_i^\dagger$ and $c_i$, and $m_{b}$ is zero everywhere except for the entries connecting $c^\dagger_i$ with $c_j$, and $c^\dagger_j$ with $c_i$. We can then write the expansion (\ref{ssep}) as
\begin{equation}
    Z = \sum_{k,\left\{b\right\}} \int\left[d\tau\right]\omega_{ij}^k {\rm Tr}\left(e^{h_{b_k}}...e^{h_{b_2}} e^{h_{b_1}}\right).
\end{equation}
Following the discussion in Chapter 1 we can show that ${\rm Tr}\left(e^{h_{b_k}}...e^{h_{b_2}}\right)$ 

$ =$ $\det$ $\left(\mathbbm{1} + e^{m_{b_k}}...e^{m_{b_2}}\right)$, which is nothing but the BSS formula (\ref{auxsum}) applied to the current problem. We will also see in Chapter 4 that there is a simple way to prove this determinant is always positive.

However, it is also possible to prove positivity based on our results we derived above. For example a generalization of the two-point correlation function example in (\ref{correlators}) leads to the result
\begin{equation}
\begin{aligned}
    \frac{{\rm Tr}\left(e^{h_{b_k}}...e^{h_{b_l}}c_i^\dagger c_j ... e^{h_{b_1}}\right)}{{\rm Tr}\left(e^{h_{b_k}}...e^{h_{b_l}}... e^{h_{b_1}}\right)} &= \left( \frac{e^{m_{b_{l-1}}}...e^{m_{b_1}}e^{m_{b_k}}...e^{m_{b_l}}}{\mathbbm{1}+e^{m_{b_{l-1}}}...e^{m_{b_1}}e^{m_{b_k}}...e^{m_{b_l}}} \right)_{ji}\\
    &= \sigma_i \sigma_j \left( \frac{\mathbbm{1}}{\mathbbm{1}+e^{m_{b_{l-1}}}...e^{m_{b_1}}e^{m_{b_k}}...e^{m_{b_l}}} \right)_{ij} \\
    &\equiv\sigma_i \sigma_j B_{ij}.
    \end{aligned}
    \label{sseid2}
\end{equation}
The $\sigma_i$ and $\sigma_j$ are the parity factors for the even sublattice and the odd sublattice, defined earlier, and $B_{ij}$ is defined as the appropriate matrix element in the second line. Also note that the product $e^{m_{b_{l-1}}}...e^{m_{b_1}}e^{m_{b_k}}...e^{m_{b_l}}$ has the $l$-th matrix furthest to the right, since $c^
\dagger_i c_j$ has been placed immediately before that matrix. From the identities in (\ref{sseid2}) it is possible to show that
\begin{equation}
    \frac{{\rm Tr}\left(e^{h_{b_k}}...e^{h_{b_l}} e^{h_{b_{\rm new}}}...e^{h_{b_1}}\right)}{{\rm Tr}\left(e^{h_{b_k}}...e^{h_{b_l}} ...e^{h_{b_1}}\right)} = V\left[\left(B_{ji}-\frac{t_{ij}}{V}\right)^2\right] \geq 0.
    \label{indcase2}
\end{equation}
In other words the ratio of a trace containing a new bond $b_{\rm new}$ to the trace without it is always positive. Note that $i$ and $j$ are the coordinates of bond $b_{\rm new}$. Combining this with the trivial ratio,
\begin{equation}
    \frac{\omega_{ij} {\rm Tr}\left(e^{h_{b_{1}}}\right)}{{\rm Tr}\left(\mathbbm{1}\right)} = \frac{t_{ij}^2}{V}\geq 0,
    \label{trivcase1}
\end{equation}
it is clear that all terms in (\ref{ssep}) must be positive. This result will be useful for our calculations in Chapter 6.

%Section headlines are {\verb \Large } and in the standard font. Compare them
%to subsections below.

%Yes, italics.  You may now dance.  Isn't it funny that upright letters are
%called ``roman'' while slanted letters are ``italic''.  That's like Italian,
%and Romans are Italians too.  What gives?  

%\subsubsection{Smaller and Smaller}

%Subsubsections are allowed, but are not numbered and don't appear in the
%table of contents.  Likewise, you can use the next level of sectioning.

%\paragraph{Paragraphs} These divisions are
%unnumbered and do not appear in the Table of Contents.
%\subparagraph{Subparagraphs} This is the finest division possible.  It's also
%unnumbered and omitted from the Table of Contents.

\section{Straightforward Extensions}

The fermion bag solution for the $t$-$V$ model in Section 3.2 can be extended to some other models in a very straightforward way. There are two simple rules to construct sign-problem-free models using the exact same proof that worked for the $t$-$V$ model. These rules also lay the groundwork for the more complicated extensions we will explore in Section 3.5.
\begin{enumerate}
\item The free term can be modified as long as (\ref{phmatrix}) can be maintained and thus $A([b,\tau])\tilde{D}$ remains real and symmetric.

\item The interaction term can be modified as long as the \textit{staggered reference configuration}, which contains a particle on the even sublattice and a hole on the odd sublattice, is respected. For example, such vertices will contain a particle number operator $n^+_i$ on the even sublattice or a hole number operator $n_i^-$ on the odd sublattice. If a vertex violates the reference configuration, then sign problems will in general be introduced. However, if violations can be introduced in a correlated fashion such that a controlled resummation over positive and negative configurations can be performed easily, then sign problems can again be solved. Couplings of the type in (\ref{model1}) are examples of such correlated couplings that violate the reference configuration yet do not cause sign problems. 
\end{enumerate}

Based on these rules, we see that the nearest neighbor interaction can be generalized to the form
\begin{equation}
H_{\rm int} = \sum_{i,j} V_{ij} \left(n_i - \frac{1}{2}\right) \left(n_j - \frac{1}{2}\right)
\end{equation}
where $V_{ij} \geq 0$ when $i$ and $j$ belong to opposite sublattice and $V_{ij} \leq 0$ when they belong to the same sublattice. It is also possible to introduce a staggered chemical potential term,
\begin{equation}
H_{\rm stagg} = -\sum_i h_i n^{s_i}_i,
\label{stag1}
\end{equation}
where $h_i \geq 0$ and $s_i$ is $+1$ on the even sublattice and $-1$ on the odd sublattice. In the next section, we will make frequent reference to this same term given in a slightly different form,
\begin{equation}
    H_{\rm stagg} = \pm \sum_i h_i \sigma_i \left( n_i-\frac{1}{2}\right).
    \label{stag2}
\end{equation}
The expression in (\ref{stag2}) with the minus sign is the same as the expression in (\ref{stag1}), up to a constant.

In addition to spin-polarized models, our solution extends easily to models with an odd number of interacting fermion species. Consider for example the following $SU(3)$ symmetric {\em attractive} Hubbard-type model involving three species of fermions on a bi-partite lattice whose Hamiltonian is given by
\begin{equation}
H = \sum_{a, \langle ij\rangle} -t \eta_{ij} \left(c^\dagger_{a,i} c_{a,j} + c^\dagger_{a,j} c_{a,i}\right) - 
V\sum_i \left(N_i - \frac{3}{2}\right)^2,
\label{model2}
\end{equation}
where $a=1,2,3$ labels the three species. The operator $N_i = n_{1,i}+n_{2,i} + n_{3,i}$ is the total particle number at the site $i$. A
straightforward extension of the discussion presented above solves the sign problem in this model also. We can see it by writing
\begin{equation}
\left(N - \frac{3}{2}\right)^2 = \frac{1}{2}\sum_{s,s'} s s'\left(n^{s}_1n^{s'}_2 + n^{s}_1n^{s'}_3 + n^{s}_2 n^{s'}_3\right)
\end{equation}
up to an overall constant. The partition function is then expanded as in (\ref{ctintagain}) but now contains a product of three traces, one for each species. Each of these traces is written as a determinant and for the same reasons as discussed in Section 3.2, the sum over the $s$ degree of freedom can be performed to get rid of diagonal terms in the matrices. Thus, only the off-diagonal terms again contribute to each determinant and we finally obtain
\begin{equation}
Z \ = \ \sum_{\left\{b\right\}}\ \int\ [d\tau]\ V^k \left\{\prod_{i=1,2,3} \det (A_i([b,\tau]))\right\}
\end{equation}
We can show that the product of the three determinants is always positive, because while interactions can violate the reference configuration on each layer, the violations always come in pairs on two different layers.

We can also add other interactions without introducing sign problems. For example, the three body interaction of the type
\begin{equation}
H = -\sum_i h_i n^{s_i}_{1,i}n^{s_i}_{2,i}n^{s_i}_{3,i},
\end{equation}
is allowed as long as $h_i \geq 0$ and $s_i$ is $+1$ on even sublattice and $-1$ on the odd sublattice.

\section{Extensions Involving Fermions and Quantum Spins}
Beyond the straightforward extensions that follow directly from the proof in Section 3.2, there are also some more complicated extensions for solving sign problems in models that involve interacting fermions and quantum spins, as we showed in \cite{PhysRevE.94.043311}. These solutions still draw from the fermion bag ideas in the Section 3.2 proof, but also draw from different techniques for the quantum spins, such as worldline pictures \cite{doi:10.1143/JPSJ.73.1379} and meron clusters \cite{Chandrasekharan:1999ys}. The models are diverse, including antiferromagnets, Kondo models, Hubbard-style models on triangular lattices, and gauge theories. The following examples illustrate how the fermion bag approach unifies various algorithmic developments, as mentioned in Section 6 of Chapter 1.

\paragraph{Fermion Bag Ideas with Worldlines:}
In this first subsection we will discuss the simplest types of these fermion-spin solutions, which emerge when we combine fermion bag ideas with worldline pictures. As a first example we can consider coupling the $t$-$V$ model to the transverse field Ising model. The Hamiltonian of the system we consider is given by
%\begin{equation}
%H = -t\sum_{\left\langle ij\right\rangle} \left(c_i^\dagger c_j + c^\dagger_j c_i \right) + V\sum_{\left\langle ij\right\rangle}\left(n_i-\frac{1}{2}\right) \left(n_j - \frac{1}{2}\right) .
%\end{equation}
%The fermion bag method originally showed us that this model for all $V\geq 0$ could be simulated free of sign problems using continuous time Monte Carlo. In addition to the $V$ interaction, we note that we may also add the following staggered chemical potential \cite{Huf14},
%\small\begin{equation}
%H_{\rm stagg} = \sum_i h_i \sigma_i \left(n_i - \frac{1}{2}\right).
%\end{equation}
%\normalsize Here $\sigma_i$ is the parity of a site, giving $+1$ for one sublattice and $-1$ for the other, and $h_i \geq 0$ for all $i$. Whether we use CT-INT, LCT-INT, or the Majorana representation, this staggered chemical potential introduces no sign problems, and the parity factor is necessary to ensure this.

%With this in mind, we next consider several models that couple bosons with the $t-V$ model.
%\subsection{Ising Model coupled with t-V Model}
%We begin by coupling an Ising spin system to the fermionic system:
\begin{equation}
    \begin{aligned}
    H = & -t \sum_{\left\langle ij\right\rangle} \left(c_i^\dagger c_j + c^\dagger_j c_i \right) + V \sum_{\left\langle ij\right\rangle}\left(n_i-\frac{1}{2}\right) \left(n_j - \frac{1}{2}\right) \\
    & - J \sum_{\left\langle ij\right\rangle} S_i^z S_j^z   + \sum_i h_i \left(n_i-\frac{1}{2}\right) S_i^x.
    \end{aligned}
    \label{ising}
\end{equation}
We will assign symbols to each term of the Hamiltonian. The first term on the right hand side is the free fermion term $H_0^f$ and the third term will be referred to as the free boson term $H_0^b$. The second term, which we refer to as $H_{\rm int}^f$, is of course the repulsive interaction between nearest neighbor fermions. The fourth term, referred to as $H_{\rm int}^{fb}$, couples fermions with bosons and mimics a fluctuating transverse field. We assume the remaining couplings $t$, $J$ and $h_i$ are real but arbitrary. 

As we already know from Section 3.4, a staggered chemical potential can be introduced without destroying the positivity of the fermionic determinant,
\begin{equation}
H_{\rm stagg} = \sum_i h_i \sigma_i \left(n_i - \frac{1}{2}\right),
\end{equation}
where $\sigma_i$ again is the parity of a site and $h_i \geq 0$ for all $i$, but in the fermion-spin interaction part of (\ref{ising}), it seems like the fluctuating quantum variable $S_i^x$ would preclude any staggered property. However, using worldlines to describe the quantum spins, we can prove that there is still no sign problem in the following CT-INT expansion,
\begin{equation}
\begin{aligned}
Z = \sum_k  \int & \left[d\tau\right] (-1)^k 
& {\rm Tr}\left(e^{-\left(\beta - \tau_k\right) H_0} H_{\rm int} e^{-\left(\tau_k -\tau_{k-1}\right) H_0} H_{\rm int}...\right),
\label{ctctintint}
\end{aligned}
\end{equation}
where we take $H_0= H_0^f + H_0^b$ and $H_{\rm int} = H_{\rm int}^f + H_{\rm int}^{fb}$. In addition, we rewrite $H_{\rm int}^{fb} = \sum_i h_i \sigma_i \left(n_i-1/2\right) \sigma_i S_i^x$, which is possible because $\sigma_i^2 = 1$. Note that in the expansion (\ref{ctctintint}) we have operators in two different spaces: the fermionic space and the spin space. Since these operators commute with each other, we can factorize the trace in each term of the expansion into a product of two traces: one trace over the spin states containing only operators in the spin space and one trace over the fermionic states containing operators only in the fermionic space. The partition function can then be written in terms of configurations labeled by $[k,b,\tau]$, as:
\begin{equation}
\begin{aligned}
Z = \sum_{l,\{k\},m,\left\{b\right\}}  \int & \left[d\tau\right] \:\:  g_f \left(\left[k,b,\tau\right]\right) \:\: g_s\left(\left[k,\tau\right]\right).
\end{aligned}
\label{ctintf}
\end{equation}
The functions $g_f \left(\left[ k,b,\tau\right]\right)$ and $g_s\left(\left[k,\tau\right]\right)$ are traces over the fermionic and spin spaces, respectively. The fermionic trace function is given by
\begin{equation}
\begin{aligned}
g_f \left(\left[k, b,\tau\right]\right) = (-1)^{m} {\rm Tr_f}\left(...\right. &\sigma_{k_l}\left(n_{k_l} -1 / 2\right)...H_{\rm int}^f(b_m) ... \\&\left. ...H_{\rm int}^f(b_1) ...\sigma_{k_1}\left(n_{k_1} -1 / 2\right), ...\right).
\end{aligned}
\label{ftrace}
\end{equation}
which contains $m$ insertions of the interaction bonds $H_{\rm int}^f(b\equiv \langle ij\rangle) = V\left(n_i - \frac{1}{2}\right) \left(n_j - \frac{1}{2}\right)$ and $l$ insertions of $\sigma_k\left(n_k -\frac{1}{2}\right)$ from the fermion-spin interactions. The presence of the free fermionic propagators $\mathrm{e}^{-\tau H_0^f}$ between these insertions are hidden in the ellipses. As we have learned from Sections 3.2 and 3.4, the expression in (\ref{ftrace}) is positive due to the presence of the $(-1)^m$ sign prefactor and the product $\sigma_{k_1}...\sigma_{k_l}$ associated with the site parities of the fermion-spin interaction pieces.

Next let us examine the spin trace terms,
\begin{equation}
\begin{aligned}
g_s\left(\left[k,\tau\right]\right) & = (-1)^l{\rm Tr_s}\left(e^{-\left(\beta - \tau_l\right)H_0^b}  h_{k_l} \sigma_{k_l} S_{k_l}^x  e^{-\left(\tau_l - \tau_{l-1}\right) H_0^b} h_{k_{l-1}} \sigma_{k_{l-1}}S_{k_{l-1}}^x ... \right. \\
&\qquad \qquad \qquad \qquad \qquad\qquad \qquad \qquad\qquad \qquad \qquad \left. ... h_{k_1} \sigma_{k_1} S_{k_1}^x e^{-\tau_{1} H_0^b}\right).
\end{aligned}
\label{strace}
\end{equation}
Here we have $l$ insertions of the interaction terms $ h_k \sigma_k S_k^x$ at the times $\tau_1,\tau_2,...,\tau_l$. Importantly, note that for each of these times, we have a corresponding insertion of $\sigma_k\left(n_k - \frac{1}{2}\right)$ at $\tau_k$ in the fermionic space. This provides correlations between the two spaces. Also note that the $\sigma_k$ cancels between the two terms as it must to reproduce $H^{fb}_{\rm int}$.

%The insertion of $\sigma_k\left(n_k -\frac{1}{2}\right)$ ensures that the trace in the fermionic space $G_f[l,\{k\},m,\{b\}] \geq 0$. Let us now argue that the trace in the spin space is also positive. We evaluate the trace in the $S^z$ basis by inserting the identity $I = \sum_{s^z} \left|s^z\right\rangle \left\langle s^z \right|$ after every insertion of the $S_k^x$. We get

In order to compute $g_s\left(\left[k,\tau\right]\right)$, we insert the identity $I = \sum_{s^z} \left|s^z\right\rangle \left\langle s^z \right|$ after every instance of $\sigma_k S_k^x$. We get 
\begin{equation}
\begin{aligned}
 g_s\left(\left[k,\tau\right]\right) =&(-1)^l \sum_{\left\{s^z\left(\tau\right)\right\}}\left\langle s^z\left(\tau_0\right)\right| e^{-\left(\beta-\tau_l\right) H_0^b}\left|s^z\left(\tau_l\right)\right\rangle \left\langle s^z\left(\tau_l\right)\right| \sigma_{k_l} S_{k_l}^x ...   \\
& ... e^{-\left(\tau_2 -\tau_{1}\right) H_0^b}  \left|s^z\left(\tau_{1}\right)\right\rangle \left\langle s^z\left(\tau_1\right)\right| \sigma_{k_{1}}S_{k_{1}}^x  e^{-\tau_1 H_0^b} \left|s^z\left(\tau_0\right)\right\rangle ,
\label{worldlinexp}
\end{aligned}
\end{equation}
%\begin{equation}
%\begin{aligned}
 %   &=\sum_{\left\{s^z(\tau),l,\{k\}\right\}} e^{-\beta E\left(s^z\left(\tau\right)\right)},\qquad\qquad\qquad\qquad\qquad\qquad\qquad\qquad \nonumber
  %  \end{aligned}
%\end{equation}
where the sum over $\left\{s^z\left(\tau\right)\right\}$ indicates a sum over all space-time spin configurations that are periodic  i.e., $s^z\left(\tau_0\right) = s^z\left(\tau_l\right)$. These configurations are usually referred to as the worldline representation of spin. Since $S_i^x = \frac{1}{2}\left(S_i^- + S_i^+\right)$, an insertion of $S_i^x$ can be viewed as an operator that flips the spin at the site $i$. On the other hand, because $H_0^b$ is diagonal in the chosen basis, propagators $\mathrm{e}^{-\tau H_0^b}$ are just positive numbers and do not change the state. Thus, the worldlines of spin configurations contain spin flips at times $\tau_1, ... \tau_{l}$.
Because they also need to be periodic at each site $i$, the insertions of $S_i^x$ come in pairs, although they may come at different times. This means the $\sigma_k$ factors in (\ref{worldlinexp}) get cancelled and $l$ must be even. Hence we see that the $g_s\left(\left[k,\tau\right]\right)$ functions defined in (\ref{strace}) must be positive for any value of $h_i$. Combined with the positive expression in (\ref{ftrace}), we see that (\ref{ising}) has no sign problem in the CT-INT formulation, as expanded in (\ref{ctctintint}).

\begin{figure}
\begin{center}
\includegraphics[width=2in]{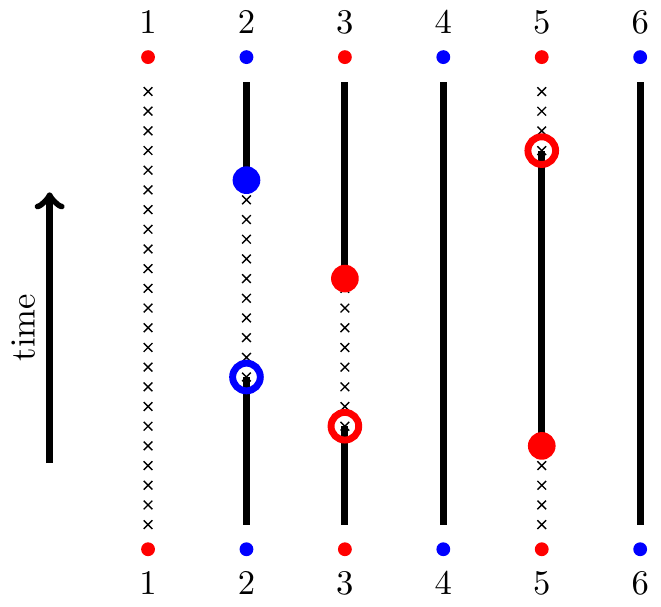}
\end{center}
\caption[Worldline diagram for the spin portion of a configuration for a model involving fermions interacting with Ising spins.]{Worldline diagram for the spin portion $[k,\tau]$ of the configuration $[k,b,\tau]$. Again the vertical axis represents continuous imaginary time, and the horizontal axis represents the spatial lattice, with blue and red dots for even and odd sites. Insertions of $S^x$, represented by the empty and filled circles, create or annihilate hardcore bosons, represented by the thick lines. This is a nonzero weight configuration, so every instance of $S^x_k$ requires a second instance of $S^x_k$ at the same site $k$.}
\label{sx4}
\end{figure}

As an alternate way of viewing this in the worldline picture, we can consider the quantum spins as hardcore bosons (with spin-up represented by particles and spin-down represented by the vacuum), and thus for every creation (annihilation) of a particular particle caused by the $S_i^x$ operator, we require a corresponding annihilation (creation) of the same particle caused by a second $S_i^x$ operator to preserve the trace. Figure \ref{sx4} shows a pictorial illustration of an allowed spin (or equivalently hardcore boson) configuration.

\begin{figure}
\begin{center}
\includegraphics[width=2in]{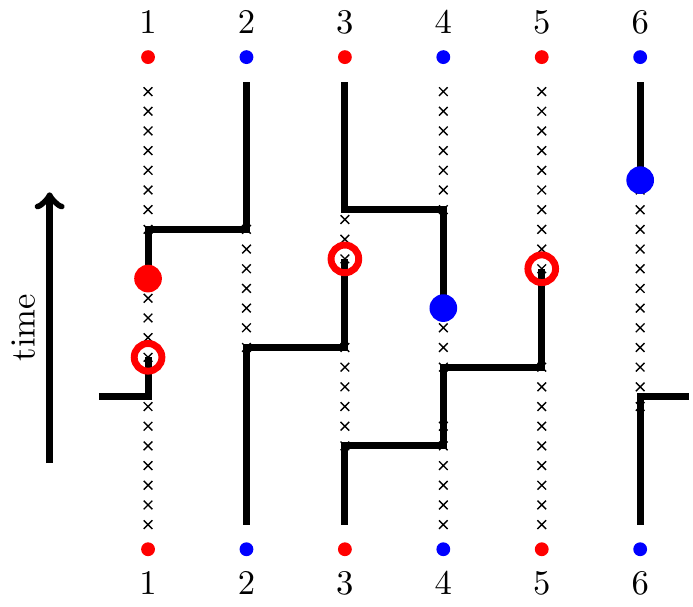}
\end{center}
\caption[Worldline diagram for the spin portion of a configuration for a model involving fermions interacting with Heisenberg spins.]{Worldline diagram for the spin portion $[k,\tau]$ of the Heisenberg fermion-spin model found in (\ref{heisen}). Insertions of $S^x$, represented by the empty and filled circles, create or annihilate hardcore bosons, represented by the thick lines. The Heisenberg interaction then allows the hardcore bosons to hop around, so while there must be an even number of $S^x$ insertions, the operators no longer need to come in pairs for each site.}
\label{heisenwl}
\end{figure}
We can use similar ideas to show that more general models which include antiferromagnetism for the spins are sign-problem-free in CT-INT. For example consider
\begin{equation}
    \begin{aligned}
    H = & -t\sum_{\left\langle ij\right\rangle} \left(c_i^\dagger c_j + c^\dagger_j c_i \right) + V\sum_{\left\langle ij\right\rangle}\left(n_i-\frac{1}{2}\right) \left(n_j - \frac{1}{2}\right) \\
    &+ J\sum_{\left\langle ij\right\rangle} \vec{S}_i \cdot \vec{S}_j + \sum_i h_i \left(n_i-\frac{1}{2}\right) S_i^x.
    \end{aligned}
    \label{heisen}
\end{equation}
We now assume that $J\geq 0$ for anti-ferromagnetism, and also set $h_i\geq 0$ for all $i$ (or equivalently $h_i\leq 0$ for all $i$). Splitting the anti-ferromagnetic piece into a free part $H_0^b=J\sum_{\left\langle ij\right\rangle} S_i^z S_j^z$ and interaction part $H_{\rm int}^b =  J\sum_{\left\langle ij \right\rangle}\left(S_i^+ S_j^- + S_i^- S_j^+\right)$, and following the steps above, we can again show that the interaction pieces $H_{\rm int}^{fb}$ still have to come in pairs for a nonzero configuration weight, but they need not come on the same site. An example worldline picture for the spin portion of a configuration is given in Figure \ref{heisenwl}. We can again show positivity, which is explained in \cite{PhysRevE.94.043311}.

\paragraph{Combining Fermion Bag Ideas with the Meron Cluster Technique:} Fermion bag ideas can also be combined with techniques involving resummation over the spin sector. This is accomplished using the meron cluster \cite{Chandrasekharan:1999ys} technique. For example, consider a model similar to (\ref{heisen}), but with a slightly modified fermion-spin coupling. The Hamiltonian of the model is given by
\begin{equation}
\begin{aligned}
H = & -t\sum_{\left\langle i,j \right\rangle} \left(c^\dagger_i c_j + c^\dagger_j c_i\right) + V\sum_{\left\langle i,j \right\rangle} \left(n_i - \frac{1}{2}\right) \left(n_j - \frac{1}{2}\right) \\
& +J \sum_{\left\langle i,j \right\rangle} \vec{S}_i \cdot \vec{S}_j - h\sum_i \sigma_i \left(S_i^x + \frac{1}{2}\right) \left(n_i - \frac{1}{2}\right).
\end{aligned}
\end{equation}
%It is tempting to split the fermion-spin coupling into two terms:
%\begin{equation}
%\sigma_i \left(S_i^x + \frac{1}{2}\right) \left(n_i - \frac{1}{2}\right)
%= \sigma_i S_i^x \left(n_i - \frac{1}{2}\right) + \frac{\sigma_i}{2} \left(n_i - \frac{1}{2}\right)
%\label{fssplit}
%\end{equation}
%and treat them separately. However, such a treatment leads to sign problems. To see this let us proceed as in the previous example except that we treat the second term on the right hand side of (\ref{fssplit}) as a new interaction that appears in the fermionic trace. Since these are similar to other interactions that already appear within the trace, $G_f$ continues to be positive. On the other hand the spin trace is given by the same equation as (\ref{gsmod2}) but multiplied by an extra factor of $(-1)^n$ due to the fact that we did not performe the unitary transformation that saved us earlier. Every insertion of $H_{\rm int}^b$ from (\ref{spinsplit}) comes with a negative sign. Performing the unitary transformation only pushes the problem into the fermionic sector and does not help.

Using again the concepts of $H_0^f,H_0^b,H_{\rm int}^f$ and $H_{\rm int}^{fb}$, as introduced above (\ref{ctctintint}), this Hamiltonian can be formulated without a sign problem if we first modify the Heisenberg anti-ferromagnetic term by adding to it an irrelevant constant and treat the whole term as a purely bosonic interaction, thus defining a purely spin interaction piece $H_{\rm int}^b$:
\begin{equation}
H_{\rm int}^b \ =\  - J \sum_{\left\langle ij \right\rangle} \ \left(\frac{1}{4}-\vec{S}_i \cdot \vec{S}_j\right).
\label{spinspinmod}
\end{equation}
Since every term containing the quantum spin variable is now treated as an interaction we set $H_0^b = 0$. With these definitions the partition function can again be written in the form (\ref{ctintf}), with the spin trace now given by
\begin{equation}
\begin{aligned}
g_s\left(\left[d,k,\tau\right]\right) ={\rm Tr}\Big( ... h (1/2+S_{k_l}^x)....H^b_{\rm int}(d_n)  ... H^b_{\rm int}(d_1) ... h (1/2+S_{k_1}^x) ...\Big). \\
\end{aligned}
\label{gsmod3}
\end{equation}
The trace depends on the $n$ insertions of nearest neighbor spin hops $H_{\rm int}^b(d\equiv \langle ij\rangle) = J(1/4-\vec{S}_i \cdot \vec{S}_j)$, and $l$ insertions of $h (1/2+S_k^x)$ with free propagators set to unit operators. The configuration is labeled with $[d,k,\tau]$. Note that we do not consider the fermionic trace because we have already discussed that it is positive. It remains to be shown that the $g_s\left(\left[d,k,\tau\right]\right)$ functions are positive. Note that if we insert a complete set $I=\sum_z \left|s\right\rangle \left\langle s \right|$, each worldline configuration can now also be negative. The solution emerges only if we compute the complete sum, which naively also seems like it may be difficult to compute in polynomial time, since it seems to contain an exponentially large number of terms. Fortunately, the same trace was encountered earlier while studying the physics of an anti-ferromagnet in a uniform magnetic field \cite{Chandrasekharan:2001ya}. It was shown that one could compute such a trace in polynomial time using the meron cluster approach, and that it is always positive. While the earlier work used a discrete time method to formulate the trace, it is straightforward to show that it is possible to work directly in continuous time \cite{Bea96}. Further details on how this works may be found in \cite{PhysRevE.94.043311}.

\paragraph{Models with Fermionic Spin:} So far, we have avoided including any fermionic spin degree of freedom. However, the class of sign-problem-free models also includes spinful fermions and even models that can be placed on nonbipartite triangular lattices, such as this model whose fermionic portion resembles the Hubbard model,
\begin{equation}
 \begin{aligned}
    H = & -\alpha\sum_{\left\langle ij\right\rangle} e^{\sum_{\sigma} 2\delta \left(c^\dagger_{i,\sigma} c_{j,\sigma} + c^\dagger_{j,\sigma} c_{i,\sigma}\right)} +\mu \sum_{i} \left(n_{i,\uparrow} + n_{i,\downarrow}\right) \\
    & -J\sum_{\left\langle ij\right\rangle} \left(S_i^x S_j^x + S_i^y S_j^y\right) + U\sum_{i}\left(n_{i,\uparrow}-\frac{1}{2}\right) \left(n_{i,\downarrow}  - \frac{1}{2}\right) S^z_i.
   \end{aligned}
   \end{equation}
Setting the free part $H_0$ to be $\mu \sum_i\left(n_{i,\uparrow} + n_{i,\downarrow}\right)$ and the interaction part $H_{\rm int}$ to include all other terms, we can expand the partition function in the CT-INT formalism without sign problems. The fermionic sector is positive because there are two identical spin layers giving us squares of traces as in (\ref{spinsquare}), and the worldline picture for spins guarantees all spin configurations are positive or zero.

As a more complicated example consider the following model with an $SU(2)$ symmetric interaction. The Hamiltonian of the model is given by
\begin{eqnarray}
H &=& -t\sum_{\left\langle ij \right\rangle, \sigma} \left(c^\dagger_{i,\sigma} c_{j,\sigma} + c^\dagger_{j,\sigma} c_{i,\sigma}\right) + h \sum_i \vec{S}_i \cdot c_i^\dagger \vec{\tau} c_i 
\nonumber \\
&&\qquad \ + \  J\ \sum_{\langle ij\rangle} \vec{S}_i\cdot\vec{S}_j,
\label{su2sym}
\end{eqnarray}
where $\sigma=\uparrow,\downarrow$ are spin degrees of freedom for spins, $\vec{\tau}$ are Pauli matrices in this space, and in the second term, $c_i^\dagger \equiv \left(\begin{array}{cc} c^\dagger_{i,\uparrow} & c_{i,\downarrow}^\dagger\end{array}\right)$. We can rewrite the interaction term between spins and fermions as
\begin{equation}
h\sum_{i} S_i^+ c_{i,\downarrow} c_{i,\uparrow} + h\sum_{i} S_i^- c_{i,\uparrow}^\dagger
 c_{i,\downarrow}^\dagger +h \sum_{i} S_i^z \left(n_{i,\uparrow} + n_{i,\downarrow}\right).
 \label{splitsf}
\end{equation}
where we have performed a particle-hole transformation and a unitary transformation that is given in \cite{PhysRevE.94.043311}. By setting the free part as $H_0 =$$ -t\sum_{\left\langle ij \right\rangle, \sigma}\left(c_{i,\sigma}^\dagger c_{j,\sigma} +\right.$ $\left. c_{j,\sigma}^\dagger c_{i,\sigma}\right) $ $+ h\sum_{i,\sigma} S_i^z n_{i,\sigma}$, the remaining two fermion-spin couplings in (\ref{splitsf}) as two separate pieces, $H_{\rm int}^{fs,a},\: a=1,2$, where $H_{\rm int}^{fs,1}(k)=h S_k^+ c_{k,\downarrow} c_{k,\uparrow}$ and $H_{\rm int}^{fs,2}(k)=hS_i^- c_{k,\uparrow}^\dagger c_{k,\downarrow}^\dagger$, and the last spin-spin interaction term in (\ref{su2sym}) plus a constant as $H_{\rm int}^b$ (defined as in (\ref{spinspinmod})), we can perform another CT-INT expansion
%Using the transformations $c_{i,\downarrow}\rightarrow\sigma_i c^\dagger_{i,\downarrow}$, $c_{i,\downarrow}^\dagger\rightarrow\sigma_i c_{i,\downarrow}$, and (\ref{utran}) we obtain the following transformed Hamiltonian (up to an overall constant):
%\small \begin{equation}
%\begin{aligned}
%H = &-t\sum_{\left\langle ij \right\rangle, \sigma} \left(c^\dagger_{i,\sigma} c_{j,\sigma}+ c^\dagger_{j,\sigma} c_{i,\sigma}\right) + h \sum_{i,\sigma} S_i^z \ n_{i,\sigma} \\
%& +h \sum_{i\in\mathcal{L}} S_i^+ c_{i,\downarrow} c_{i,\uparrow} + h \sum_{i\in\mathcal{L}} S_i^- c_{i,\uparrow}^\dagger c_{i,\downarrow}^\dagger \\
%&  - J\ \sum_{\left\langle ij \right\rangle} \Big(\frac{1}{4}  - S_i^z S_j^z \Big)  - \frac{J}{2} \sum_{\left\langle ij \right\rangle} \left( S_i^+ S_j^- + S_i^- S_j^+\right).
%\end{aligned}
%\label{sumodel}
%\end{equation}
%We treat all the terms on the right hand side of the above equation in the first line as the free fermionic Hamiltonian $H_0^f$. The two terms in the second line are treated as two different fermion-spin couplings $H_{\rm int}^{fs,a},a=1,2$. The terms in the last line are treated as $H_{\rm int}^b$ as in the previous section. Performing the usual CT-INT expansion we obtain,

\begin{equation}
\begin{aligned}
Z = & \sum_{l^1,\left\{k^1\right\},l^2,\left\{k^2\right\},m,\left\{d\right\}}  \int  \left[d\tau\right] \ (-1)^{l^{(1)}+l^{(2)}} 
 {\rm Tr}\Big(.... H_{\rm int}^b(d_m)... \\
&  ... H_{\rm int}^{fs,1}(k^1_{l^{\left(1\right)}})
 ... H_{\rm int}^{fs,2}(k^2_{l^{\left(2\right)}}) ... H_{\rm int}^b(d_1)... H_{\rm int}^{fs,2}(k^2_{1})...\Big),
\label{ctintfinal}
\end{aligned}
\end{equation}
where the ellipses stand for the free fermion propagators. Inside the trace we have $l^{(1)}$ insertions of $H_{\rm int}^{fs,1}(k)$, $l^{(2)}$ insertions of
$H_{\rm int}^{fs,2}(k)$, and $m$ insertions of the bond operator $H_{\rm int}^b(d=\langle ij\rangle) =  J \Big(1/4 -S_i^z S_j^z\Big)  + J/2\left( S_i^+ S_j^- + S_i^- S_j^+\right)$.  Due to spin and fermion number conservation, it is clear we must have $l^{(1)} = l^{(2)}$. These trace terms differ from the previous ones in that they cannot be factored into products each consisting of a trace over the fermion space and a trace over the spin space. However if we evaluate the spin trace in the $S^z$ basis, then it is still clear that the spin configurations add no additional sign to (\ref{ctintfinal}). After evaluation of the spin trace, we are left with the following
\begin{equation}
\begin{aligned}
Z = \sum_{l^1,\left\{k^1\right\},l^2,\left\{k^2\right\},m,\left\{d\right\}}  \int & \left[d\tau\right] \  (J/2)^m(-h)^{l^{(1)}+l^{(2)}}
\sum_{\left\{s_i(\tau)\right\}} {\rm Tr}_f\Big(.... c_{k^1_1,\downarrow}c_{k^1_1,\uparrow} \\
& ... c_{k^2_1,\uparrow}^\dagger c_{k^2_1,\downarrow}^\dagger
 ...  c_{k^2_{l^{(2)}},\uparrow}^\dagger c_{k^2_{l^{(2)}},\downarrow}^\dagger... c_{k^1_{l^{(1)}},\uparrow}c_{k^1_{l^{(1)}},\downarrow}...\Big),
\end{aligned}
\end{equation}
where the spin trace appears now as a sum over $\left\{s_i(\tau)\right\}$ bosonic fields, while the fermion trace still appears in the expression. Since the fermion spins do not mix with each other and appear symmetrically, the fermion trace factors into two identical terms: one is a trace over the spin up space and the other over the spin down space. Each of these can be expressed as a determinant of a matrix $M\left([k^1,k^2,d,\tau,s_i(\tau)]\right)$ that depends on the spin configuration. The exact expression for $M\left([k^1,k^2,d,\tau,s_i(\tau)]\right)$ can be obtained using the usual Wick's theorem \cite{Negele:1988vy}. Thus, we finally obtain the expression
\begin{equation}
Z = \sum_{l^1,\left\{k^1\right\},l^2,\left\{k^2\right\},m,\left\{d\right\}}   \int  \left[d\tau\right]  (J/2)^m (h)^{l^{(1)}+l^{(2)}}  \sum_{\left\{s_i(\tau)\right\}} \left(\det M\left([k^1,k^2,d,\tau,s_i(\tau)]\right)\right)^2,
\end{equation}
and the sign problem is solved.

Closely related to the above model is the well known Kondo-lattice model at half filling. The Hamiltonian of this model is given by
\begin{eqnarray}
H &=& -t\sum_{\left\langle ij \right\rangle, \sigma} \left(c^\dagger_{i,\sigma} c_{j,\sigma} + c^\dagger_{j,\sigma} c_{i,\sigma}\right) + h \sum_{i\in {\cal L}} \vec{S}_i \cdot c_i^\dagger \vec{\tau} c_i 
\end{eqnarray}
where the fermions interact with a lattice of impurities located at the sites $i\in{\cal L}$. While this problem is also solvable with the usual auxiliary field Monte Carlo method \cite{PhysRevLett.83.796}, alternative approaches shed new light on the problem and could help find a solution to the difficult sign problem that exists away from half filling, where the Kondo lattice model is considered as the microscopic model to understand heavy fermion systems \cite{PhysRevLett.57.877}.

\paragraph{Gauge Theories interacting with Fermions:} By combining fermion bags with the worldline picture, we can even study $Z_2$ gauge theories interacting with matter fields, such as the following model,
\begin{equation}
\begin{aligned}
H  =& -t \sum_{\left\langle ij\right\rangle}\left( c_i^\dagger \sigma_{ij}^3 c_j + c_j^\dagger \sigma_{ij}^3 c_i \right)+ \sum_{{\rm plaquettes}} \sigma_a^3 \sigma_b^3 \sigma_c^3 \sigma_d^3\\ &\pm V\sum_{\left\langle ij\right\rangle} \left(n_i - 1/2\right) \left(n_j-1/2\right)  \sigma_{ij}^1 .
\end{aligned}
\end{equation}
Some models of this type have been recently found to containing interesting physics \cite{PhysRevX.6.041049,Gazit2017}. Further details on this solution may be found in \cite{Huffman:2016cgh}.

\chapter{The Majorana Approach}
\label{hello}

{\small\textit{``There is always another way to say the same thing that doesn't look at all like the way you said it before.'' \\
--Richard P. Feynman}}
\newline
\newline
\section{Majorana Solution for the $t$-$V$ Model}
\label{majsol}
In Chapter 3 we showed how fermion bag ideas helped us to find the solution to the sign problem for the $t$-$V$ model and its extensions using the CT-INT approach. Since our results were published, a new way to prove positivity, partially motivated by our solution, was discovered by Li, Jiang, and Yao \cite{Li:2014tla}. The essential idea was to use the Majorana representation to analyze the problem. We can always write the fermionic creation and annihilation operators at every lattice site in terms of Hermitian Majorana operators,
\begin{equation}
c_i = \frac{1}{2} \left(\gamma_i + i \bar{\gamma}_i \right), \quad c_i^\dagger = \frac{1}{2} \left(\gamma_i - i \bar{\gamma}_i \right),
\label{transmaj}
\end{equation}
which satisfy the following anticommutation relations:
\begin{equation}
    \left\{\bar{\gamma}_i,\bar{\gamma}_j\right\} = \left\{\gamma_i,\gamma_j\right\} = 2\delta_{ij}, \qquad \left\{\bar{\gamma}_i,
    \gamma_j\right\}=0.
    \label{commute}
\end{equation}
When the $t$-$V$ model (\ref{model}) is expressed in terms of Majorana operators, Li, Jiang, and Yao showed that there is also a sign-problem-free way to formulate the model so that an auxiliary field QMC method can be designed. Since then these ideas have been used to expand the class of models known to be sign-problem-free \cite{Wei:2016sgb,Li:2016gte}, and it is also possible to combine the Majorana approach with fermion bag ideas to further extend this class \cite{PhysRevE.94.043311}. Some underlying mathematical structure has also been found to partially unify the various ideas \cite{Wang:2015vha}.

Here we show how the Majorana representation provides an alternative proof of the absence of the sign problem in the CT-INT expansion. Using the relations in (\ref{transmaj}), the $t$-$V$ Hamiltonian (\ref{model}) can be expressed in terms of Majorana operators as
\begin{equation}
H = \sum_{\left\langle ij \right\rangle} \left[\frac{it_{ij}}{2} \left(\bar{\gamma}_i \gamma_j + \bar{\gamma}_j \gamma_i \right) + \frac{V}{4} \bar{\gamma}_i \gamma_j \bar{\gamma}_j \gamma_i \right],
\end{equation}
where again we are summing a bipartite lattice and for every nearest neighbor pair $\left\langle ij \right\rangle$, $i$ is on the even sublattice and $j$ is on the odd sublattice (or $A$ and $B$ sublattices on a honeycomb lattice, respectively). Redefining $\gamma_i\rightarrow -\gamma_i$ for even sites $i$, we obtain
\begin{equation}
H = \sum_{\left\langle ij \right\rangle} \left[\frac{it_{ij}}{2} \left(\bar{\gamma}_i \gamma_j + \gamma_i \bar{\gamma}_j \right) - \frac{V}{4} \left(i \bar{\gamma}_i \gamma_j\right)  \left( i\gamma_i \bar{\gamma}_j\right) \right].
\label{layers}
\end{equation}
Due to the bipartite nature of the lattice, we see that the Hamiltonian (\ref{layers}) is constructed with two independent sets of Majorana operators: $S_\mu=\left\{\bar{\gamma_i},\gamma_j\right\}$ and $S_\nu=\left\{\gamma_i,\bar{\gamma}_j\right\}$, for $i,j\in\left\langle ij \right\rangle$. Since all operators in $S_\mu$ commute with all operators in $S_\nu$, taking $H_{\rm int,b=\left\langle ij \right\rangle} =\frac{V}{4} \left(i \bar{\gamma}_i \gamma_j\right)  \left( i\gamma_i \bar{\gamma}_j\right)$, we can in fact write the CT-INT expansion as
\begin{equation}
\begin{aligned}
    Z&= \sum_{k,\left\{b\right\}}  \int \left[d\tau\right] {\rm Tr}\left(e^{-\left(\beta-\tau_k\right)H_{0}} H_{\rm int,b_k}...e^{-\tau_1 H_{0}}\right) \\
    &=\sum_{k,\left\{b\right\}} \int \left[d\tau\right]{\rm Tr_\mu}\left(e^{-\left(\beta-\tau_k\right)H_{0}^{\mu}}H_{\rm int,b_k}^{\mu}...e^{-\tau_1 H_{0}^{\mu}}\right) {\rm Tr_\nu}\left(e^{-\left(\beta-\tau_k\right)H_{0}^{\nu}}H_{\rm int,b_k}^{\nu}...e^{-\tau_1 H_{0}^{\nu}}\right),
    \end{aligned}
    \label{prod}
\end{equation}
where in the second line we have split the trace into the $\mu$ and $\nu$ spaces. Here the free Hamiltonian pieces are given by $H_0^\mu = \sum_{\left\langle ij \right\rangle} \frac{it_{ij}}{2} \bar{\gamma}_i \gamma_j$ and $H_0^\nu = \sum_{\left\langle ij \right\rangle} \frac{it_{ij}}{2} \gamma_i \bar{\gamma}_j$, and the interaction parts are given by $H^\mu_{\rm int,b=\left\langle ij \right\rangle} = \frac{iV}{2} \bar{\gamma}_i \gamma_j$ and $H^\nu_{\rm int,b=\left\langle ij \right\rangle} = \frac{iV}{2} \gamma_i \bar{\gamma}_j$. The traces over the $\mu$ and the $\nu$ spaces can be viewed as being performed in a new fermionic space with creation and annihilation operators defined as
\begin{equation}
    b^{\mu \dagger}_i = \frac{\gamma_{i+\hat{e}_1} - i\bar{\gamma}_i}{2}, \quad b^{\mu}_i = \frac{\gamma_{i+\hat{e}_1} + i\bar{\gamma}_i}{2}, \quad b^{\nu \dagger}_i = \frac{\gamma_i - i\bar{\gamma}_{i+\hat{e}_1}}{2}, \quad b^{\nu }_i = \frac{\gamma_i + i\bar{\gamma}_{i+\hat{e}_1}}{2},
    \label{bspace}
\end{equation}
with $i$ even and $\hat{e}_1$ a specific unit vector towards one of the nearest neighbors of $i$ (for example, the nearest neighbor in the positive $x$ direction on a square lattice). Note that for a model with $N$ sites, there are $N/2$ creation operators in the $\mu$ space and $N/2$ creation operators in the $\nu$ space, and these operators obey the usual fermionic anticommutation relations with each other.
%\begin{equation}
%b^{\alpha \dagger}_{{i\left(1\right)}} ... b^{\alpha \dagger}_{{i\left(n\right)}} \left|\tilde{0}\right\rangle \otimes b^{\beta \dagger}_{{i\left(1\right)}} ... b^{\beta \dagger}_{{i\left(m\right)}} \left|\tilde{0} \right\rangle ,
%\end{equation}
%where the operators $b^{\alpha \dagger}_i$ and $b^{\beta \dagger}_i$ in one representation may be written as

The key feature to note about each trace product in (\ref{prod}) is that all of the operators in the $\mu$-space are identical to those in the $\nu$-space trace. Thus, in a similar manner to (\ref{spinsquare}), we can rewrite (\ref{prod}) as
\begin{equation}
\begin{aligned}
   Z
    &=\sum_{k,\left\{b\right\}} \int\left[d\tau\right]\left[{\rm Tr_\mu}\left(e^{-\left(\beta-\tau_k\right)H_{0}^{\mu}}H_{\rm int,b_k}^{\mu}...e^{-\tau_1 H_{0}^{\mu}}\right)\right]^2.
    \end{aligned}
    \label{square}
\end{equation}
By inverting the relations in (\ref{bspace}) and rewriting $H_0^\mu$ and $H^\mu_{\rm int,b=\left\langle ij \right\rangle}$ in terms of $b^{\mu\dagger}$ and $b^\mu$ operators we can show that the single trace over the $\mu$ space in (\ref{square}) is real, and thus the weights of all $\left[b,\tau\right]$ configurations in the CT-INT expansion are positive.

\section{Alternative Expansion for $t$-$V$ Model}
In Section 2 of Chapter 3 we introduced an alternative expansion for the partition function of the $t$-$V$ model that was also sign-problem-free and consisted of only local $H_{b}$ operators:
\begin{equation}
    Z = \sum_{k,\left\{b\right\}} \int \left[d\tau\right] \left(-1\right)^k {\rm Tr}\left(H_{b_k} H_{b_{k-1}} ... H_{b_1}\right),
    \label{ssechap}
\end{equation}
where
\begin{equation}
        H_{b}= -t_{i j}(c_{i}^\dagger c_{j} + c_{j}^\dagger c_{i}) + V\left(n_{i} - \frac{1}{2}\right)\left(n_{j}-\frac{1}{2}\right) - \frac{t_{i j}^2}{V}.
\end{equation}
The positivity of the terms in this expansion can also be proven in a simple way using the Majorana representation of the previous section, which we will explain below.

The operators $H_b$ in terms of Majorana fermions (again taking $\gamma_i\rightarrow -\gamma_i$ for even $i$), are given by (see (\ref{layers}))
\begin{equation}
H_{b} =  \frac{it_{ij}}{2}\left(\bar{\gamma}_i \gamma_j + \gamma_i \bar{\gamma}_j \right) - \frac{V}{4} \left(i \bar{\gamma}_i \gamma_j\right) \left(i \gamma_i \bar{\gamma}_j\right) -\frac{t_{ij}^2}{V}.
\end{equation}
These operators can then be factored in the following way:
\begin{equation}
H_b =  \frac{-t_{ij}^2}{V} \left(1 - \frac{iV}{2t_{ij}} \bar{\gamma}_i \gamma_j \right) \left(1 - \frac{iV}{2t_{ij}} \gamma_i \bar{\gamma}_j \right).
\label{trans}
\end{equation}
Following steps similar to those explained in Section \ref{majsol}, we get two commuting factors in every term of the sum that are identical according to the $i$ and $j$ labels. So again--as long as the lattice is bipartite--we can write the partition function as
\begin{equation}
\begin{aligned}
Z=\sum_{k,\left\{b\right\}} \int \left[d\tau\right] \left(\frac{t_{ij}^2}{V}\right)^k\left[{\rm Tr_\mu}\left(\left(1 - \frac{iV}{2t_{ij}} \bar{\gamma}_{i_k} \gamma_{j_k} \right)...\left(1 - \frac{iV}{2t_{ij}} \bar{\gamma}_{i_1} \gamma_{j_1} \right)\right)\right]^2 .
    \label{ssepos}
    \end{aligned}
\end{equation}
Again we know every term in the partition function must be positive because inverting the relations in (\ref{bspace}) to express the trace over $\mu$-space in terms of $b^{\mu \dagger}$ and $b^\mu$ removes all imaginary $i$ factors from the operators and shows that the trace over $\mu$ must be real. Thus its square is positive and the expansion is sign-problem-free.

\section{Computing the Fermionic Trace: The BSS Formula}
While we can use the Majorana representation to easily show that both the expansions in Sections 4.1 and 4.2 are sign-problem-free, we still have not considered how to compute the weights for these expansions. While the Majorana representation has proven to be a valuable tool for showing positivity, there is no need to stay in that representation once we have done that. In fact, it is often nice to go back to the original basis because there are useful formulas for computing the trace terms in the ordinary fermionic basis, such as the BSS formula \cite{PhysRevD.24.2278}, which works for any trace over operators in the following form
\begin{equation}
    {\rm Tr}\left(e^{O_k} ... e^{O_2} e^{O_1}\right),
        \label{initexp}
\end{equation}
where $O_n=\sum_{ij} c^\dagger_i \left(o_n\right)_{ij} c_j$ is bilinear in fermionic operators. Below we will motivate the BSS formula and discuss how to compute the terms in (\ref{ssechap}) using this formula.

As a starting point, let us consider computing the following expression
\begin{equation}
    {\rm Tr}\left(e^{O_1}\right), \qquad O_1 = \sum_{ij} c_i^\dagger \left(o_1\right)_{ij} c_j.
\end{equation}
Assuming a diagonalizable $o_1$, we can write $o_1 = U D U^\dagger$, where $D_{ij}=\lambda_i \delta_{ij}$ is the diagonal eigenvalue matrix, and $U$ is a unitary matrix. Introducing $d^\dagger,d$ operators such that $c^\dagger_i = \sum_j d^\dagger_j U^\dagger_{ji}$ and $c_i = \sum_j U_{ij} d_j$, we have
\begin{equation}
    {\rm Tr}\left(e^{O_1}\right) = {\rm Tr}\left(e^{\sum_{i}\lambda_i d^\dagger_i d_i}\right) = \prod_i \left(1 + e^{\lambda_i}\right) = \det\left(\mathbbm{1} + e^{o_1}\right).
    \label{finresult}
\end{equation}
%\begin{equation}
 %   {\rm Tr}\left(e^{O_1}\right) = {\rm Tr}\left(e^{\sum_{i}\lambda_i d^\dagger_i d_i}\right) =  {\rm Tr}\left[\prod_i\left(\mathbbm{1}+ \left(e^{\lambda_i} - \mathbbm{1}\right)n^d_i\right)\right],
%\end{equation}
%where $n^d_i = d^\dagger_i d_i$. The last equality comes from the fact that $n^d_i = \left(n^d_i\right)^k $ for all $k$. We can then factor the trace into traces over single particle subspaces, and evaluate each trace independently to obtain
%\begin{equation}
 %   {\rm Tr}\left(e^{O_1}\right) = \prod_i{\rm Tr_i}\left[\left(\mathbbm{1}+ \left(e^{\lambda_i} - \mathbbm{1}\right)n^d_i\right)\right] = \prod_i \left(1 + e^{\lambda_i}\right) = \det\left(\mathbbm{1} + e^{o_1}\right).
  %  \label{finresult}
%\end{equation}
In this way we see that we have gone from the trace on the left of (\ref{finresult}) to an $N\times N$ determinant on the right of (\ref{finresult}), which is given using the matrix $o_1$. The matrix $\mathbbm{1}$ is the $N\times N$ identity matrix. This result is generalizable to the $k$ factors of matrices found in (\ref{initexp}) as
\begin{equation}
\begin{aligned}
    {\rm Tr}\left(e^{O_k} ... e^{O_2} e^{O_1}\right) &= \det\left(\mathbbm{1} + e^{o_k}...e^{o_2} e^{o_1}\right)\\
    &\equiv \det\left(\mathbbm{1} + B_k...B_2 B_1\right),
    \end{aligned}
    \label{bssform}
\end{equation}
and it is this result that is known as the BSS formula. In the last line we introduce the matrices $B = e^{o}$, which have been the symbols traditionally used in QMC methods that employ the BSS formula. If we combine this formula with the representation (\ref{expham}), given by
\begin{equation}
    H_{b} = -\sum_{\left\langle ij \right\rangle} \omega_{ij} e^{2\alpha_{ij}\left(c_i^\dagger c_j + c_j^\dagger c_i\right)},
\end{equation}
where $\omega_{ij} = t_{ij}^2/\left(V\left(1-\left(V/2t_{ij}\right)^2\right)\right)\geq0$, and for $\alpha_{ij}$ we have the expressions $\sinh2\alpha_{ij} =\left(V/t_{ij}\right)/\left(1-\left(V/2t_{ij}\right)^2\right)$ and $\cosh2\alpha_{ij} =\left(1+\left(V/2t_{ij}\right)^2\right)/\left(1-\left(V/2t_{ij}\right)^2\right)$, we can write (\ref{ssechap}) as
\begin{equation}
\begin{aligned}
    Z &= \sum_{k,\left\{b\right\}} \int \left[d\tau\right] \left(\prod_{b}\omega_b\right) \det\left(\mathbbm{1}+e^{h_{b_k}} e^{h_{b_{k-1}}} ... e^{h_{b_1}}\right),
    \end{aligned}
    \label{trbss}
\end{equation}
where $\left(h_{b=ij}\right)_{lm} = 2\alpha_{ij} \left(\delta_{il}\delta_{jm} +\delta_{jl}\delta_{im}\right)$ is a matrix with only two nonzero entries.

\section{Computing the Fermionic Trace: The Pfaffian Approach}

In this section we consider an alternative way to calculate (\ref{ssechap}), by deriving a formula for the fermionic trace in (\ref{ssepos}), given generically by
\begin{equation}
{\rm Tr_\mu}\left[\left(1-ia_{i_k j_k}\bar{\gamma}_{i_k} \gamma_{j_k}\right) ...\left(1-ia_{i_2 j_2}\bar{\gamma}_{i_2} \gamma_{j_2}\right)\left(1-ia_{i_1 j_1}\bar{\gamma}_{i_1} \gamma_{j_1}\right)\right],
\label{int}
\end{equation}
in terms of a Pfaffian of an antisymmetric matrix that can be easily constructed. We begin this proof by noting that the following quantity for a general $n$,
\begin{equation}
   \mathcal{S} = \frac{{\rm Tr_\mu}\left(\bar{\gamma}_{i_n} \gamma_{j_n}...\bar{\gamma}_{i_2} \gamma_{j_2}\bar{\gamma}_{i_1} \gamma_{j_1}\right)}{{\rm Tr_{\mu}}\left(1\right)},
    \label{majtr}
\end{equation}
can only be $\pm 1$ or $0$. This is because $\bar{\gamma}^2_i=\gamma_i^2=1$, and the trace vanishes unless every $\bar{\gamma}_i$ (or $\gamma_i$) operator comes in pairs. If one of these operators is not paired, we can move it around in a cycle, anti-commuting with all other operators and show that ${\rm Tr_\mu}(...) = -{\rm Tr_\mu}(...)$.

When the terms in (\ref{int}) are expanded, we get traces of the form
\begin{equation}
    \left(-i\right)^n \prod_{m=1}^n a_{i_m j_m} {\rm Tr_\mu}\left(\bar{\gamma}_{i_n} \gamma_{j_n}...\bar{\gamma}_{i_2} \gamma_{j_2}\bar{\gamma}_{i_1} \gamma_{j_1}\right).
    \label{traceexppiece}
\end{equation}
Let us then first focus on computing the trace (\ref{traceexppiece}) as a Grassmann integration. For this purpose we introduce a Grassmann variable $\bar{\xi}_{i,t}$ for every $\bar{\gamma}_{i}$ occurring at location $t$ and another $\xi_{j,t}$ for every $\gamma_{j}$ appearing at location $t$ in the trace. Note that $t$ here is a location index, in contrast to $\tau$ which would label a continuous time value. Note also that there are two Grassmann variables with the index $t$ that label the time at which the bond enters. We can define the Grassmann integration measure at each spatial site $i$ and $j$ as
\begin{equation}
[d\bar{\xi}]_i \equiv \prod_{t \in i,\ \mathrm{time\: ordered}}\ d\bar{\xi}_{i,t},\ \ \  
[d \xi]_j \equiv ...  \prod_{t \in j,\ \mathrm{time\: ordered}}
d\xi_{j,t}
\end{equation}
where the differentials are time ordered with early times coming on the right and later times coming on the left. The full integration measure in terms of these site measures would then be
\begin{equation}
    \prod_i \left[d\bar{\xi}\right]_i \prod_j \left[d\xi\right]_j.
    \label{sitetimeordered}
\end{equation}

There is a sign ambiguity for this measure if any of the site labels $i$ or $j$ appear an odd number of times, but the Grassmann integrals in these cases will be zero anyway, as shown below. Note that any of the site labels $i$ or $j$ appearing an odd number of times is also the condition that must be met for $\mathcal{S}$ to be zero. It is also useful to define a second way of writing the full integration measure, which is by placing the $d\bar{\xi}_{i,t}$ and $d\xi_{i,t}$ in the same order as their corresponding $\bar{\gamma}_i$ and $\gamma_i$ operators in the trace from (\ref{traceexppiece}). We label this measure $\left[d\bar{\xi}d\xi\right]$, where
\begin{equation}
    \left[d\bar{\xi}d\xi\right] = d\bar{\xi}_{i_n,t_n} d\xi_{j_n,t_n}...\bar{\xi}_{i_2,t_2} d\xi_{j_2,t_2}\bar{\xi}_{i_1,t_1} d\xi_{j_1,t_1}.
    \label{timeordered}
\end{equation}
Now both $\bar{\gamma},\gamma$ operators and  $\bar{\xi},\xi$ Grassmann numbers anticommute with each other, so (assuming an even number of every $i$ and every $j$) if we wanted to take the differentials in (\ref{timeordered}) and commute them so that their order matched that of (\ref{sitetimeordered}), the overall extra sign introduced would in fact match that of (\ref{majtr}). The only difference in commutation that could occur would be if a $\bar{\gamma}_i$ or $\gamma_i$ operator needed to commute with another $\bar{\gamma}_i$ or $\gamma_i$ operator. Those operators would commute instead of anticommute (whereas the Grassmann differentials always anticommute), but we do not ever encounter this situation, since the differentials in (\ref{sitetimeordered}) remain time-ordered for each individual site. Thus we have (for even $i$, $j$):
\begin{equation}
        \mathcal{S} \prod_i \left[d\bar{\xi}\right]_i \prod_j \left[d\xi\right]_j = \left[d\bar{\xi}d\xi\right].
        \label{signmeasure}
\end{equation}

Using the measure definitions in (\ref{sitetimeordered}), it is possible to verify that for even $n$
\begin{equation}
    \left(-i\right)^{n}\sqrt{a_{ij}\left(t_1\right)....a_{ij}\left(t_n\right)}\:\:  = \int \left[d\bar{\xi}\right]_i e^{\sum_{t< t'}\sqrt{a_{ij}\left(t'\right)a_{ij}\left(t\right)} \bar{\xi}_{i,t'}\bar{\xi}_{i,t}},
    \label{coordi}
\end{equation}
and
\begin{equation}
    \sqrt{a_{ij}\left(t_1\right)....a_{ij}\left(t_n\right)}\:\:  = \int \left[d\xi\right]_j e^{-\sum_{t< t'}\sqrt{a_{ij}\left(t'\right)a_{ij}\left(t\right)} \xi_{j,t'}\xi_{j,t}}.
    \label{coordj}
\end{equation}
For odd $n$ the expressions are zero. Here we define $t_1 < t_2 < ... < t_n$, and note these times are for coordinates $i$ in (\ref{coordi}) and coordinates $j$ in (\ref{coordj}). The notation $a_{ij}\left(t\right)$ refers to the $a_{ij}$ constant located at the time $t$. The details for the proof of these identities can can be found in Appendix B. Combining the expressions (\ref{coordi}) and (\ref{coordj}) for all sites tells us that (again assuming even $i$, $j$)
\begin{equation}
\begin{aligned}
    \left(-i\right)^n&\prod_{m=1}^n a_{i_m j_m} =   
    \int \prod_i [d\overline{\xi}]_i \prod_j [d\xi]_j\ \\
    &\qquad\qquad\qquad\times\prod_i e^{\sum_{t<t'}\sqrt{a_{ij}\left(t'\right)a_{ij}\left(t\right)} \bar{\xi}_{i,t'} \bar{\xi}_{i,t}} \prod_j e^{-\sum_{t<t'}\sqrt{a_{ij}\left(t'\right)a_{ij}\left(t\right)}  \xi_{j,t'} \xi_{j,t}}.
    \end{aligned}
    \label{prodoversites}
\end{equation}
We can then combine (\ref{prodoversites}) with (\ref{signmeasure}) to obtain for the following expression for (\ref{traceexppiece}):
\begin{equation}
\begin{aligned}
    &\left(-i\right)^k\prod_{m=1}^n a_{i_m j_m}{\rm Tr_\mu}\left(\bar{\gamma}_{i_n} \gamma_{j_n}...\bar{\gamma}_{i_2} \gamma_{j_2}\bar{\gamma}_{i_1} \gamma_{j_1}\right)\\
   & = {\rm Tr_\mu}(1) 
    \int \left[d\overline{\xi} d\xi\right]\ \prod_i e^{\sum_{t<t'}\sqrt{a_{ij}\left(t'\right)a_{ij}\left(t\right)} \bar{\xi}_{i,t'} \bar{\xi}_{i,t}} \prod_j e^{-\sum_{t<t'}\sqrt{a_{ij}\left(t'\right)a_{ij}\left(t\right)}  \xi_{j,t'} \xi_{j,t}}.
    \end{aligned}
    \label{keyid}
\end{equation}
We know this holds for a nonzero left-hand-side due to the nonzero $\mathcal{S}$ condition of (\ref{signmeasure}). However, upon inspection, it also holds in the cases of odd $i$ and $j$ values and thus a zero left-hand-side. In these cases we know that the right-hand-side also is zero due to the zero conditions given for the integrals in (\ref{coordi}) and (\ref{coordj}).

Equation (\ref{keyid}) thus enables us to create a Grassmann integral for any of the terms in the expansion of (\ref{int}). The extension of such a formula for the full trace (\ref{int}) in a single Grassmann integral is then straightforward. Each of the identity operators $\mathbbm{1}$ in (\ref{int}) for a corresponding $\bar{\gamma}_i \gamma_j$ can be represented by an insertion of $e^{-\bar{\xi}_{i,t} \xi_{j,t}}$. The full expression for the trace over $\mu$-space is then
\begin{eqnarray}
&&{\rm Tr_\mu}\Big[\left(1-ia_{i_k j_k}\bar{\gamma}_{i_k} \gamma_{j_k}\right) ...\left(1-ia_{i_2 j_2}\bar{\gamma}_{i_2} \gamma_{j_2}\right)\left(1-ia_{i_1 j_1}\bar{\gamma}_{i_1} \gamma_{j_1}\right)\Big]
\nonumber \\
&& = {\rm Tr_{\mu}}\left(1\right)\int [d\overline{\xi} d\xi]\ 
\ \mathrm{e}^{-\sum_t \bar{\xi}_{i,t} \xi_{j,t} + 
\sum_{i,\left(t <t'\right)} m_{t't}\bar{\xi}_{i,t'} \bar{\xi}_{i,t} 
- \sum_{j,\left(t <t'\right)} m_{t't} \xi_{j,t'} \xi_{j,t}}
\nonumber \\
&& = {\rm Tr_{\mu}}\left(1\right) \int [d\xi] \ \mathrm{e}^{-\frac{1}{2} \xi^T A \xi}\ =\ {\rm Tr_{\mu}}\left(1\right) \mathrm{Pf}(A),
\label{trpfaf}
\end{eqnarray}
where in the last line we have simply written the Grassmann integral schematically involving a 2k valued Grassmann vector $\xi$ which contains both $\overline{\xi}_{i,t}$ and $\xi_{j,t}$, and a $2k \times 2k$ anti-symmetric matrix $A$ whose matrix elements can be obtained from the second equation. The factor $m_{t't}$ is simply $\sqrt{a_{ij}\left(t'\right) a_{ij}\left(t\right)}$.

As is well known and can be seen in Appendix C using Grassmann integrals, ${\rm Pf}^2 \left(A\right)=\det A$ for an antisymmetric matrix, so to get the weights of the elements of (\ref{ssepos}), we simply have
\begin{equation}
\left[{\rm Tr_\mu}\left(\left(1 - ia_{i_k j_k} \bar{\gamma}_{i_k} \gamma_{j_k} \right)...\left(1 - ia_{i_1 j_1} \bar{\gamma}_{i_1} \gamma_{j_1} \right)\right)\right]^2={\rm Tr_{\mu}}\left(1\right)\det A,
\label{pfafdet}
\end{equation}
for the matrix defined in (\ref{trpfaf}).

We have checked our formula (\ref{pfafdet}) against the BSS formula (\ref{bssform}) numerically for many configurations during the testing of our computer codes and they do agree in all the cases.

\chapter{The Hamiltonian Fermion Bag Algorithm}
\textit{\small``I like to learn. That's an art and a science.''\\
--Katherine Johnson}
\newline
\newline
\section{Model Applicability}
We are now ready to introduce a fermion bag algorithm that is directly applicable to the Hamiltonian picture. We will use the SSE-inspired expansion in terms of local degrees of freedom (\ref{sseeq}) as discussed in Chapter 2, and compute the fermionic trace using the BSS formula (\ref{bssform}) which is discussed in Chapter 4. Our algorithm will be applicable to a wide variety of models where the Hamiltonian can be written as $H = \sum_{b=\left\langle ij\right\rangle} H_{b}$, with
\begin{equation}
H_{b} \ =\ -\omega_{ ij}\ 
\mathrm{e}^{2\alpha_{ ij}\ 
\sum_{a=1}^{N_f} \big({c^a_i}^\dagger c^a_{j }+ {c^a_{j}}^\dagger c^a_i\big)}.
\label{fact}
\end{equation}
Here $i$ and $j$ label nearest neighbor spatial lattice sites on a bipartite lattice, as before. The operators ${c^a_i}^\dagger$ and $c^a_i$ are fermionic creation and annihilation operators at the site $i$ with flavors $a=1,2..,N_f$. The couplings of the model are defined through the real constants $\omega_{ij} > 0$ and $\alpha_{ ij}$.
%In the discussions below we focus on the $N_f=1$ model on a two dimensional square lattice with periodic boundary conditions and $L$ sites in each direction with $N=L^2$. However, they can be extended to any value of $N_f$ and all bi-partite lattice models where the sites connected to the bond $\langle x,d\rangle$ lie on different sub-lattices.

Although the Hamiltonians in (\ref{fact}) are unconventional, they contain rich physics. They may be considered de-\textit{sign}-er Hamiltonians: chosen so that there is no sign problem and the idea of fermion bags is applicable, and most importantly containing interesting quantum critical points in the same universality class as models with more conventional Hamiltonians \cite{RevCondMat}. For a fixed $N_f$ it is possible to prove that (\ref{fact}) is invariant under an $O(2N_f)$ flavor symmetry in addition to the usual lattice symmetries, some of which may be broken spontaneously at quantum critical points  \cite{PhysRevX.6.041049}. The application is not limited to only one type of $H_b$ term, as well. For example, in addition to the terms of the form (\ref{fact}), the Hamiltonian could also contain single site $H_{b}$ terms of the form
\begin{equation}
    H_{\rm int, i} = \pm \frac{U}{4}e^{i\pi n_{i,\uparrow}} e^{i\pi n_{i,\downarrow}}.
\end{equation}
This is the interaction for the Hubbard model at half-filling and introduces no sign problem.

As discussed in (\ref{expham}) and (\ref{oatv}) of Chapter 3, there is a way to write the $t$-$V$ model in the form (\ref{fact}). In Chapter 6 we will use our algorithm on the $t$-$V$ model, but the algorithm is applicable to all the models given in this section.

%when $N_f=1$ our model is equivalent (up to an constant) to the $t-V$ model
%\begin{equation}
%H_{ij} \ =\ - t_{ij}\left(c_i^\dagger c_{j} + c_{i}^\dagger c_j\right)
% + V \left(n_i -\frac{1}{2}\right) \left(n_j -\frac{1}{2}\right),
%\end{equation}
%where $V > 0$, if we set $\omega_{ ij }=t_{ij}^2/(V (1-\left(V/2t_{ij}\right)^2))$, and $\alpha_{ ij}$ such that $\cosh2\alpha_{ij}=(1+\left(V/2t_{ij}\right)^2)/(1-\left(V/2t_{ij}\right)^2)$, and $\sinh2\alpha_{ij} = (V/t_{ij})/(1-\left(V/2t_{ij}\right)^2)$ \cite{PhysRevB.93.155117}. If we put the model on a square lattice and define $t_{ij} = t\eta_{ij}$, where $\eta_{ i,i+\hat{e}_x} = 1$ and $\eta_{i,i+\hat{e}_y} = (-1)^{i_x}$ for even $i$, the model describes interacting two dimensional massless Hamiltonian staggered fermions \cite{PhysRevD.16.3031}. These $t_{ij}$ values are also known as the $\pi$-flux lattice and we will use the $t$-$V$ model on this lattice to test out the algorithm in Chapter 6.

\section{Fermion Bags in the Algorithm}
\begin{figure}[t]
\begin{center}
  \includegraphics[width=6cm]{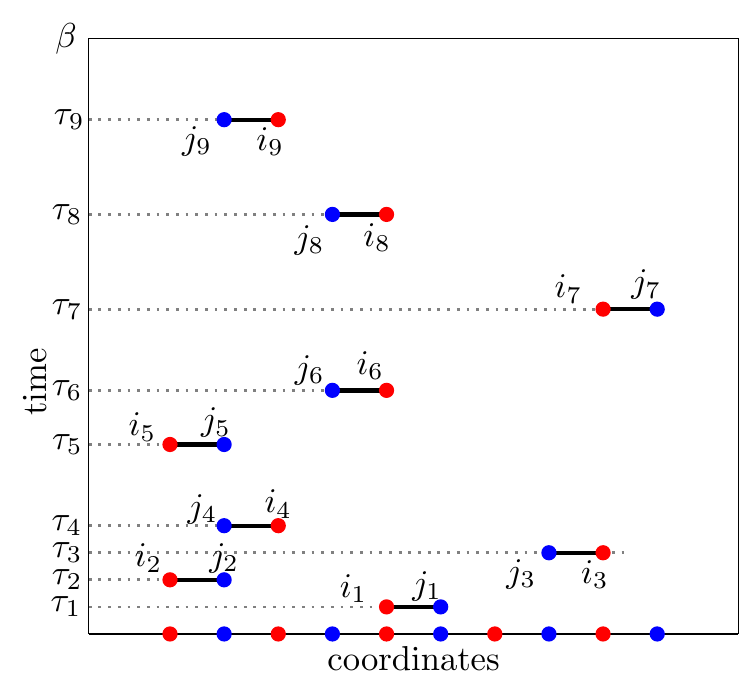}
  \end{center}
  \caption[An example configuration from the SSE expansion of the $t$-$V$ model.]{An example configuration $[b,\tau]$, from the expansion (\ref{partition}). The horizontal axis represents the spatial lattice, while the vertical axis is continuous imaginary time. Every bond $b$ in the expansion has two sites, $i$ and $j$. The sites are represented by two different colored dots to represent the two different sublattices in the bipartite lattice.}\label{exp}
\end{figure}

In Chapter 2 we discussed how the SSE-inspired CT-INT expansion of the partition function can lead to a localized fermion bag picture. Here we will explicitly show this in the context of a fermionic Hamiltonian consisting of exponentiated bilinear operators that involve only nearest neighbor terms, such as (\ref{fact}). Let us begin with the expression for the partition function, coming from (\ref{sseeq}):
\begin{equation}
Z\ =\ \sum_{k,\left\{b=\langle ij\rangle\right\}}\ \int \ [d\tau]
\ (-1)^k\: {\rm Tr}
\Big(H_{b_k}\ ...\ H_{b_2}\ H_{b_1}\Big),
\label{partition}
\end{equation}
where there are $k$ insertions of the bond Hamiltonian $H_{b}$ inside the trace at times $\tau_1\leq \tau_2 \leq ...\leq \tau_k$. The symbol $[d\tau]$ represents the $k$ time-ordered integrals and $\left\{b=\langle ij\rangle\right\} = \{b_1=\langle i_1 j_1\rangle,b_2=\langle i_2 j_2\rangle,...b_k=\langle i_kj_k\rangle\}$ represents the configuration of bonds at different times. Let us label each of these bond configurations as $[b,\tau]$. An illustration of such a bond configuration is shown in Fig.~\ref{exp}.

\begin{figure*}[t]
\begin{center}
  \includegraphics[height=6cm]{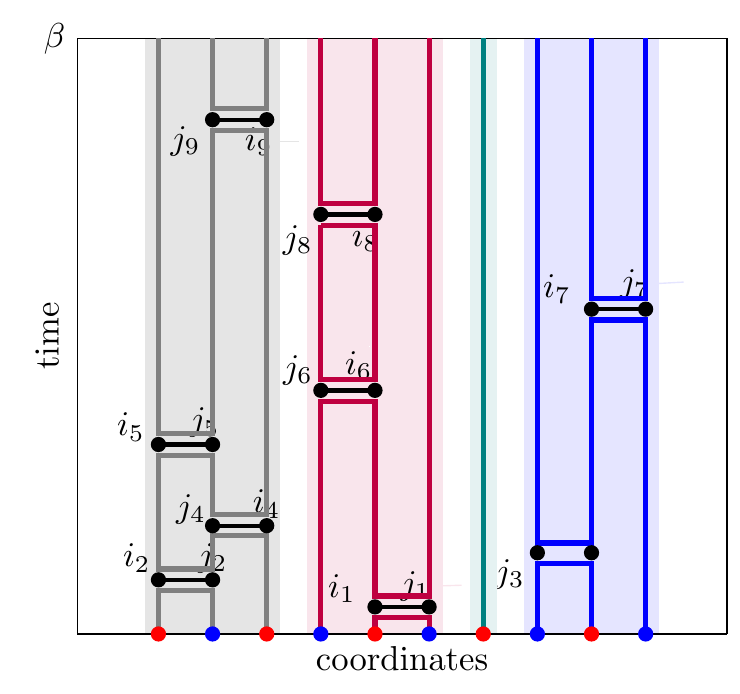}
  \includegraphics[height=6cm]{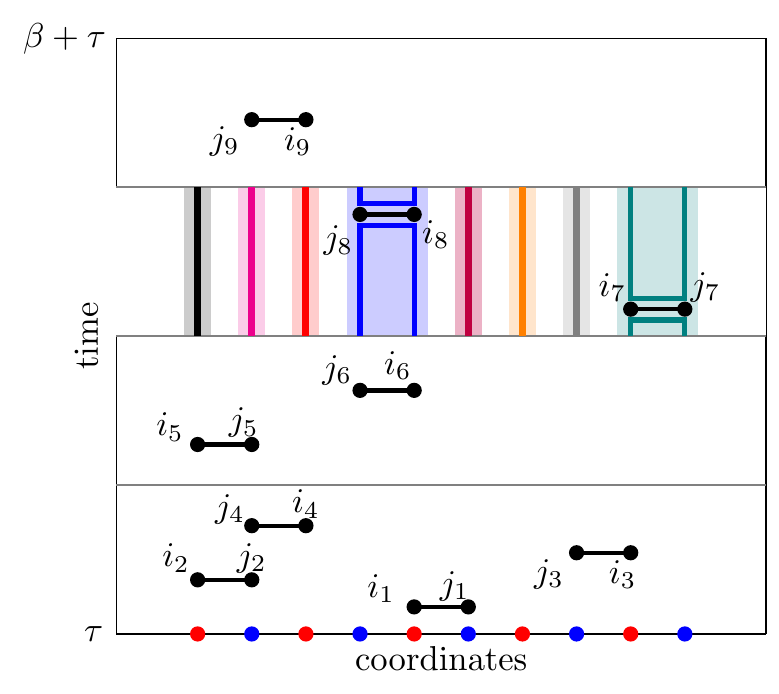}
  \end{center}
  \caption[How time-slices define fermion bags in the SSE expansion of the $t$-$V$ model.]{The bonds in the configuration when all of $\beta$ is one time-slice form four fermion bags between $t=0$ and $t=\beta$, as shown to the left. When four time-slices are used to divide $\beta$, there are more fermion bags, as shown to the right with the bags in one time-slice colored.}\label{regions1}
\end{figure*}

Each bond represents the operator $H_{b}$ that is present inside the trace in (\ref{partition}). We can imagine $H_{b}$ as creating a quantum entanglement between the fermions at $i$ and $j$. Thus, all spatial sites connected by bonds to each other at various times become entangled with each other. Such a group of entangled sites can be defined as a fermion bag. Note that lattice sites that are not connected to any bonds form their own fermion bags. For the bond configuration in Figure~\ref{exp} we identify four fermion bags as shown in the left side of Figure~\ref{regions1}. When two bonds $b=\left\langle ij\right\rangle$ and $b'=\left\langle i'j'\right\rangle$ do not share a site between them, the bond Hamiltonians commute, i.e.,
$\left[H_{b}, H_{b'}\right] = 0$. This implies that the weight of the bond configuration can be written as a product of weights from fermion bags.

It is possible to show that the space-time density of bonds is a physical quantity related to the energy density of the system \cite{PhysRevB.93.155117}. Hence for a fixed value of the coupling $V$ we expect a fixed density of bonds. This implies that we can use the temperature as a parameter to control the size of fermion bags. At high temperatures we will have fewer bonds and many small fermion bags, but as the temperature is lowered, fermion bags will begin to merge to form a single large fermion bag. This suggests that at some optimal temperature the fermion bags may efficiently break up the system into smaller regions that do not depend on the system size. Even at low temperatures, we may be able to divide the imaginary time axis into many time-slices and update a single time-slice efficiently. This is illustrated in the right side of Figure~\ref{regions1}, where the imaginary time extent of the configuration shown in the left side of the same figure is divided into four time-slices and in the shaded time-slice there are eight fermion bags, instead of four. We will use this concept of time-slices to define fermion bags that are small enough for efficient calculations.

\section{Scaling}
Before we get into the algorithmic details, we give a brief note on the scaling of our algorithm and similar Hamiltonian fermionic algorithms used for strong couplings. These algorithms require a calculation of a determinant for the fermion weights, and typically for strong coupling the BSS formula offers the best scaling for the determinant. As seen in (\ref{bssform}), the determinants are of $N\times N$ matrices, and the scaling for such a computation, which can be accomplished by $LU$ factorization, is $O\left(N^3\right)$. In practice, the update probabilities are based on ratios of determinants, which can in fact be written as determinants of much smaller matrices. However, obtaining the entries for these small update matrices will require getting the inverse of an $N\times N$ matrix, which still scales as $O\left(N^3\right)$. Thus this scaling typically cannot be avoided.

In addition, updates must occur throughout the entire trace, and so the number of updates necessary to bring a configuration to equilibration will scale as $O\left(\beta\right)$. Combined with the time taken to perform the matrix inversions, fermionic algorithms based on the BSS formula, such as Auxiliary Field Quantum Monte Carlo and our algorithm, scale as $O\left(\beta N^3\right)$. In the following sections we will discuss the operations we perform, their scaling, and explain why the fermion bag approach is faster than traditional auxiliary field methods, although both scale similarly.

\section{Algorithm and Updates}
We now discuss the Hamiltonian fermion bag algorithm in detail. In a QMC algorithm, the ultimate goal is to make a measurement, which typically involves some kind of correlation observable $C$. We will focus on an equal time correlation function which means we will need to compute the expectation value
%In the context of the $t$-$V$ model on a $\pi$-flux lattice, the correlation observable 
\begin{equation}
\left\langle C \right\rangle = \frac{{\rm Tr}\left(C e^{-\beta H}\right)}{{\rm Tr}\left(e^{-\beta H}\right)},
    \label{correlation}
\end{equation}
%\begin{equation}
%\langle C\rangle = \sigma_{\left(0,0\right)}\sigma_{\left(L/2,0\right)}{\rm Tr}\left(\left(n_{\left(0,0\right)}-1/2\right) \left(n_{\left(L/2,0\right)}-1/2\right) e^{-\beta H}\right)/{\rm Tr}\left(e^{-\beta H}\right),
%\label{obs}
%\end{equation}
where C is a product of two operators at equal time. In Chapter 6 we will discuss the specific correlation observable we measure for the $t$-$V$ model. In order for us to measure equal time correlation functions, we choose a time $\tau_0$ to insert the correlation function operator C. Thus, in our algorithm we generate new types of configurations labeled with $([b,\tau];\tau_0)$ in two sectors: the partition function sector ($n=0$) of weight $\Omega_0([b,\tau];\tau_0)$, without the observable, and the observable sector ($n=1$) of weight $f\Omega_1([b,\tau];\tau_0)$, in the presence of the operator $C$. These weights are then defined by
\begin{equation}
\begin{aligned}
\Omega_n([b,\tau];\tau_0) &= {\rm Tr}\left[H_{b_k} ...C_n... H_{b_2} H_{b_1}\right]
\label{confweights}
\end{aligned}
\end{equation}
Here $0 \leq \tau_0 \leq \beta$ is a time where the operator $C_n$ is introduced. In the partition function sector $C_0 = I$ (the identity operator) and in the observable sector $C_1 = C$ (the equal time correlation function operator). The observable $\left\langle C\right\rangle$ in terms of these weights in the two sectors is
\begin{equation}
    \left\langle C\right\rangle = \frac{\sum_{\left\{b\right\},\tau_0}\int \left[d\tau\right]\Omega_1([b,\tau];\tau_0)}{\sum_{\left\{b'\right\},\tau_0'}\int\left[d\tau'\right]\Omega_0([b',\tau'];\tau_0')}.
\end{equation}
A reweighting factor $f > 0$ is chosen so that the two sectors can be sampled with roughly equal probabilities. We record the number
 \begin{equation}
 {\cal N}\ =\ \frac{\Omega_1([b,\tau];\tau_0)}{\Omega_0([b,\tau];\tau_0) + f\Omega_1([b,\tau];\tau_0)}
 \label{nobs}
 \end{equation}
for each configuration generated by a sweep (defined below). It is straightforward to show that $\langle C \rangle = \langle {\cal N}\rangle/(1-f\langle {\cal N}\rangle)$.

We use four different types of updates to generate the configurations $([b,\tau];\tau_0)$ in the two sectors:
\begin{enumerate}
\item The \textit{sector-update} flips only the sector from $n=0$ to $n=1$ and vice versa.
\item The \textit{bond-update} changes the entire bond configuration $[b,\tau]$ to $[b',\tau']$ while keeping $\tau_0$ and the sector number $n$ fixed. 
\item The \textit{time-update} changes the imaginary time location $\tau_0$ where the observable is measured. This is only done in the $n=0$ sector to save time. 
\item The \textit{move-update} changes the times where the bonds occur as long as the bond operators $H_b$ commute. This does not change the overall weight. While the time ordering of bonds in general will change, the time ordering of bonds which share sites in common will be preserved.
\end{enumerate}
We define a sweep when we accomplish all the four updates a certain number of times in a particular order.
Out of these four updates the two most time intensive updates are the sector-update and the bond-update, as we discuss below. The bond-update is performed only once per sweep. The time-update and move-update do not require us to calculate any weight ratios and are easy to implement. We make the time-update, move-update, and sector-update a fixed number of times per sweep depending on the lattice size.

For the bond updates we need to compute the ratio
\begin{equation}
    R = \frac{\Omega_n\left([b,\tau];\tau_0\right)}{\Omega_n\left([b',\tau'];\tau_0\right)},
    \label{ratior}
\end{equation}
to calculate the transition probabilities in the Metropolis accept/reject step. The sector update is a special case of the bond update and requires a similar calculation. Here we only focus on the details of the bond updates since it can easily be modified to perform the sector update. Using the BSS formula (\ref{bssform}), given in Chapter 4, we can show
\begin{equation}
\Omega_n\left([b,\tau];\tau_0\right)=\det\left(\mathbbm{1}_N+B_{b_k}...O_n...B_{b_2} B_{b_1}\right),
\end{equation}
where $\mathbbm{1}_N$, $B_{b}$ and $O_n$ are all $N\times N$ matrices with rows and columns labeled by spatial lattice sites. The matrix $\mathbbm{1}_N$ is the identity matrix, while $B_{b}$ is the identity matrix except in a $2\times 2$ block labeled by the rows and columns of the sites that touch the bond $b=\langle ij\rangle$. Finally, the matrix $O_n$ depends on the sector $n$. $O_0$ is given by $O_0=\mathbbm{1}_N$ and $O_1$ depends on the correlation function operator $C$.
%Within this block, $B_{ij}$ takes the form 
%\begin{equation}
%{\cal B}_{ij} = \left(\begin{array}{cc} \cosh2\alpha_{ij} & \sinh2\alpha_{ij} \\ \sinh2\alpha_{ij} & \cosh2\alpha_{ij} \end{array}\right).
%\end{equation}

%For the particular correlator given in (\ref{obs}), $(O_1)_{ij}=\delta_{ij}- 2\delta_{i,(0,0)} - 2\delta_{i,(L/2,0)}$.

\begin{figure*}[t]
\begin{center}
  \includegraphics[height=6cm]{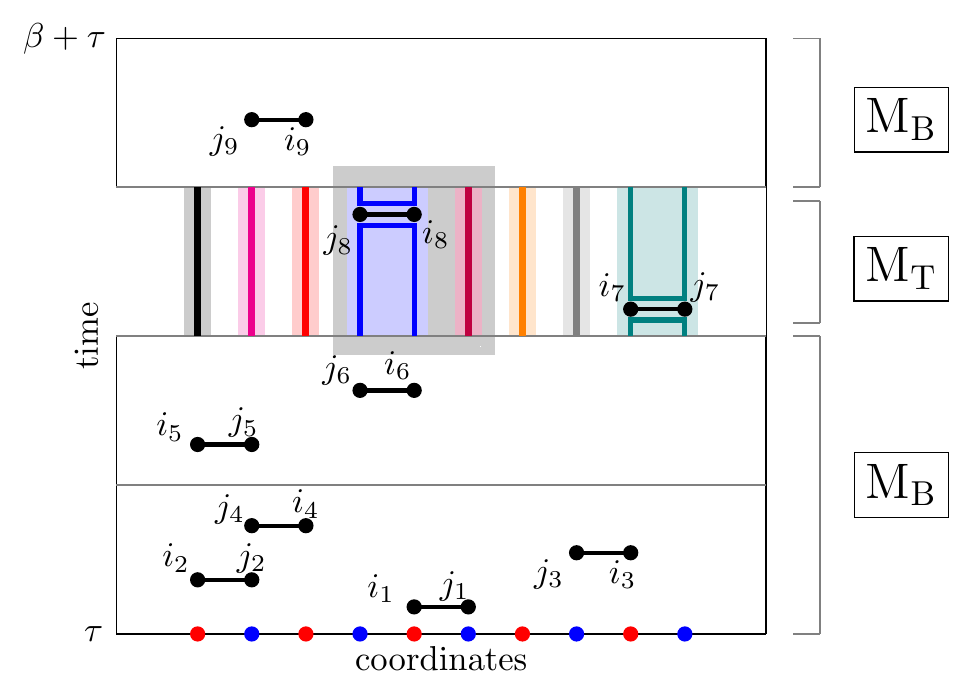}
  \end{center}
  \caption[An illustration of update regions for the Hamiltonian fermion bag algorithm.]{This is the same configuration with four time-slices found in Figure \ref{regions1}. The algorithm moves through the time-slices sequentially, updating one at a time. The fermion bags are highlighted for the timeslice currently being updated in this figure. Within each time-slice, spatial blocks are randomly chosen and updated one at a time. The gray square represents one such spatial block. The $M_T$ and $M_B$ regions used in (\ref{rdefn}) and (\ref{largem}) are defined to the right of the box.}\label{regions}
\end{figure*}

In order to perform the updates, we divide the configuration space into time-slices of an appropriate width $w_\tau$ with $\tau_0$ chosen to be at the beginning of the first time slice. We then update bonds within each time-slice sequentially. During the update of a time-slice we define two $N\times N$ matrices: the \textit{background matrix} $M_B$ (which is a product of all of the $B_{ij}$ matrices outside the selected time-slice and $O_n$), and the \textit{time-slice matrix} $M_T$, which is the product of all the $B_{ij}$ matrices within the time-slice being updated. Figure~\ref{regions} shows what contributes to $M_B$ and $M_T$. When the configuration of bonds within the time-slice is changed then only $M_T$ changes to $M'_T$. The ratio $R$ from (\ref{ratior}) is given by
\begin{equation}
R \ =\ \frac{\det(\mathbbm{1}_N + M_B M_T')}
{\det(\mathbbm{1}_N + M_B M_T)}.
\label{rdefn}
\end{equation}
For reasons explained below, within a time-slice update we further subdivide the update to make what we will call \textit{block updates}. For each block update we randomly choose a spatial block containing a relatively small number of sites compared to the entire spatial lattice (highlighted in gray for Figure \ref{regions}), and focus on updating the bonds only within that block.  We will make enough block updates within a time-slice to theoretically cover the entire spatial lattice at least once before moving onto the next time-slice. While $R$ in (\ref{rdefn}) is the ratio of the proposed update configuration weight and the current configuration weight, we can define $R_{\rm curr}$ to be such a ratio for a current configuration during the block update and a background configuration at the beginning of the block update, and $R_{\rm new}$ to be such a ratio for a proposed configuration during a block update and the same background configuration. The weight ratio that we need is then found from $R_{\rm new}/R_{\rm curr}$ (the same concept used in (\ref{wtbgrat}) for fermion bags in the Lagrangian formalism). Since we usually have already found $R_{\rm curr}$ from a previous update proposal, the new update calculation usually consists of only one calculation of $R_{\rm new}$.

These ratios will in fact be determinants of smaller matrices than $N\times N$, because it can be shown that we can write (\ref{rdefn}) as
\begin{equation}
R \ =\ \frac{\det(\mathbbm{1}_N + M_B M_T')}
{\det(\mathbbm{1}_N + M_B M_T)} \ =\ \det\left(\mathbbm{1}_N + G_B\Delta \right),
\label{largem}
\end{equation}
where we have defined two new $N\times N$ matrices: $G_B = \left(\mathbbm{1}_N+M_B M_T\right)^{-1} M_B M_T$ and $\Delta = \left(M_T^{-1} M'_T - \mathbbm{1}_N\right)$. Since the bond matrices $B_{b}$ in different fermion bags commute, it is easy to verify that $\Delta$ is non-zero only within rows and columns which contain spatial sites connected to fermion bags that change. Thus during a block-update the size of the matrix $\Delta$ cannot be greater than the sum of the sites in all the fermion bags that touch the sites within the block. We refer to this set of sites, which can be larger than the spatial block size, as a \textit{super-bag} and denote its size as $s$. Since $\Delta$ is non-zero only in an $s\times s$ block matrix, it is easy to show that the computation of $R_{\rm curr}/R_{\rm new}$ (the ratio of the weight of the current or new configuration with that of the background configuration that existed at the time when the block update began) using (\ref{largem}), reduces to the computation of the determinant of an $s\times s$ matrix.

Within a block-update, it is clear that the matrices $M_T'$ that are found in $\Delta$ (\ref{largem}) must be updated often, once for each accept/reject proposal. For ease of computation and to ensure stability, the quantity we update is actually $F = G_B {M_T}^{-1} {M_T}'$, and the determinant we calculate is
\begin{equation}
R = \det\left(\left[\mathbbm{1}-G_B + F\right]_{s\times s}\right) = \left|\det\left(\left[\left(\mathbbm{1}-G_B\right)\mathcal{Q}^T + \mathcal{R}\right]_{s\times s} \right) \right|,
\label{qrrat}
\end{equation}
where we are using the $RQ$ factorization of $F$ into an upper triangular matrix $\mathcal{R}$ and an orthogonal matrix $\mathcal{Q}$, as in \cite{inbook}. Only the ${M_T}'$ matrices have to be updated each time, so we store an $RQ$ factorization of the $G_B {M_T}^{-1}$ product for the block update.

When we need to move to the next block in the same timeslice, $G_B$ is also updated according to the super-bag $S$ of the current block. Here we use the Woodbury identity to make the update only of order $O\left(sN^2\right)$. Assuming $G_B = \left(\mathbbm{1} + M_B M_T\right)M_B M_T$ before the update,
\begin{equation}
\begin{aligned}
&\mathbbm{1} -G_B' =\left(\mathbbm{1}+M_B M +  M_B M \left(M_T^{-1} M_T' - \mathbbm{1}\right)\right)^{-1}\\
&=\mathbbm{1}-G_B \\
&+\left[G_B\right]_{N\times s} \left(\left[ \mathbbm{1}-\mathcal{G}_T - G_B+ 2 \mathcal{G}_T G_B \right]_{s\times s}\right)^{-1}\left[\left(\mathbbm{1} -2\mathcal{G}_T \right)\right]_{s\times s} \left[\left(\mathbbm{1}-G_B\right)\right]_{s\times N}
\label{woodinv}
\end{aligned}
\end{equation}
where $M_T$ and ${M_T}'$ are matrix products in timeslice $T$ for the configurations that go with $G_B$ and $G_B'$, respectively, and $\mathcal{G}_T = \left(\mathbbm{1}+M_T^{-1} {M_T}'\right)^{-1}M_T^{-1} {M_T}'$. The symbol $\left[\;\right]_{s\times s}$ means only the rows and columns belonging to the super-bag $S$ are used, with $\left[\;\right]_{N\times s}$ and $\left[\;\right]_{s\times N}$ forming matrices from columns belonging to $S$ and rows belonging to $S$, respectively. The expression (\ref{woodinv}) is in a slightly different form than the typical Woodbury formula and is written specifically in terms of these $G_B$ and $\mathcal{G}_T$ quantities because doing so improves the numerical stability of the calculation.

%In Section III of the paper we mention using four different updates to generate the configurations $([x,d,t];t_0)$ in the two sectors. These updates consist of: (1) \textit{Sector-update:} We flip the sector $n \rightarrow 1-n$ while keeping $([x,d,t];t_0)$ fixed. (2) \textit{Move-update:} Since bond insertions commute with each other when they do not share a lattice site we can move all the bonds in time as long as two non-commuting operator insertions do not cross each other. We try to move roughly the same number of bonds moves as there are bonds in an equilibrated configuration. During this step $t_0$ and $n$ remain fixed. (3) \textit{Time-update:} $t_0 \leftrightarrow t_0'$ while keeping the bond configuration $[x,d,t]$. We perform this update only in the $n=0$ sector where it is trivial. (4) \textit{Bond-update:} This is the most time consuming update where we attempt to change the entire bond configuration $[x,d,t] \leftrightarrow [x',d',t']$ while keeping $t_0$ and $n$ fixed.
\section{Stabilization of Calculations}
The remaining challenge lies in calculating and updating $G_B$ from one time-slice to the next, which is well-known to be unstable if computed naively \cite{inbook}. The problem stems from the fact that the matrices $M_B$, $M_T$, and $M_{T'}$ cannot be constructed naively as a product of the block matrices $B_b$ and then added to an identity matrix at the end before we take an inverse or determinant, since the calculation involves sums of terms that can be orders of magnitude apart. In a typical auxiliary field Monte Carlo method the product is accomplished using the singular value decomposition (SVD) or Gram-Schmidt decomposition of the individual $B$ matrices that are contained in these matrices. This is time consuming and it would be helpful to avoid it as much as possible. In our case, since each $B_{b}$ is only non-trivial in a $2\times 2$ block, we can indeed multiply many of them at a time without worrying about stabilization issues. We call each such grouping as a partial product $M_p$. However the process of combining these partial products will still need some kind of stabilization in principle. There are three main numerical instabilities that we have to deal with: (1) computing $G_B$ and then updating it when we move on to a different time-slice, (2) updating $G_B$ between block-updates within the same time-slice, and (3) updating $\Delta$ for each configuration update and ensuring a stable determinant. We have already discussed how we deal with (2) and (3), by giving the special Woodbury update form we use in (\ref{woodinv}) and the RQ factorization we use in (\ref{qrrat}). We will now discuss how we deal with (1) in some detail. We have found that we can accomplish a stable calculation of $G_B$ without using SVDs, as discussed below.
%Our stabilization procedures follow. For the sector-update and the bond-update, which as mentioned earlier are the more expensive update types, we need to compute the ratio $R$, defined by
%\begin{equation}
%R \ =\ \det(\mathbbm{1}_N + M_B M_T')/\det(\mathbbm{1}_N + M_B M_T) \ =\ \det\left(\mathbbm{1}_N + G_B\Delta \right),
%\label{ratR}
%\end{equation}
%where $G_B$ is $G_B = \left(\mathbbm{1}_N+M_B M_T\right)^{-1} M_B M_T$ and $\Delta = \left( M_T^{-1} M_T' - \mathbbm{1}_N\right)$.

First note that given two matrices, $M_1$ and $M_2$, we have the identity
\begin{equation}
\begin{aligned}
\left( \mathbbm{1}+\right.&\left.M_1  M_2\right)^{-1} 
= \left(\mathbbm{1}+M_2\right)^{-1} \\ 
\times & \left(\left(\mathbbm{1}+M_1\right)^{-1} \left(\mathbbm{1}+M_2\right)^{-1}\right.
\left. + \left(\mathbbm{1}+M_1\right)^{-1}M_1 M_2\left(\mathbbm{1}+M_2\right)^{-1}\right)^{-1} \left(\mathbbm{1}+M_1\right)^{-1}.
\label{inv}
\end{aligned}
\end{equation}
Further it is convenient to know that $\left( \mathbbm{1}+ M \right)^{-1} M =  \mathbbm{1} - (\mathbbm{1}+M)^{-1}$. This means we can construct $(1+M_1 M_2)^{-1}$
from $(1+M_1)^{-1}$ and $(1+M_2)^{-1}$. Hence we can build $G_B$ from partial versions labeled as $G_T=(1+M_T)^{-1}M_T$ associated with each time slice $T$. The matrix $G_T$ in turn is obtained by combining the partial matrices $G_p= (1+M_p)^{-1}M_p$ within a time slice, where the matrices $M_p$ are the partial products we explained above. In fact the last step can be performed more efficiently using the idea of fermion bags. For each time-slice, we first focus our efforts in constructing $G_T$ at the fermion bag level. Within each fermion bag we build up partial $G_p$ matrices (built from $M_p$ partial groupings), and combine them to create $G_f$, which is the total contribution from that bag. Thus each $G_f$ is a matrix with $f$ rows and $f$ columns corresponding to the fermion bag sites. We can then combine the $G_f$ matrices into a $G_T$ matrix, which has distinct blocks according to the fermion bags. Thus, while $G_B$ is an $N\times N$ matrix, we can use the idea of fermion bags along with the identity (\ref{inv}) to reduce the number of operations. While this unfortunately does not reduce the scaling of the algorithm as compared to a traditional auxiliary field Monte Carlo method which scales as $O\left(\beta N^3\right)$, it does significantly reduce the slope.

As we build $G_B$, the partial forms of $G_B$ consisting of various combinations of the $G_T$ matrices are already stored either in computer memory or on the hard disk. The stored partial forms allow us to make fast updates to $G_B$ when we move sequentially through the time-slices by only having to combine one or two matrices per timeslice update. This allows us to keep the linear $\beta$ scaling. More details on how the storage scheme works can be found in \cite{Wang:2015rga}.

\chapter{Results for the $t$-$V$ Model}
{\small\textit{``A very large part of space-time must be investigated, if reliable results are to be obtained.'' \\
--Alan Turing}}
\newline
\newline
\section{Quantum Critical Behavior in the $t$-$V$ Model}
%To test the efficiency of the Hamiltonian fermion bag algorithm, we have used it to study the quantum critical behavior of the $t$-$V$ model on a $\pi$-flux lattice, given by
We finally discuss how we have used the fermion bag algorithm discussed in Chapter 5 to study the quantum critical behavior of the $t$-$V$ model Hamiltonian (\ref{model}). More specifically, the Hamiltonian we have studied is defined on a square lattice with a $\pi$-flux on each plaquette (introduced in (\ref{hammatrix})). We can write it as
\begin{equation}
    H = -t\sum_{\left\langle ij \right\rangle} \eta_{ij}\left(c_i^\dagger c_j + c_j^\dagger c_i \right) + V\sum_{\left\langle ij \right\rangle} \left(n_i-\frac{1}{2}\right) \left(n_j -\frac{1}{2}\right).,
    \label{tvmodel6}
\end{equation}
where $\eta_{i,i+\hat{e}_x} = 1$ for all $i$ and $\eta_{i,i+\hat{e}_y}= (-1)^{i_x}$ for even $i$. As discussed in Chapter 3, this model is expected to undergo a second-order quantum phase transition at a critical coupling $V=V_c$, between a semimetal phase for small $V$ and an insulator phase for large $V$ with a charge density wave ordering where $\sigma_i \left(n_i-1/2\right)\neq 0$. The symbol $\sigma_i$ is the parity factor, introduced in (\ref{phtran}), and is +1 if the point $i$ is on the even sublattice and -1 if the point $i$ is on the odd sublattice. The transition is expected to belong to the chiral Gross-Neveu universality class with one flavor of four-component Dirac fermions.

Using renormalization group analysis it is possible to show that second-order phase transitions can be characterized by critical exponents \cite{Zinn-Justin:2280881}. In the bosonic sector, there are usually two independent leading exponents, which we can take to be $\nu$ and $\eta$. The $\nu$ exponent describes the power law behavior of the correlation length near the critical coupling:
\begin{equation}
    \xi\:\: \propto \:\: \frac{1}{\left|\left(V-V_c\right) / V_c\right|^\nu}.
    \label{corrl}
\end{equation}
The $\eta$ exponent describes the critical behavior of the order parameter correlation function $G\left(r\right)$, which decays algebraically with the distance $r$ as
\begin{equation}
    \left\langle G\left(r\right)\right\rangle \:\:\propto  \:\:\frac{1}{r^{d-2+\eta}},
    \label{corrf}
\end{equation}
where $d$ is the space-time dimension of the system \cite{continentino_2017}. Here we assume that the critical point is relativistically invariant, which is expected to be true for the $t$-$V$ model.

For the $t$-$V$ model, the order parameter that describes the phase transition is the charge density wave order characterized by the local operator $\sigma_i \left(n_i-1/2\right)$. To detect order in this operator, we measure the equal time correlation function of the operator $C$ at a distance L/2 defined as
\begin{equation}
    C=\sigma_{\left(0,0\right)}\sigma_{\left(L/2,0\right)}\left(n_{\left(0,0\right)}-\frac{1}{2}\right)\left(n_{\left(L/2,0\right)}-\frac{1}{2}\right).
    \label{tvcorrelate}
\end{equation}
How we can measure this operator in our QMC was introduced in Chapter 5 (see (\ref{correlation}) and the discussion following it.) The ordered pairs in the subscripts in (\ref{tvcorrelate}) label the $x$ and $y$ coordinates of the sites on the square lattice, with $\left(0,0\right)$ being the origin, and $\left(L/2,0\right)$ being $L/2$ units in the $x$ direction from the origin, where $L$ is width of the entire lattice.

Since a single correlation length dominates the critical region, we can combine (\ref{corrl}) and (\ref{corrf}) with dimensional analysis and derive a finite size scaling relation for our correlation observable defined in (\ref{tvcorrelate}):
\begin{equation}
    \left\langle C\left(\left(L/2,0\right)\right) \right\rangle 
    \:\:= \:\: \frac{1}{L^{1+\eta}} f\left(\left(V-V_c\right)L^{1/\nu}/t\right).
    \label{scalerel}
\end{equation}
The expression (\ref{scalerel}) will play a key role in our determination of the critical exponents.

\section{Testing the Algorithm for the $t$-$V$ Model}
Before making measurements on large lattices, we needed to test both the stabilization aspects and the QMC updates for our algorithm. We discussed in Chapter 5 how one of the main steps in the algorithm involves computing ratios (\ref{ratior}), $R$, of configuration weights, $\Omega_n([b,\tau];\tau_0)$, as found in (\ref{confweights}). 

While these weights are most efficiently computed using the BSS formula (\ref{bssform}), the computation needs to be stabilized and the stabilization techniques we developed (described in Sections 4 and 5 of Chapter 5) need to be tested. Fortunately, as we have discussed in Chapter 4, the weights in (\ref{confweights}) can also be computed as determinants of $2k\times 2k$ antisymmetric matrices (\ref{pfafdet}), where $k$ is the number of bond insertions. Although the calculation of each determinant with the Pfaffian formula is more time consuming than with the BSS formula, it is stable. We have computed $R$ by both these methods on small lattices and confirmed that they agree to very high accuracy.

\begin{table}[t]
\begin{tabular*}{\linewidth}{@{\extracolsep{\fill}}|c|c|c|c|c|}
\hline
\multicolumn{5}{|c|}{\textbf{2$\times$2 Lattice, V=1.200}} \\
\hline
& $\beta=1.0$ & $\beta=2.0$ & $\beta=4.0$ & $\beta=8.0$  \\
\hline
\hline
MC (f=10) & 0.09838(16) & 0.14265(15) & 0.15091(9) & 0.15105(7) \\
\hline
\hline
MC (f=50) & 0.09872(23) & 0.14242(19) & 0.15082(13) & 0.15107(11) \\
\hline
\hline
Exact & 0.098550... & 0.142590... & 0.150801... & 0.150939... 
\\
\hline
\end{tabular*}\\
\begin{tabular*}{\linewidth}{@{\extracolsep{\fill}}|c|c|c|c|c|}
\hline
\multicolumn{5}{|c|}{\textbf{2$\times$2 Lattice, V=1.304}} \\
\hline
& $\beta=1.0$ & $\beta=2.0$ & $\beta=4.0$ & $\beta=8.0$  \\
\hline
\hline
MC (f=10) & 0.10290(17) & 0.14562(13) & 0.15286(9) & 0.15320(7) \\
\hline
\hline
MC (f=50) & 0.10261(25) & 0.14557(18) & 0.15287(13) & 0.15305(11) \\
\hline
\hline
Exact & 0.102948... & 0.145738... & 0.152973... & 0.153078... 
\\
\hline
\end{tabular*}\\
\begin{tabular*}{\linewidth}{@{\extracolsep{\fill}}|c|c|c|c|c|}
\hline
\multicolumn{5}{|c|}{\textbf{4$\times$4 Lattice, V=1.200}} \\
\hline
& $\beta=1.0$ & $\beta=2.0$ & $\beta=4.0$ & $\beta=8.0$  \\
\hline
\hline
MC (f=10) & 0.03117(8) & 0.05955(12) & 0.07781(15) & 0.07997(14) \\
\hline
\hline
MC (f=50) & 0.03127(8) & 0.05948(13) & 0.07780(16) & 0.07994(16) \\
\hline
\hline
Exact & 0.031285... & 0.059458... & 0.077769... & 0.080009... 
\\
\hline
\end{tabular*}\\
\begin{tabular*}{\linewidth}{@{\extracolsep{\fill}}|c|c|c|c|c|}
\hline
\multicolumn{5}{|c|}{\textbf{4$\times$4 Lattice, V=1.304}} \\
\hline
& $\beta=1.0$ & $\beta=2.0$ & $\beta=4.0$ & $\beta=8.0$  \\
\hline
\hline
MC (f=10) & 0.03819(10) & 0.07272(15) & 0.08959(17) & 0.09060(16) \\
\hline
\hline
MC (f=50) & 0.03798(13) & 0.07293(16) & 0.08934(18) & 0.09087(18) \\
\hline
\hline
Exact & 0.038105... & 0.072760... & 0.089511... & 0.090672... 
\\
\hline
\end{tabular*}
\\
\caption{Monte Carlo measurements (MC) and exact calculation values (Exact) on small lattices for different parameter combinations.}
\label{tablecheck}
\end{table}

We have also tested the correctness of the QMC updates by comparing the results for the observable $\langle C\rangle$ obtained from the Monte Carlo algorithm against exact diagonalization calculations. Table \ref{tablecheck} summarizes the results on 2$\times$2 and 4$\times$4 lattices at couplings $V=1.200,1.304$ for inverse temperature values of $\beta=1.0,2.0,4.0,8.0$. We have also verified that the value of the reweighting factor $f$ does not affect the observable. We show our results for $f=10.0$ and $50.0$. The table shows that the Monte Carlo results agree with the exact calculations within errors as expected. Additionally, similar tests were performed at $V=1.304$ for $f=1.0,2.0,$ and $f=20.0$, and again the exact results were within the errors of the Monte Carlo results.

\begin{figure}[t]
    \centering
    \includegraphics[width=8cm]{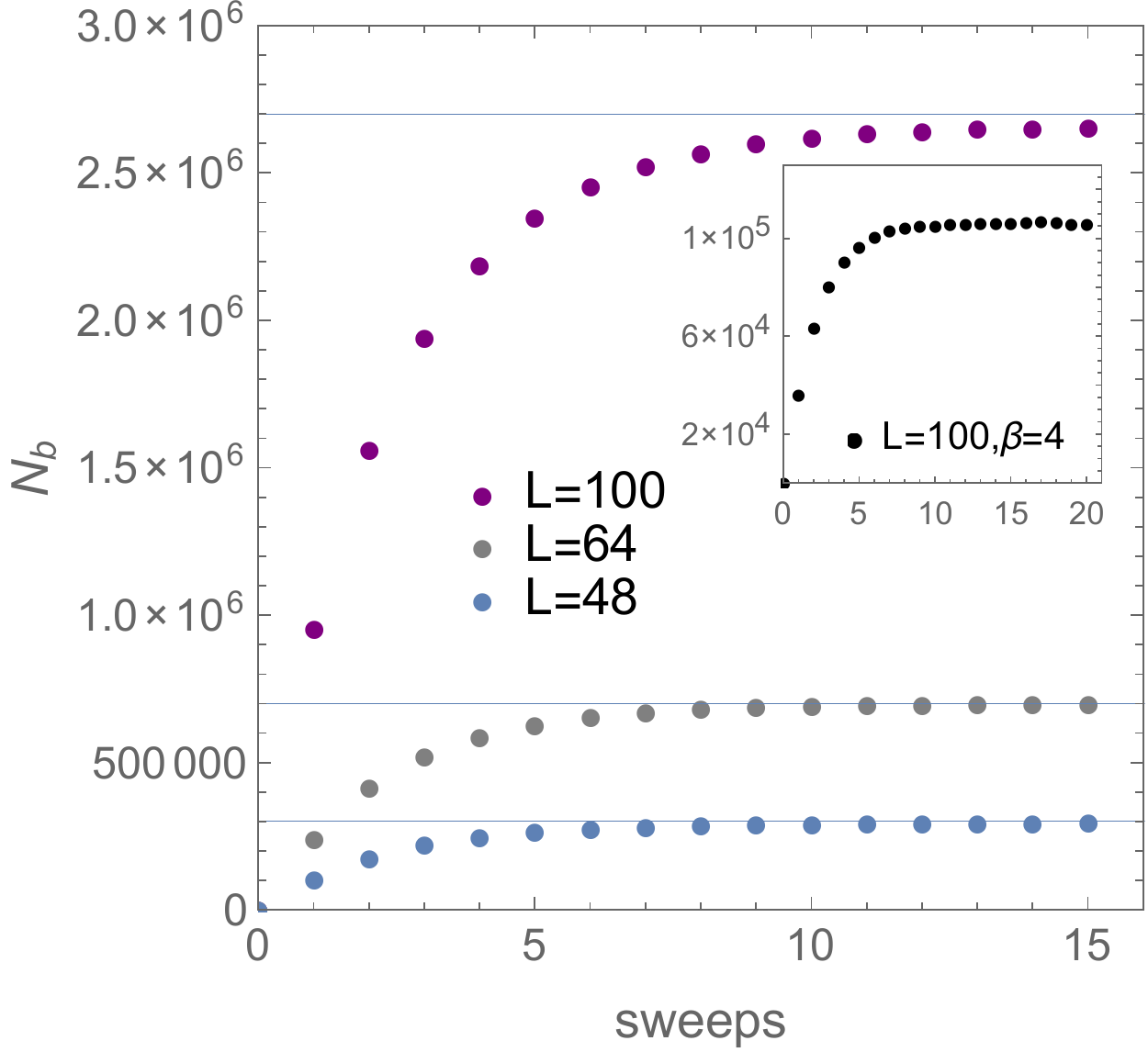}
    \caption[Plot showing $\beta=L$ equilibration for the $t$-$V$ model.]{Plot showing $\beta=L$ equilibration of the total number of bonds $N_b$ in a bond configuration starting from zero, as a function of Monte Carlo sweeps. The horizontal lines show the expected equilibrated values. The times for a single bond-update on a single core are approximately 30 days for $L=100$, 30 hours for $L=64$, and 4 hours for $L=48$. Inset shows equilibration at $L=100$, $\beta=4$.}
    \label{fig:equil}
\end{figure}

To see how far we could push our new algorithm, we looked at its results for some very large lattices. Note that our algorithms are designed so that the maximum fermion bag size in the block updates remains independent of the lattice size near the critical point. For the $t$-$V$ model we tested this by taking $\beta=4.0$ and dividing the imaginary time direction into $16$ time-slices, and studied the fermion bag size as a function of the lattice size. For equilibrated configurations of $L=48, 64$ and $100$, the average maximum fermion bag size within a time-slice was about $30$ sites--independent of $L$. Further tests suggest that the optimal temperature (time-slice size) is roughly $0.25$. Because of this, we set the timeslice size ($w_\tau$ from Chapter 5) to $0.25$ for all calculations. This implies that the block updates usually involve computing determinants of 30 x 30 matrices independent of the lattice size.

In Figure~\ref{fig:equil} we show equilibration of $N_b$ (the total number of bonds in a configuration) as a function of sweeps for $\beta=L=48,64,100$ ($2,304$, $4,096$, and $10,000$ site lattices, respectively) at the coupling $V=1.304 t$, which was predicted to be the critical coupling in \cite{1367-2630-16-10-103008}. Although the $L=100$ data has not equilibrated, there is no bottleneck (see inset of Fig.~\ref{fig:equil}). We estimate the bond density at equilibrium to be $N_b/\beta L^2 \approx 2.7$, which means at $L=\beta=100$ we will have roughly $2.7$ million bonds after equilibration, which appears consistent with the figure. A single sweep requires roughly a month on a single 3GHz CPU core.

\begin{figure}
\begin{center}
\includegraphics[width=6cm]{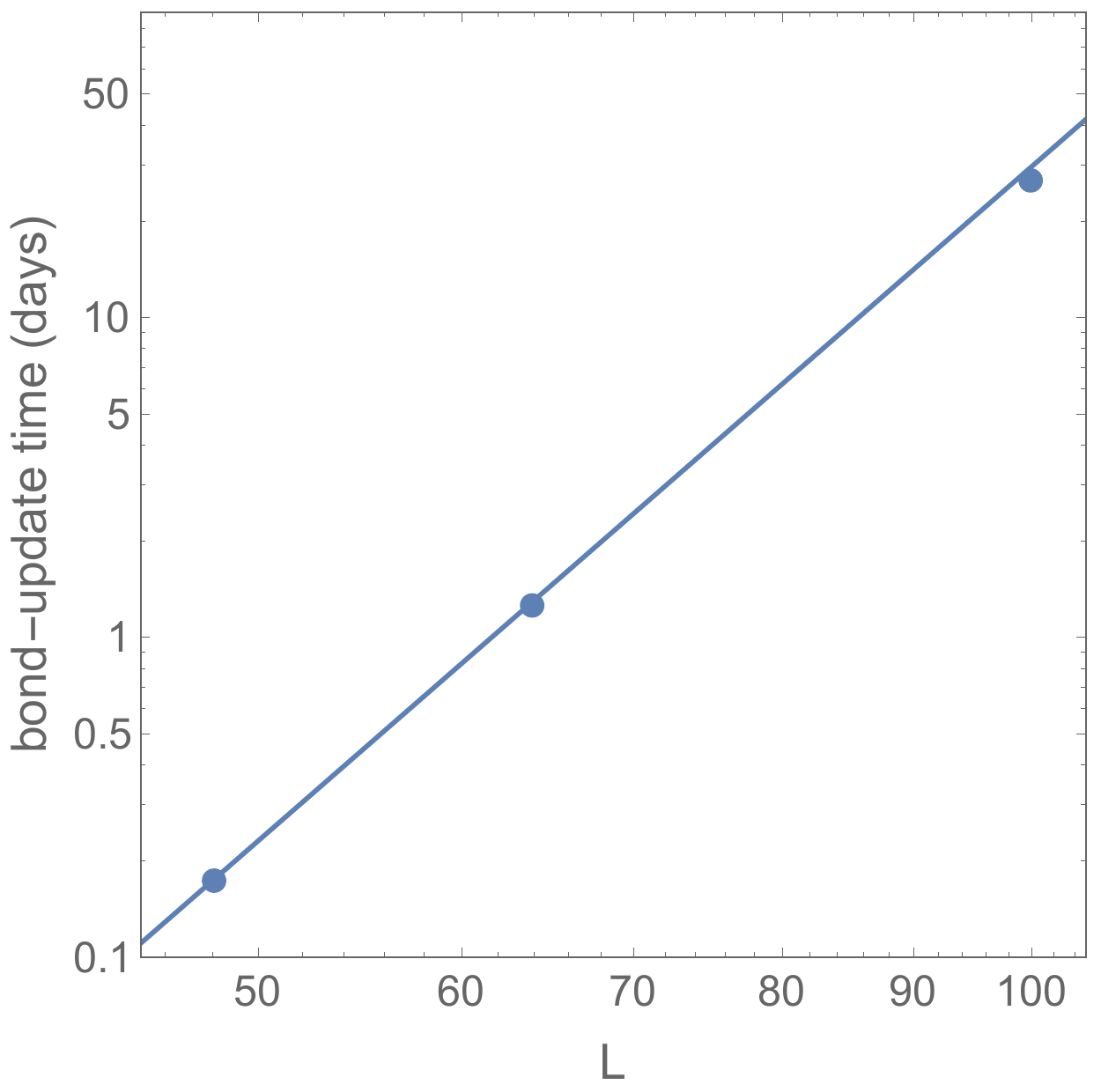}
\end{center}
\caption[Plot showing $L^7$ scaling for algorithm updates.]{Plot showing the time to complete a single \textit{bond-update} (in days) for $L=48,64,100$ with $\beta=L$ at $V/t=1.304$. The solid line is a plot of $\tau = 3 \times 10^{-13} L^7$ (days).}
\label{scales}
\end{figure}
\begin{figure}[t]
\minipage{0.3\textwidth}
  \includegraphics[width=\linewidth]{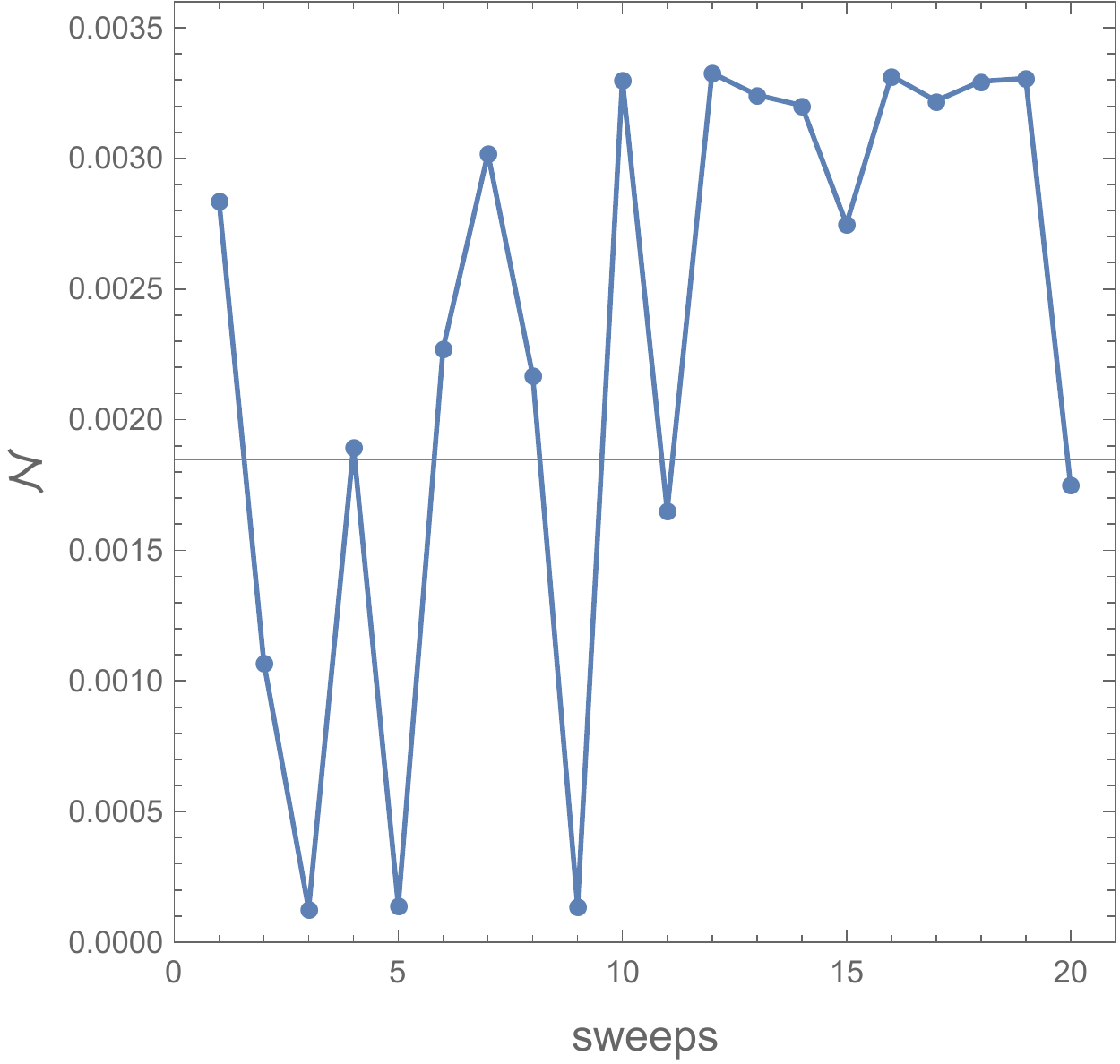}
\endminipage\hfill
\minipage{0.3\textwidth}
  \includegraphics[width=\linewidth]{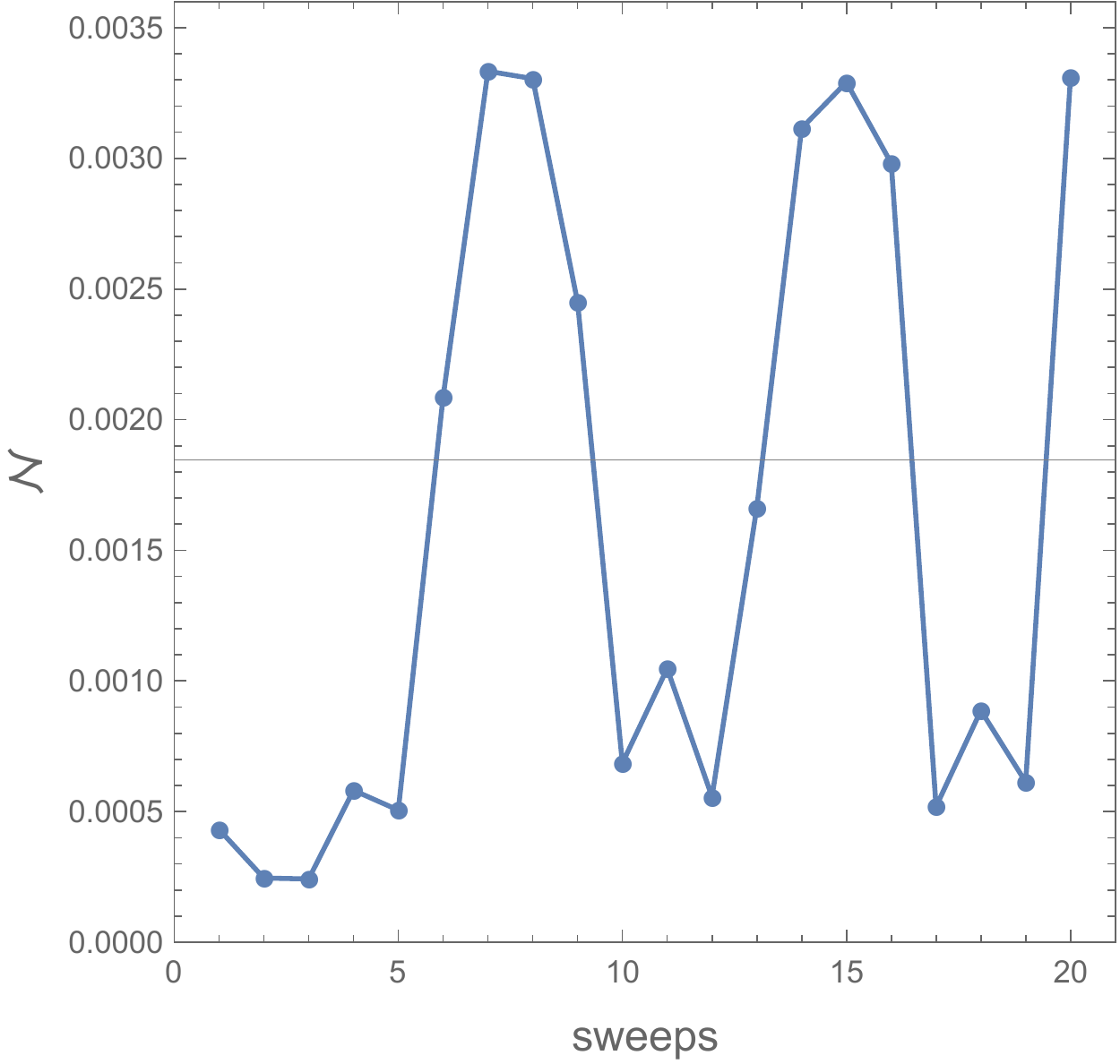}
\endminipage\hfill
\minipage{0.3\textwidth}%
  \includegraphics[width=\linewidth]{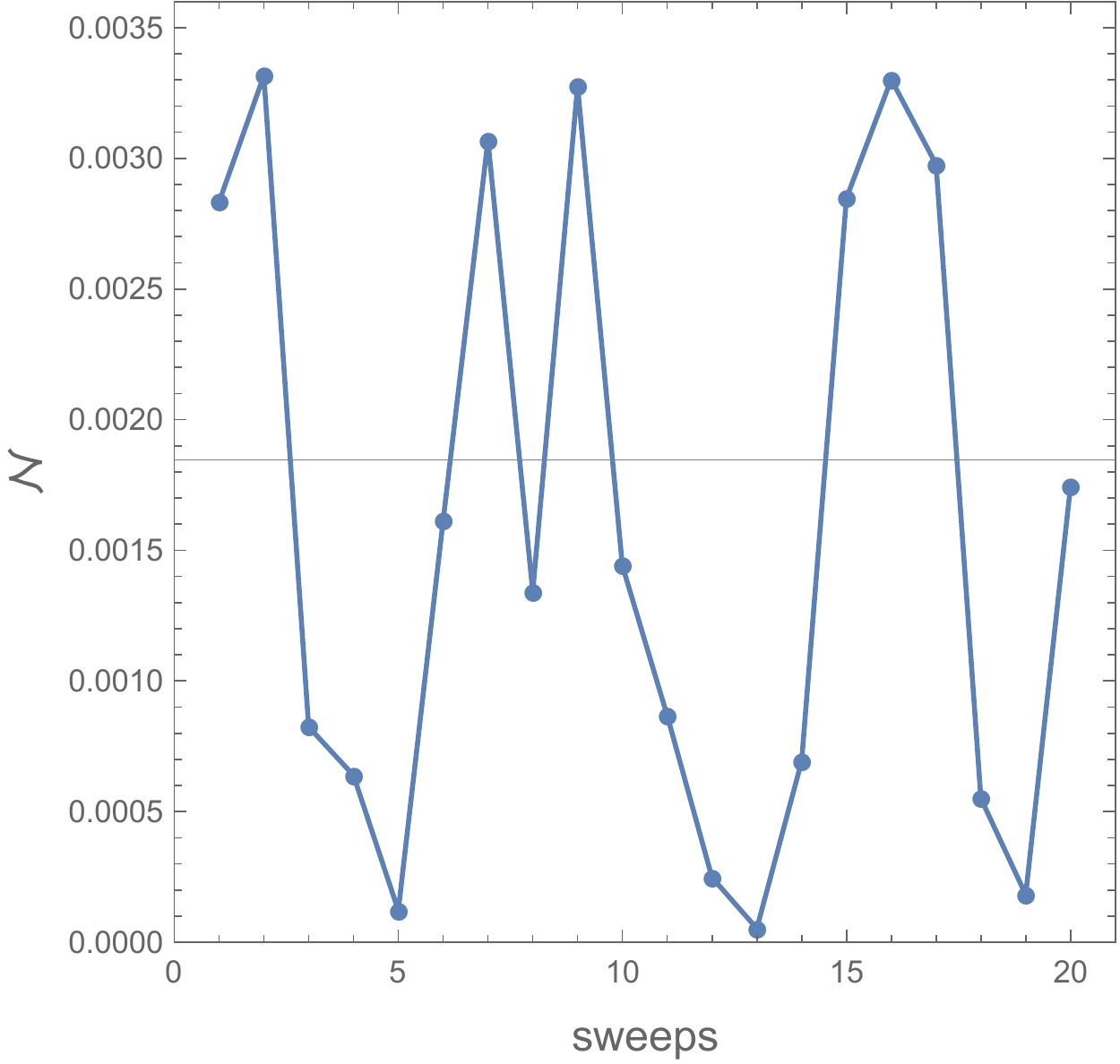}
\endminipage
 \caption[Measurements of $\mathcal{N}$ taken from different threads starting from the same equilibrated configuration.]{Measurements of $\mathcal{N}$ taken from different threads starting from the same equilibrated configuration for $V=1.304$ at $L=48$. The first two measurements for each thread have been discarded. The full average from the data, $.0019$, is given by the gray line in each plot. The first plot has an average of $.0023$, the second an average of $.0016$, and the third an average of $.0014$.}\label{configurations}
\end{figure}

We have confirmed that the computation time of the bond update as a function of the lattice size and imaginary time scales as expected. In Figure \ref{scales}, we show the $O(\beta N^3)$ scaling of time for a complete \textit{bond-update}. In particular we plot the bond-update time $\tau_b$ (in days) as a function of $L$ for three different lattice sizes at the coupling $V/t=1.304$ close to the critical point. Since $\beta=L$ we expect a scaling of $O(L^7)$. The solid line in the figure is the plot of $\tau_b = 3 \times 10^{-13} L^7$ (days) and roughly passes through all the points. We have compared the performance of our Hamiltonian Fermion Bag algorithm with that of an Auxiliary Field Quantum Monte Carlo algorithm, which as discussed in Chapter 5, also scales as $O\left(\beta N^3\right)$. The results are given in Appendix D.

In order to compute our observable accurately, we need to generate a large number of statistically independent configurations. For small lattices it is easy to run many independent processes from a configuration without any bonds and let them each equilibrate independently before taking data. We then generate about $2\times 10^4$ equilibrated configurations for statistics. For large ($L=48,64$) lattices, where it is difficult to equilibrate the configurations, we start with 10 independent equilibrated configurations. We then make 100 copies of each of the equilibrated configuration and distribute them over 1000 cores. We then start an independent Monte Carlo thread on each of these cores. We used two computing resources outside of our local cluster: The Open Science Grid and Bridges at the Pittsburgh Supercomputing Center through XSEDE to obtain 1000 cores. In Figure~\ref{configurations} we plot the Monte Carlo fluctuations of three such threads starting from the same equilibrated configuration with different random number sequences. Note that the observable ${\cal N}$ defined in (\ref{nobs}) seems to become decorrelated within a few sweeps. We generate $20$ sweeps of data on each of the $1000$ threads for our statistics, and compute averages after throwing away the first few sweeps.

\begin{table}[b]
\begin{tabular*}{\linewidth}{@{\extracolsep{\fill}}|c|c|c|c|}
\hline
\multicolumn{4}{|c|}{\textbf{Chiral Ising Universality Class Results}} \\
\hline
 Model & Method & $\nu$ & $\eta$  \\
\hline
\hline
Honeycomb & QMC (CT-INT) \cite{1367-2630-16-10-103008} & 0.80(3) & 0.302(7) \\
\hline
\hline
$\pi$-flux & QMC (CT-INT) \cite{1367-2630-16-10-103008} & 0.80(6) & 0.318(8) \\
\hline
\hline
 Honeycomb & QMC (AUX) \cite{1367-2630-17-8-085003} & 0.77(3) & 0.45(2) \\
 \hline
\hline
 $\pi$-flux & QMC (AUX) \cite{1367-2630-17-8-085003} & 0.79(4) & 0.43(2) \\
  \hline
\hline
 Honeycomb & QMC (CT-INT) \cite{Hesselmann:2016tvh}  & 0.74(4) & 0.275(25)
\\
  \hline
\hline
 Gross-Neveu & $4-\epsilon$ (third-order) \cite{Mihaila:2017ble}  & 0.858 & 0.463
\\
\hline
\hline
 Gross-Neveu & $4-\epsilon$ (fourth-order) \cite{Zerf:2017zqi}  & 0.91 & 0.4969
\\
\hline
\hline
 Gross-Neveu & FRG \cite{PhysRevLett.86.958,PhysRevB.66.205111}  & 0.927 & 0.525
\\
\hline
\hline
Gross-Neveu & Conformal Bootstrap \cite{Iliesiu2018}  & 1.316 & 0.551
\\
\hline
\hline
 Gross-Neveu & 1/N Expansion \cite{PhysRevB.66.205111}  & 0.738 & 0.635
\\
\hline
\end{tabular*}
\\
\caption[Critical exponents according to various continuum and QMC methods.]{Critical exponents according to various  continuum and QMC methods. Organized similarly to the compilation in \cite{sorella}.}
\label{tab:compare}
\end{table}

\section{Results}
An exciting aspect of our sign problem solution for the $t$-$V$ model was that it opened the door to the first quantum Monte Carlo calculations for the critical behavior in the Chiral Ising universality class with a single four-component Dirac fermion. In response to the fact that we had proven that QMC could be applied successfully to the $t$-$V$ model, there were multiple QMC studies carried out for this model within the first year of our paper. Table \ref{tab:compare} summarizes the QMC results along with those in the continuum that have been obtained up until the present. The values for the critical exponents $\nu$ and $\eta$, which were discussed in Section 6.1, are given for each of the calculations.

There are a couple things to note from the table. First, while the QMC results seem to be in general agreement regarding the $\nu$ value, there is significant variation for the $\eta$ value. The auxiliary field methods (AUX) in the table, which are discrete time methods, have $\eta$ significantly larger than the continuous time (CT-INT) methods. On the other hand the auxiliary field methods go to larger lattices ($1152$ sites \cite{1367-2630-17-8-085003} for the honeycomb lattice, $484$ sites \cite{1367-2630-17-8-085003} on the $\pi$-flux lattice) than the continuous time methods ($882$ sites \cite{Hesselmann:2016tvh} and $450$ sites \cite{1367-2630-16-10-103008} on the honeycomb lattice, $400$ sites \cite{1367-2630-16-10-103008} on the $\pi$-flux lattice). Additionally, the QMC methods for the most part seem to have lower values for both $\eta$ and $\nu$ than the continuum techniques. The QMC (AUX) methods in \cite{1367-2630-16-10-103008} were able to reproduce the lower $\eta$ values of the CT-INT calculations on smaller lattices. This suggested that there were substantial finite size effects, and our hope was that perhaps we could use the Fermion Bag algorithm for the $t$-$V$ model to reach larger lattices and resolve some of this conflict.

\begin{table}[b]
\begin{center}
\begin{tabular}{|l||l|l|}
\hline
 $V/t$ & $L=12$ & $L=16$ \\
\hline
$1.200$ & 0.01008(10) & 0.00527(7)  \\
\hline
$1.250$ & 0.01410(14) & 0.00833(10) \\
\hline
$1.270$ & 0.01586(16) & 0.00997(11) \\
\hline
$1.280$ & 0.01694(16) & 0.01086(12) \\
\hline
$1.296$ & 0.01910(17) & 0.01310(15) \\
\hline
$1.304$ & 0.01990(14) & 0.01354(13)  \\
\hline
$1.350$ & 0.02582(23) & 0.02012(20) \\
\hline
$1.400$ & 0.03413(30) & 0.02903(30) \\
\hline 
\end{tabular}
\vspace{.5cm}
\caption[Our Monte Carlo results for the $t$-$V$ model: small lattices.]{Our small lattice Monte Carlo results for the $t$-$V$ model (\ref{tvmodel6}) on a square lattice with $12\leq L \leq 16$ and $\beta=L$. This data was not included in the scaling fit.}
\label{datatable1}
\end{center}
\end{table}

\begin{table}[t]
\begin{center}
\begin{tabular}{|l||l|l|l|l|l|}
\hline
 $V/t$ & $L=20$ & $L=24$ & $L=32$ & $L=48$ & $L=64$ \\
\hline
$1.200$ & 0.00298(3) & 0.00184(3) &
0.00080(1) & $\qquad -$ & $\qquad -$ \\
\hline
$1.250$ & 0.00545(6) & 0.00380(5) & 0.00204(2) & 0.00074(2) & $\qquad -$ \\
\hline
$1.270$ & 0.00699(8) & 0.00517(7) & 0.00315(4) & 0.00151(3) & 0.00085(1) \\
\hline
$1.280$ & 0.00787(9) & 0.00590(9) & 0.00377(4) & 0.00204(3) & 0.00130(2)  \\
\hline
$1.296$ & 0.00946(10) & 0.00740(9) 
& 0.00512(6) & 0.00339(5) & $\qquad -$ \\
\hline
$1.304$ & 0.01022(8) & 0.00844(9) 
& 0.00611(6) & 0.00423(5) & $\qquad -$ \\
\hline
$1.350$ & 0.01705(16) & 0.01522(16) 
& 0.01426(18)* & $\qquad -$ & $\qquad -$\\
\hline
$1.400$ & 0.02707(20) & 0.02630(35) & 0.02637(38)* & $\qquad -$ & $\qquad -$\\
\hline
\end{tabular}
\vspace{.5cm}
\caption[Our Monte Carlo results for the $t$-$V$ model: large lattices.]{Our large lattice Monte Carlo results for the $t$-$V$ model (\ref{tvmodel6}) on a square lattice with $20\leq L \leq 64$ and $\beta=L$. The starred data was not included in the scaling fit.}
\label{datatable}
\end{center}
\end{table}

%With these tests performed, we present the data that we obtained for the $t$-$V$ model on the $\pi$-flux lattice. We have computed the critical exponents at the quantum phase transition between the massless and the massive fermion phases. These critical exponents are expected to belong to the Ising Gross-Neveu universality class with $N_f=1$ four-component Dirac fermions  \cite{PhysRevLett.86.958,Mihaila:2017ble,Zerf:2017zqi}.

All our data obtained for our observable is tabulated in Tables \ref{datatable1} and \ref{datatable}. In all of these results we set $\beta=L$. We expect the following asymptotic behavior for our specific observable, $C$:
\begin{equation}
    \left\langle C \right\rangle \sim \left\{ \begin{array}{ccc} L^{-4}&, & V < V_c \\
    1&, & V > V_c\\
    L^{-1-\eta}&, & V = V_c
    \end{array} \right. ,
    \label{piecescale}
\end{equation}
and near the critical point we can get a more detailed formula:
\begin{equation}
\langle C \rangle = \frac{1}{L^{1+\eta}} f\left(\left(V-V_c\right)L^{1/\nu}/t\right), \qquad V\approx V_c.
\label{scaleapprox}
\end{equation}
The last relation was derived in the first section of this chapter \cite{Cardy:1988ag,PhysRevB.89.094516}.

\begin{figure}
\begin{center}
  \includegraphics[width=8cm]{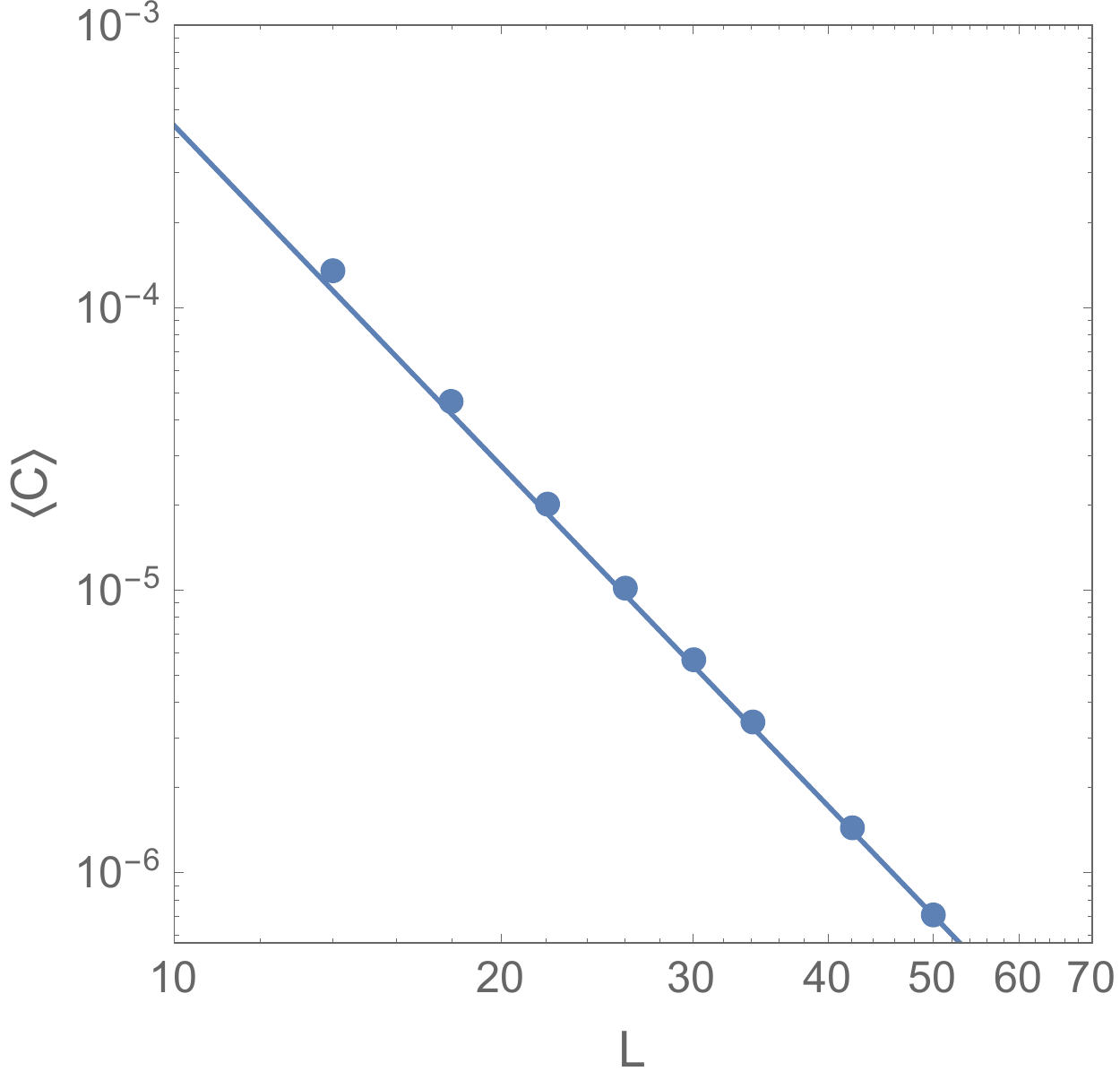}
  \end{center}
  \caption[Exact data at zero coupling.]{Exact data for the $t$-$V$ model at $V=0.0$. The correlation function goes as $L^{-4}$. The solid line shows the slope for $L^{-4}$ behavior.}\label{free}
\end{figure}

We now proceed to check our Monte Carlo results for consistency with (\ref{piecescale}) and (\ref{scaleapprox}). First, in the region where $V< V_c$, we expect to see an asymptotic behavior of $\left\langle C \right\rangle \sim L^{-4}$. This occurs because for the free theory there is no scale in the problem and we can use dimensional analysis in the Lagrangian formalism to show that the conditions imposed by the dimensionless action ensure that a two-point correlation function must have dimension $L^{-4}$. Figure \ref{free} shows data from an exact calculation at $V=0$. The $L^{-4}$ asymptotic behavior is clear.

\begin{figure}
\begin{center}
  \includegraphics[width=8cm]{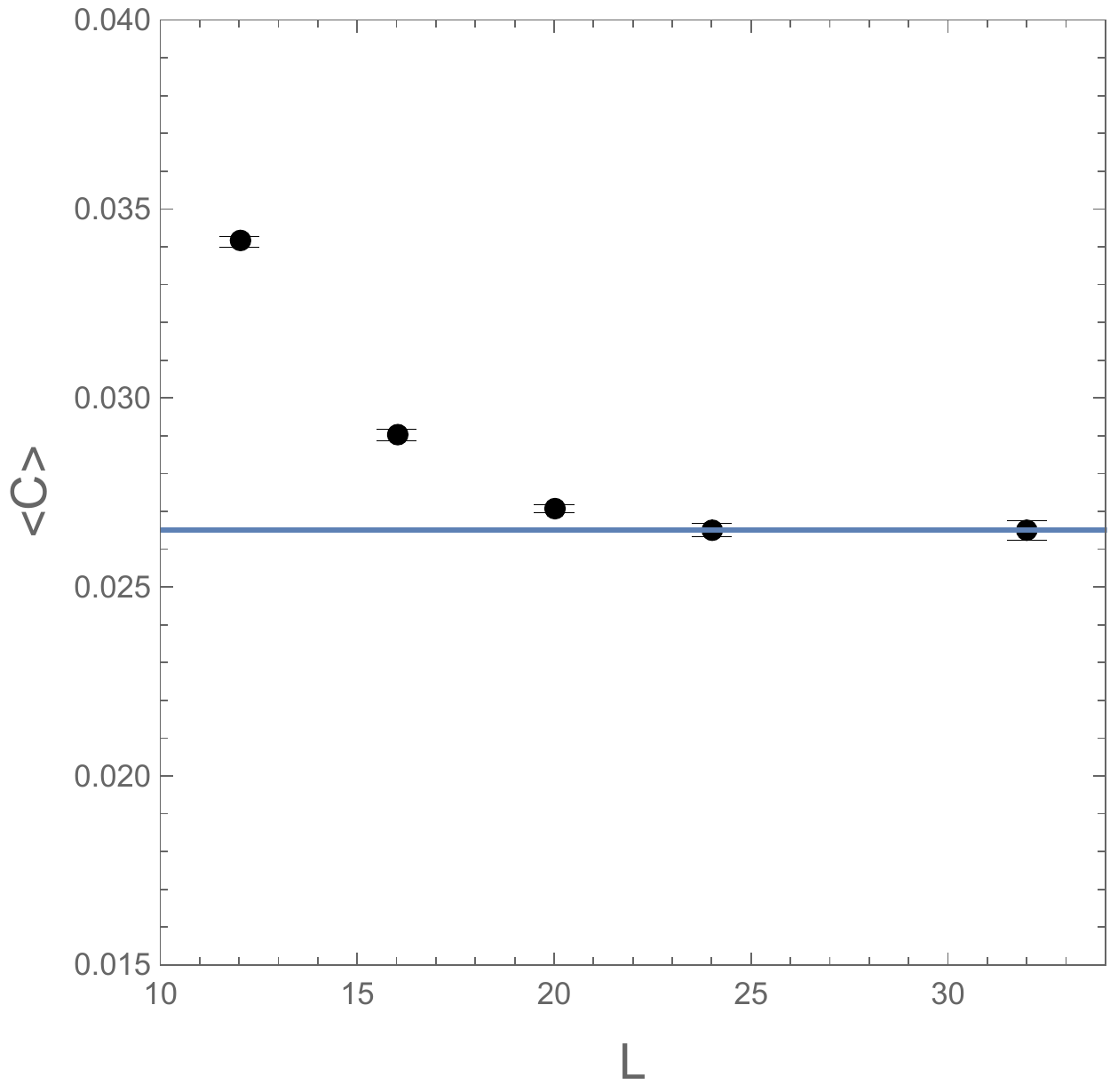}
  \end{center}
  \caption[Exact data at large coupling.]{Monte Carlo data for the $t$-$V$ model at $V=1.4$. The correlation function goes as a constant at large $L$, as expected. The solid line highlights this constant behavior.}\label{brokenphase}
\end{figure}

On the other hand, when $V> V_c$, we expect to see $\left\langle C \right\rangle$ go as a constant. This occurs because in a local theory we expect correlation functions to behave as
\begin{equation}
  \lim_{i\rightarrow {\rm large}}  \left\langle O_0 O_i \right\rangle \sim \left\langle O_0 \right\rangle \left\langle O_i \right\rangle,
\end{equation}
where the subscripts $0$ and $i$ are spatial locations, and $O_0$ and $O_i$ are bilinear operators. This was true for $V < V_c$ because $\left\langle \sigma_i \left(n_i -1/2\right) \right\rangle = 0$, and $\left\langle C \right\rangle$ scaled as $L^{-4}$. However, when $V > V_c$, $\left\langle \sigma_i \left(n_i -1/2\right) \right\rangle \neq 0$, which implies that $\left\langle C \right\rangle$ must become a constant independent of $L$. Figure \ref{brokenphase} shows data from our Monte Carlo results for $V=1.4$. We indeed see that $\left\langle C\right\rangle$ goes to a constant in this plot.

\begin{figure}
\begin{center}
  \includegraphics[width=8cm]{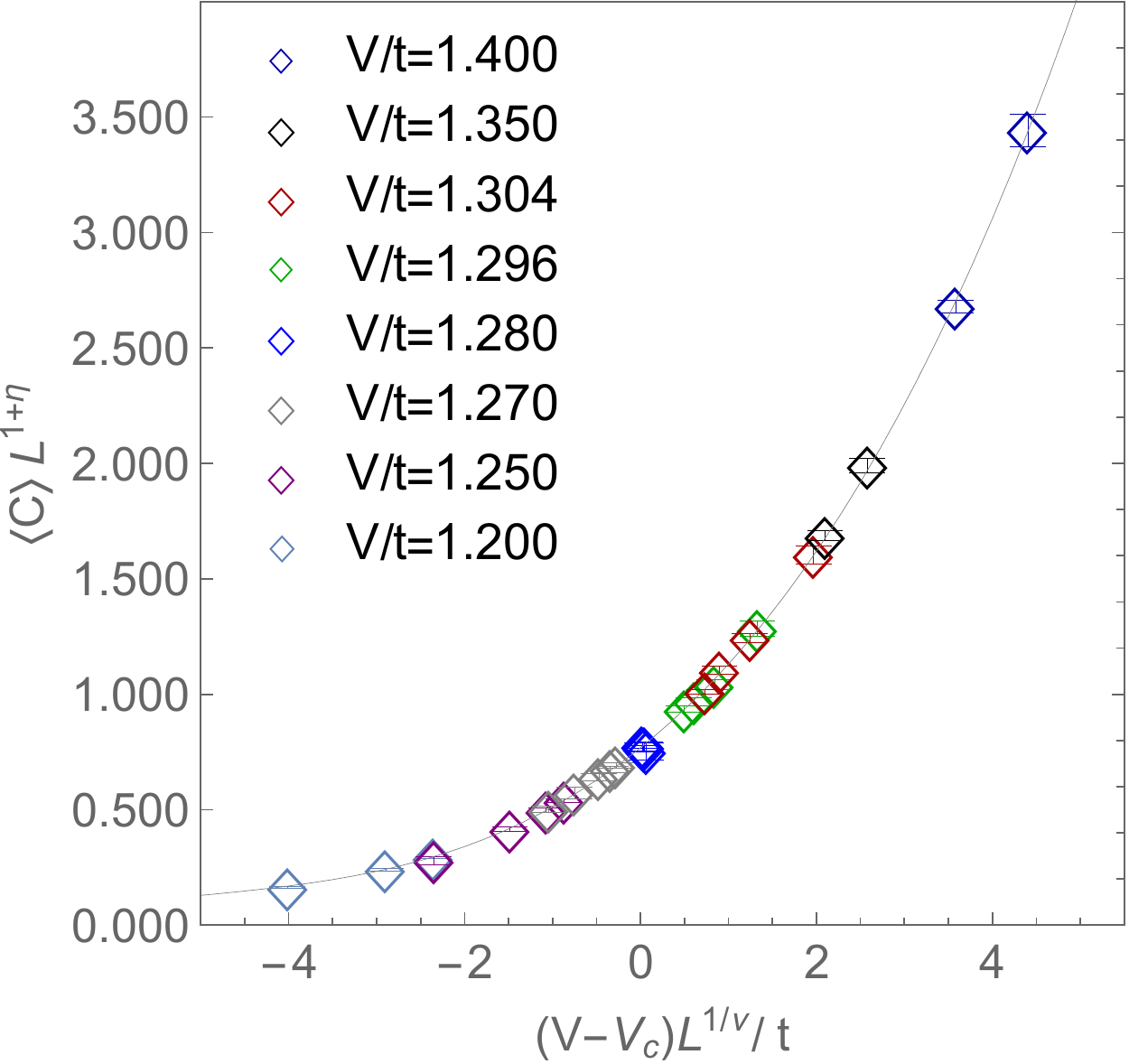}
  \end{center}
  \caption[Critical scaling plot.]{Critical scaling plot showing our Monte Carlo data scaled with $\eta=0.51$, $\nu=0.89$, $V_c=1.281 t$. The solid line shows $f(x)=0.72 + 0.29x+0.051x^2+0.0034x^3$.}\label{fits}
\end{figure}

Finally we focus on our results for $\left\langle C\right\rangle$ near the critical point and try to fit them to the equation (\ref{scaleapprox}). Approximating $f(x)=f_0+f_1 x + f_2 x^2 + f_3 x^3$, we have performed a seven parameter combined fit of the data without special symbols in Table \ref{datatable}. From the fit we obtain $\eta=0.51(3)$, $\nu=0.89(1)$, $V_c=1.281(2) t$, $f_0=0.72(6)$, $f_1=0.29(2)$, $f_2=0.051(5)$ and $f_4=0.0034(5)$. The $\chi^2/DOF$ for the fit is $0.90$. We show the data and the fit in Figure~\ref{fits}. The theoretical $4-\epsilon$ expansion exponent predictions for three and four loops, seen in Table \ref{tab:compare}, are compatible with our results \cite{Mihaila:2017ble,PhysRevLett.86.958}.

\begin{figure}
\begin{center}
  \includegraphics[width=7cm]{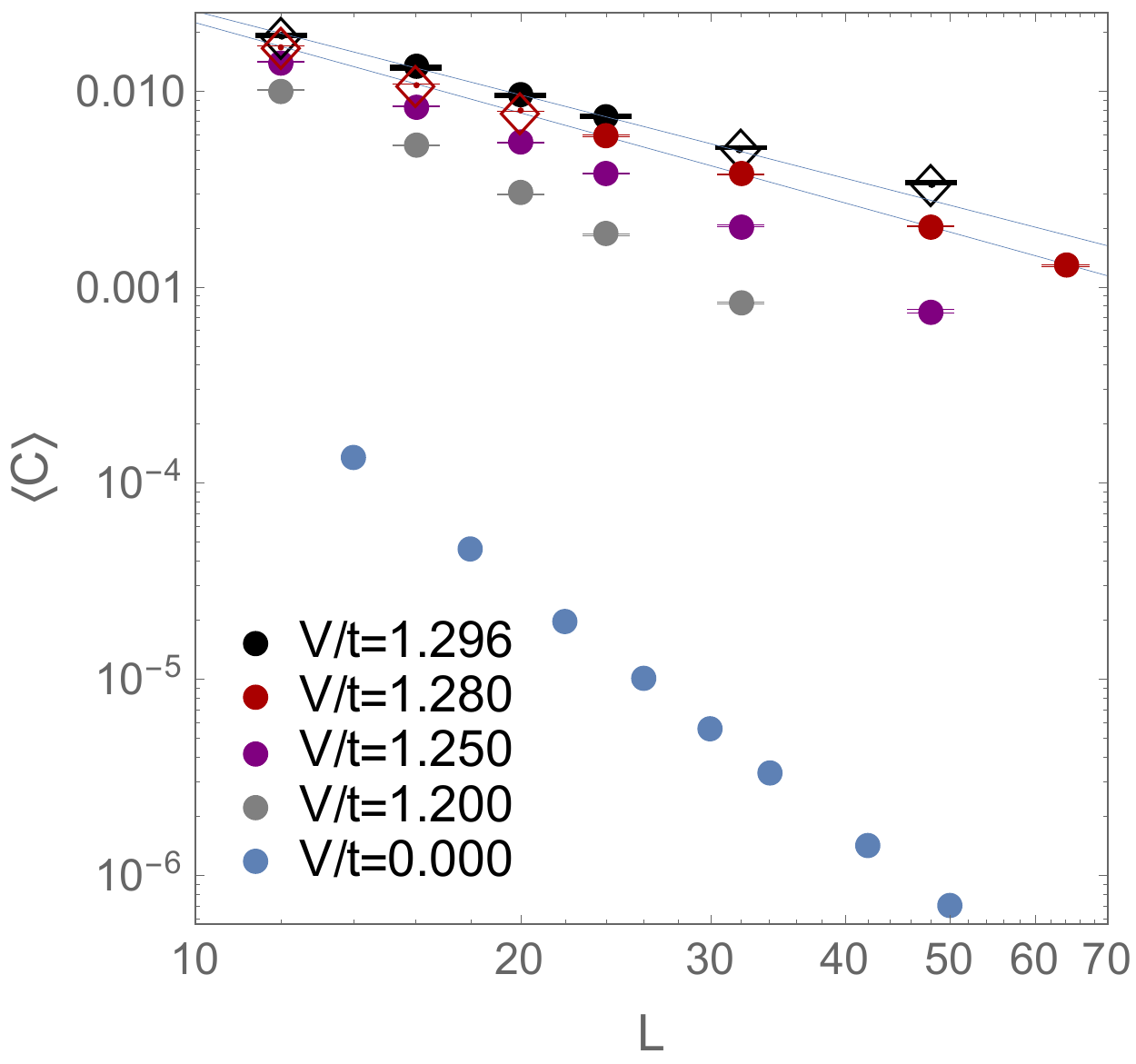}
  \includegraphics[width=7cm]{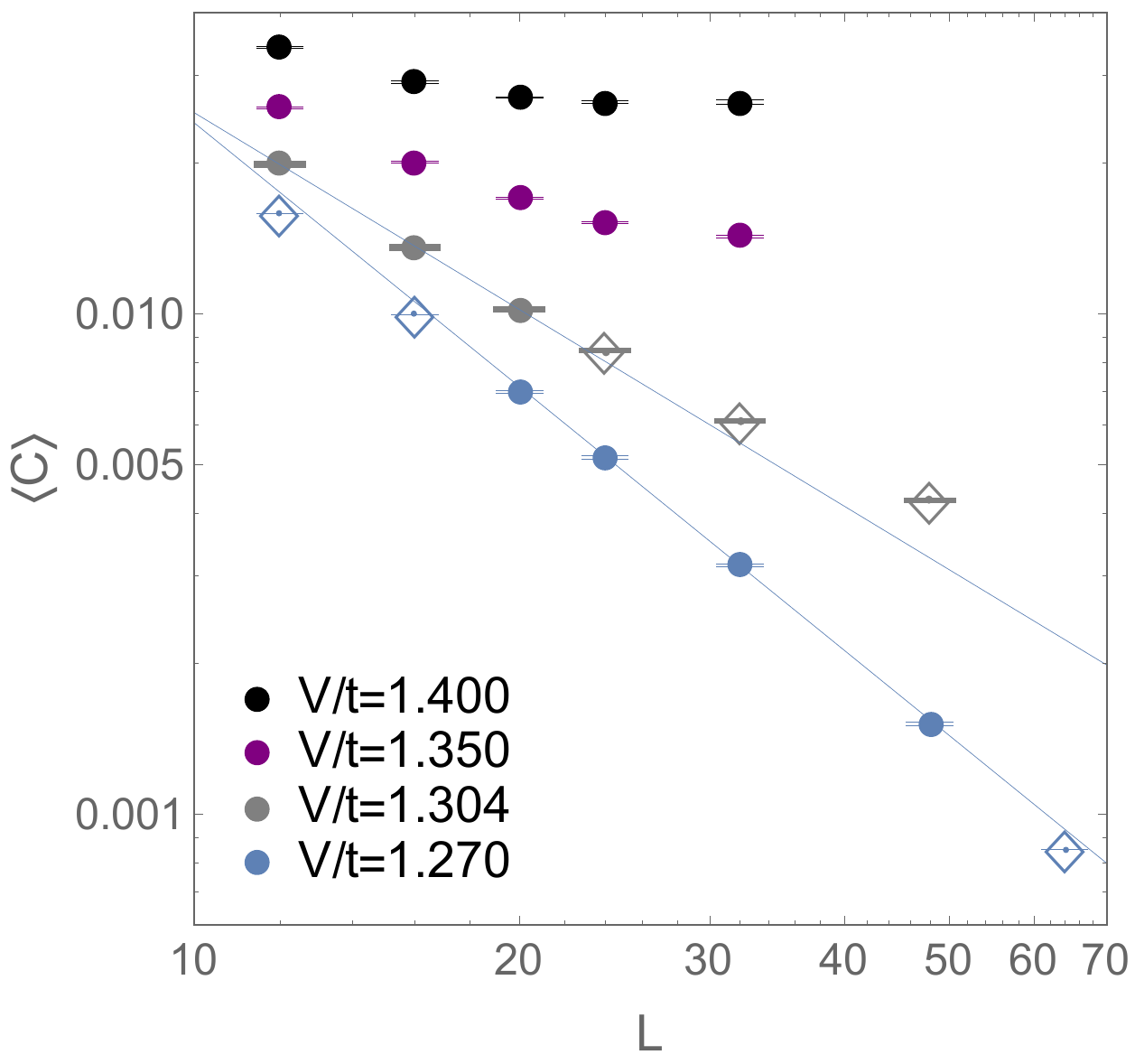}
  \end{center}
 \caption[Plots of $\langle C\rangle$ as a function of $L$.]{Plots of $\langle C\rangle$ as a function of $L$ (with $\beta=L$) at various values of $V$. The left plot shows that $\langle C\rangle$ scales as $L^{-4}$ at $V=0$ as expected, and the solid line shows the best fits for $\langle C\rangle \sim L^{-(1+\eta)}$ at couplings $V=1.296t$ and $V=1.280t$. The right plot shows how $\langle C\rangle$ saturates to a constant at $V=1.4 t$, and the solid lines shows the best fit for $\langle C\rangle \sim L^{-(1+\eta)}$ at couplings $V=1.304t$ and $V=1.270t$. Empty diamonds are used to mark data not included in a fit. See text for further explanation of the plot.}\label{vcscals2}
\end{figure}

As discussed in Section 6.1, this phase transition has been studied earlier by QMC on smaller lattices by two groups. We now understand why the critical point and the critical exponents measured were in disagreement with each other. The  first calculation on a $\pi$-flux lattice, row 2 in Table \ref{tab:compare}, was performed on lattices up to $N=400$ sites and it was found that $V_c=1.304(2)$, $\eta=0.318(8)$ and $\nu=0.80(6)$ \cite{1367-2630-16-10-103008}. In a later $\pi$-flux calculation, row 4 in Table \ref{tab:compare}, lattices up to $N=484$ sites were used and it was found that $V_c=1.296(1)$, $\eta=0.43(2)$ and $\nu=0.79(4)$ \cite{1367-2630-17-8-085003}. Our results are compatible with both of these as shown in Figure \ref{vcscals2}. We believe the small lattices and
the range of fits considered led the authors to the erroneous conclusions. For example, if we assume $V_c/t=1.296$ or $1.304$ and fit our data to the form $L^{-(1+\eta)}$, after dropping larger values of $L$ we get $\eta=0.41(4)$ and $\eta=0.31(4)$ respectively with a reasonable $\chi^2/DOF$ (see Figure~\ref{vcscals2}). However, the fits fail dramatically if data from $L=32$ and $L=48$ is included. 

Our results are obtained from lattice sizes that are up to eight times larger than earlier studies and suggest a lower critical point and so a higher value for the exponent $\eta$. The value of $\nu$ also seems slightly higher. On one hand at $V = 1.270 t$ the data fits well for larger lattice sizes from $L=20$ up to $L=48$ and gives us $\eta=0.74(2)$ with a reasonable $\chi^2/DOF$. However, once we include data from $L=64$ at $V=1.270t$ in the fit, even if we drop $L=20$, the $\chi^2/DOF$ increases to 3.85, and so with the inclusion of large lattices we are able to conclude that $V=1.270t$ is below the critical coupling. Figure \ref{vcscals2} shows this deviation at $V=1.270t$. Now when we do this fit for $V=1.280t$ from $L=20$ to $L=64$, we keep a low $\chi^2/DOF$, and for the data $L=24$ to $L=64$ we get an estimated $\eta=.52(2)$, consistent with our combined fit. This fit is also shown in Figure \ref{vcscals2}. Everything is consistent with our results for the critical coupling of $V=1.281(2)t$ as discussed above.

\chapter{Conclusions and Summary}
{\small\textit{``My interpretation can only be as inerrant as I am, and that's good to keep in mind.'' \\
--Rachel Held Evans}}\\
\newline
\newline
In this thesis we have adapted the idea of fermion bags to the Hamiltonian picture and this has helped us in two ways: First, in finding new ways to solve sign problems, thus allowing QMC studies of the critical behavior in strongly correlated systems that previously could not be studied nonperturbatively. Second, in developing new efficient algorithms for studying large systems.

To summarize the first point, fermion bag ideas helped us find a solution for the $t$-$V$ model, which involves only one layer of strongly interacting fermions--odd numbers of layers have traditionally been more difficult to simulate. This particular sign problem had remained unsolved for almost thirty years since its first study \cite{PhysRevB.32.103}. Fermion bag ideas have since led to solutions for other fermionic models with an odd number of layers and/or models involving fermions interacting with quantum spins, as well as extensions to simple $\mathbb{Z}_2$ lattice gauge theories that interact with fermions. We found that a key advantage to fermion bag ideas is that they can be combined with other techniques such as worldline methods and meron cluster techniques to solve a problem. The Majorana representation, which was discovered later and partially motivated by our work, has helped in extending the solution to many new models as well.

As to the second point, we have found fermion bag ideas to be a useful guiding principle for developing new efficient QMC algorithms in the Hamiltonian picture. By looking for ways to group local spatial degrees of freedom and only update one or two of these groups at a time, we have developed in this work an algorithm that not only draws from the strengths of state-of-the-art Hamiltonian algorithms--such as matching the $O\left(\beta N^3\right)$ scaling of auxiliary field methods and possessing the continuous time abilities of the LCT-INT methods \cite{Wang:2015rga}--but also offers its own unique advantages, such as the ability to make very fast global updates like the \textit{move updates}, which are only $O(1)$ each in computational time. Additionally, the fermion bag concept allows for faster stabilization in the algorithm.

We have demonstrated that this algorithm works well for the $\pi$-flux $t$-$V$ model and used the algorithm to study the model's quantum critical behavior. Using lattices with up to $64^2$ spatial sites in our work, we have computed the critical exponents of the chiral Gross-Neveu universality class containing a single four-component Dirac fermion more reliably than before. We find $\eta = .51(3)$, and $\nu =.89(1)$ and our results are compatible with those from recent $4-\epsilon$ expansion results in the continuum \cite{Zerf:2017zqi}. These exponents have values that are shifted higher compared to the previous QMC studies, which did not include the large lattices used in our work. We can reproduce the results from previous QMC studies if we remove the results from our large lattices.

%The phase transition is in the Chiral Ising universality class, and prior to the sign problem solution this class had only been studied by continuum techniques, but in the aftermath several QMC studies were preformed by other groups. We used our algorithm to calculate the $\nu$ and $\eta$ exponents for this transition, and compared the values to the calculations from other QMC studies. While there was disagreement on the value for the $\eta$ exponent in particular from previous QMC studies, we have gone to lattices for the $\pi$-flux model that are eight times the size of those for the previous calculations, and shown that while we can reproduce the $\nu$ and $\eta$ values for the smaller lattice calculations if we exclude our larger lattices, we do in fact arrive at different $\nu$ and $\eta$ values when we include our largest lattices. These new $\eta$ and $\nu$ values seem to be consistent with those from recent $4-\epsilon$ calculations in the continuum.
 
There are multiple interesting directions for expanding our work, and a few of them are given below. While many QMC studies have been done for the Chiral Heisenberg universality class and now the Chiral Ising universality class as well, which consist of Dirac fermions acquiring mass by breaking $SU(2)$ and $\mathbb{Z}_2$ symmetries, respectively, there is still more need to establish the behavior for the Chiral XY universality class, where Dirac fermions acquire mass by breaking a $U(1)$ symmetry \cite{Li2017}. There are models assumed to be in this class that can be simulated without sign problems, and they appear to be amenable to our new Hamiltonian fermion bag techniques. This offers a straightforward extension to our work.

While we have found one useful way to group degrees of freedom in the Hamiltonian picture, there may be other fermion bag inspired ways to approach the Hamiltonian picture that lead to sign problem solutions and new algorithms. Our work connecting the fermion bag approach with the meron-cluster approach \cite{PhysRevE.94.043311} still needs to be developed and understood more fully, especially since the meron-cluster approach offers very favorable computational scaling ($O(N)$ updates) compared to our fermion bag work so far. By combining these approaches we have found ways to extend our original sign problem solution to additional models, including models that involve interacting fermions and quantum spins. We should be able to compute critical exponents on large lattices for these models, which can exhibit phases that include semimetal and massive CDW phases for the fermions, along with antiferromagnetic, ferromagnetic, and disordered phases for the quantum spins. The critical behavior of the fermions and quantum spins is coupled due to the fermion-spin interaction, and it would be interesting to characterize the resulting transitions.

Finally, for certain models involving fermions interacting with $\mathbb{Z}_2$ gauge theories where current methods face difficult critical slowing down \cite{PhysRevX.6.041049,Gazit2017}, it seems our algorithm may offer certain advantages. These theories offer exotic transitions where the fermions undergo symmetry breaking while the gauge field undergoes a confinement at the same time, and may be experimentally realizable with cold atom systems \cite{0034-4885-79-1-014401}. The QMC studies thus far suggest these transitions are continuous and consistent with being second-order, but there is not yet an established understanding for any transitions of this type, and larger lattice studies would provide a rigorous way to confirm their existence and characterize them.

In this thesis we have provided a starting point for the exploration of fermion bag ideas in the Hamiltonian picture, but there remains much room for the further development of these ideas, which show potential for both uncovering and studying new quantum critical points. It is exciting to contemplate what the future for nonperturbative studies may hold.

%==============================================================================

%-----------------------------------------------------------------------------%
% APPENDICES -- OPTIONAL. These are just chapters enumerated by Appendix A,
%                Appendix B, Appendix C...
%-----------------------------------------------------------------------------%
\appendix
\chapter{Grassmann Number Identities}

Grassmann numbers are key for writing fermionic path integrals in quantum mechanics. Their basic property can be motivated from seeking the eigenstate $\left|\psi\right\rangle$ of the fermionic annihilation operator (known as a \textit{coherent state}):
\begin{equation}
    c \left|\psi\right\rangle = \psi \left|\psi\right\rangle .
\end{equation}
The eigenvalue $\psi$ is the Grassmann numbers, and from here it becomes clear that because $c c =0$ (from the anticommutation relations), we must have
\begin{equation}
    \psi^2 = 0.
   \label{squaredids}
\end{equation}
These numbers must also anticommute with each other and with fermionic creation and annihilation operators. The form of the coherent state can then be understood to be $\left|\psi\right\rangle =\left|0\right\rangle - \psi \left|1\right\rangle$. For the creation operator there is a corresponding Grassmann number $\bar{\psi}$ that defines a corresponding coherent state $\left\langle\bar{\psi}\right| =\left\langle 0 \right| - \left\langle 1\right| \bar{\psi}$. More details on the derivation of and the physics related to these numbers can be found in \cite{Shankar}.

From (\ref{squaredids}) we have the key property that any function of Grassmann numbers can be expanded as
\begin{equation}
    f\left(\psi\right) = f_0 + f_1 \psi,
    \label{grassmanfcn}
\end{equation}
since all higher powers are zero.

Grassmann integrals are all defined based on two key identities:
\begin{equation}
    \begin{aligned}
    \int  d\psi \: \psi &= 1 \\
    \int d\psi \: 1 &= 0.
    \end{aligned}
   \label{gsints}
\end{equation}

From (\ref{gsints}) combined with the anticommutation of Grassmann numbers we get
\begin{equation}
    \begin{aligned}
    \int d\bar{\psi} d\psi \: \psi \bar{\psi} &= 1 \\
    \int d\bar{\psi} d\psi \: \bar{\psi} \psi &= -1.
    \end{aligned}
   \label{gsints2}
\end{equation}

The following identities then hold for integrals of exponentiated Grassmann numbers:
\begin{equation}
\begin{aligned}
        &\int d\bar{\psi} d\psi \: e^{-a\bar{\psi} \psi} = a\\
       &\int \left[d\bar{\psi} d\psi\right] \: e^{-\sum_{xy}\bar{\psi}_x A_{xy}\psi_y} = \det A,
        \end{aligned}
        \label{gsxpints}
\end{equation}
where $a$ and $A_{xy}$ are just numbers, and $A_{xy}$ is an element of an $N\times N$ matrix. The notation $\left[d\bar{\psi} d\psi\right]$ is defined as $d\bar{\psi}_1 d\psi_1 ...d\bar{\psi}_N d\psi_N$.

Finally, the trace of an operator $O$ is given by
\begin{equation}
    {\rm Tr} \left(O\right) = \int d\bar{\psi} d\psi \left\langle -\bar{\psi}\right| O \left|\psi\right\rangle e^{-\bar{\psi}\psi}.
    \label{gstrace}
\end{equation}
\chapter{Writing a Trace as a Pfaffian}
In Section 4 of Chapter 4, we proved that the Quantum Monte Carlo weights in (\ref{ssepos}) could be written as the squares of Pfaffians. In doing so, we used the following identities for even $n$:
\begin{equation}
    \left(-i\right)^{n}\sqrt{a_{ij}\left(t_1\right)....a_{ij}\left(t_n\right)}\:\: \bar{\gamma}_{i} .... \bar{\gamma}_i =  \int \left[d\bar{\xi}\right]_i e^{\sum_{t< t'}\sqrt{a_{ij}\left(t'\right)a_{ij}\left(t\right)} \bar{\xi}_{i,t'}\bar{\xi}_{i,t}},
    \label{firstone}
\end{equation}
and
\begin{equation}
    \sqrt{a_{ij}\left(t_1\right)....a_{ij}\left(t_n\right)}\:\: \gamma_{j} .... \gamma_j =  \int \left[d\xi\right]_j e^{-\sum_{t< t'}\sqrt{a_{ij}\left(t'\right)a_{ij}\left(t\right)} \xi_{j,t'}\xi_{j,t}},
    \label{secondone}
\end{equation}
and stated that for odd $n$ the integrals are zero. Here we prove why these identities hold.

First it is trivial to see why each integral with odd $n$ is zero. If $n$ is odd, then the measure will have an odd number of differentials. However, the exponential always contains pairs of Grassmann numbers. Thus the only nonzero terms in an expansion of the exponential will have fewer Grassmann numbers than there are differentials, and the integral will be zero.

Now we turn to the nontrivial case of even $n$. We will find that (\ref{secondone}) will follow directly from the proof of (\ref{firstone}). It is useful to introduce a label for the right-hand-side of the integral in (\ref{firstone}). We define $P_n$ for even $n$ as
\begin{equation}
    P_n = \int \left[d\bar{\xi}\right]_i e^{\sum_{t< t'}\sqrt{a_{ij}\left(t'\right)a_{ij}\left(t\right)} \bar{\xi}_{i,t'}\bar{\xi}_{i,t}}/\sqrt{a_{ij}\left(t_1\right)....a_{ij}\left(t_n\right)}.
\end{equation}
This will be a useful quantity in the following proof by induction, where we will seek to show that $P_n = \left(-1\right)^n$, as (\ref{firstone}) states. We begin with the base cases for (\ref{firstone}). For a single pair of Majorana operators on the left-hand-side ($n=2$), the right hand side of (\ref{firstone}) reads as
\begin{equation}
\begin{aligned}
\sqrt{a_{ij}\left(t_1\right)a_{ij}\left(t_2\right)} P_2 =& \int d\bar{\xi}_{i,t_2} d\bar{\xi}_{i,t_1} e^{\sqrt{a_{ij}\left(t_1\right)a_{ij}\left(t_2\right)} \bar{\xi}_{i,t_2} \bar{\xi}_{i,t_1}}\\
=&  -\sqrt{a_{ij}\left(t_1\right)a_{ij}\left(t_2\right)}\: \\ =&(-i)^2\sqrt{a_{ij}\left(t_1\right)a_{ij}\left(t_2\right)} \:.
\end{aligned}
\end{equation}
This is certainly in agreement with (\ref{firstone}), and it also tells us that
\begin{equation}
    P_2 = (-i)^2 .
    \label{pidentity}
\end{equation}
Moving on to two pairs of Majorana operators on the left hand side, the right hand side gets a little more complicated. We show it because it makes the generic $n$th step clearer:
\begin{equation}
\begin{aligned}
&\sqrt{a_{ij}\left(t_1\right)a_{ij}\left(t_2\right)a_{ij}\left(t_3\right)a_{ij}\left(t_4\right)} P_4= \int d\bar{\xi}_{i,t_4} d\bar{\xi}_{i,t_3} d\bar{\xi}_{i,t_2} d\bar{\xi}_{i,t_1}\\
&\qquad\qquad\times e^{\sqrt{a_{ij}\left(t_1\right)a_{ij}\left(t_2\right)} \bar{\xi}_{i,t_2} \bar{\xi}_{i,t_1}} e^{\sqrt{a_{ij}\left(t_1\right)a_{ij}\left(t_3\right)} \bar{\xi}_{i,t_3} \bar{\xi}_{i,t_1}} e^{\sqrt{a_{ij}\left(t_1\right)a_{ij}\left(t_4\right)} \bar{\xi}_{i,t_4} \bar{\xi}_{i,t_1}}\\
&\qquad\qquad\times e^{\sqrt{a_{ij}\left(t_2\right)a_{ij}\left(t_3\right)} \bar{\xi}_{x_i,t_3} \bar{\xi}_{i,t_2}} e^{\sqrt{a_{ij}\left(t_2\right)a_{ij}\left(t_4\right)} \bar{\xi}_{i,t_4} \bar{\xi}_{i,t_2}} e^{\sqrt{a_{ij}\left(t_3\right)a_{ij}\left(t_4\right)} \bar{\xi}_{i,t_4} \bar{\xi}_{i,t_3}}\\
&\qquad\qquad= \int d\bar{\xi}_{i,t_2} d\bar{\xi}_{i,t_1} e^{\sqrt{a_{ij}\left(t_1\right)a_{ij}\left(t_2\right)} \bar{\xi}_{i,t_2} \bar{\xi}_{i,t_1}} \sqrt{a_{ij}\left(t_3\right)a_{ij}\left(t_4\right)} P_2\\
&\qquad\qquad - \int d\bar{\xi}_{i,t_3} d\bar{\xi}_{i,t_1} e^{\sqrt{a_{ij}\left(t_1\right)a_{ij}\left(t_3\right)} \bar{\xi}_{i,t_3} \bar{\xi}_{i,t_1}} \sqrt{a_{ij}\left(t_2\right)a_{ij}\left(t_4\right)}P_2\\
&\qquad \qquad +\int d\bar{\xi}_{i,t_4} d\bar{\xi}_{i,t_1} e^{\sqrt{a_{ij}\left(t_1\right)a_{ij}\left(t_4\right)} \bar{\xi}_{i,t_4} \bar{\xi}_{i,t_1}} \sqrt{a_{ij}\left(t_2\right)a_{ij}\left(t_3\right)} P_2\\
&\qquad\qquad =\sqrt{a_{ij}\left(t_1\right)a_{ij}\left(t_2\right)a_{ij}\left(t_3\right)a_{ij}\left(t_4\right)}\: P_2 \left(P_2 - P_2 +P_2\right) \\
&\qquad\qquad= (-i)^4\sqrt{a_{ij}\left(t_1\right)a_{ij}\left(t_2\right)a_{ij}\left(t_3\right)a_{ij}\left(t_4\right)} \: .
\end{aligned}
\end{equation}
%The pattern continues for $n\left(x_i\right) = 6$,
%\begin{equation}
%\begin{aligned}
%I_6 =&\left(-1\right)^{6/2} \int d\xi_{x_i,t\left(x_i,1\right)} d\xi_{x_i,t\left(x_i,2\right)} d\xi_{x_i,t\left(x_i,3\right)} d\xi_{x_i,t\left(x_i,4\right)} d\xi_{x_i,t\left(x_i,5\right)} d\xi_{x_i,t\left(x_i,6\right)}\\
%&\times e^{\alpha \xi_{x_i,t\left(x_i,1\right)} \xi_{x_i,t\left(x_i,2\right)}} e^{\alpha \xi_{x_i,t\left(x_i,1\right)} \xi_{x_i,t\left(x_i,3\right)}} e^{\alpha \xi_{x_i,t\left(x_i,1\right)} \xi_{x_i,t\left(x_i,4\right)}} e^{\alpha \xi_{x_i,t\left(x_i,1\right)} \xi_{x_i,t\left(x_i,5\right)}}\\
%&\times e^{\alpha \xi_{x_i,t\left(x_i,1\right)} \xi_{x_i,t\left(x_i,6\right)}} e^{\alpha \xi_{x_i,t\left(x_i,2\right)} \xi_{x_i,t\left(x_i,3\right)}} e^{\alpha \xi_{x_i,t\left(x_i,2\right)} \xi_{x_i,t\left(x_i,4\right)}} e^{\alpha \xi_{x_i,t\left(x_i,2\right)} \xi_{x_i,t\left(x_i,5\right)}} \\
%&\times e^{\alpha \xi_{x_i,t\left(x_i,2\right)} \xi_{x_i,t\left(x_i,6\right)}} e^{\alpha \xi_{x_i,t\left(x_i,3\right)} \xi_{x_i,t\left(x_i,4\right)}} e^{\alpha \xi_{x_i,t\left(x_i,3\right)} \xi_{x_i,t\left(x_i,5\right)}} e^{\alpha \xi_{x_i,t\left(x_i,3\right)} \xi_{x_i,t\left(x_i,6\right)}} \\
%&\times e^{\alpha \xi_{x_i,t\left(x_i,4\right)} \xi_{x_i,t\left(x_i,5\right)}} e^{\alpha \xi_{x_i,t\left(x_i,4\right)} \xi_{x_i,t\left(x_i,6\right)}} e^{\alpha \xi_{x_i,t\left(x_i,5\right)} \xi_{x_i,t\left(x_i,6\right)}} \\
%& = I_2 \left(I_4 - I_4 + I_4 -I_4 + I_4\right) %= \alpha^3.
%\end{aligned}
%\end{equation}
Again, we find agreement with the statement in (\ref{firstone}), and we also learn that
\begin{equation}
    P_4 = \left(-i\right)^4.
\end{equation}
In general, we find that by assuming the following expression for $P_n$:
\begin{equation}
    P_n = (-i)^n ,
\end{equation}
we then find the following expression for the left-hand side of (\ref{firstone}):
\begin{equation}
\begin{aligned}
\sqrt{a_{ij}\left(t_1\right)...a_{ij}\left(t_{n+2}\right)} P_{n+2}
&= \sqrt{a_{ij}\left(t_1\right)...a_{ij}\left(t_{n+2}\right)} \: P_{2} \left(P_{n} - P_{n} ... + P_{n}\right) \\
&= \sqrt{a_{ij}\left(t_1\right)...a_{ij}\left(t_{n+2}\right)} \: P_{2} P_{n} \\
&= (-i)^{n+2}\sqrt{a_{ij}\left(t_1\right)...a_{ij}\left(t_{n+2}\right)} \: ,
\label{finalstep}
\end{aligned}
\end{equation}
and thus we get that $P_{n+2}= (-i)^{n+2}$, and we have proven that the left-hand side of (\ref{firstone}) is indeed the same as its right-hand-side. In (\ref{finalstep}) there are $n-1$ terms of $P_{n}$, and all cancel except for one. The alternating signs come from commuting $d\xi_{i,t_m}$ over next to $d\xi_{i,t_1}$, where $m\in\left\{2,3,...,n\right\}$, because the commutation is either past an even number of numbers or an odd number of numbers. The identity in (\ref{secondone}) is proven in a nearly identical way. The negative signs in the exponents keep all intergals positive in that case, however.
\chapter{Relation of Pfaffian to Determinant}

The following is a proof that $ {\rm Pf}^2 M = \det A$, where $A$ is the antisymmetric part of the matrix $M$, using Grassmann numbers. An introduction to Grassmann numbers was given in Appendix A.

We define a set of Grassman variables $\xi_1, \xi_2, ...\xi_{2n}$. For a $2n\times 2n$ dimensional matrix $M$, the Pfaffian is given by the following Grassman integral:
\begin{equation}
{\rm Pf}\left(M\right) = \int \left[d\xi\right] e^{- \frac{1}{2}\xi^T M \xi} ,
\end{equation}
where $\left[d\xi\right] = d\xi_1 d\xi_2 ... d\xi_{2n}$ and $\xi^T$ is given by $\left(\xi_1, \xi_2, ...,\xi_{2n}\right)$.

Defining two more sets of Grassman variables, $\bar{\psi}_1,\bar{\psi}_2,...,\bar{\psi}_{2n}$ and $\psi_1,\psi_2,...\psi_{2n}$, the determinant of $M$ is given (from (\ref{gsxpints})) by
\begin{equation}
\det M = \int \left[d\bar{\psi} d\psi\right] e^{-\bar{\psi}^T M \psi},
\end{equation}
where $\left[d\bar{\psi} d\psi\right]=d\bar{\psi}_1 d \psi_1...d\bar{\psi}_{2n} d\psi_{2n}$, $\bar{\psi}^T=\left(\bar{\psi}_1,\bar{\psi}_2,...,\bar{\psi}_{2n}\right)$, and $\psi^T = \left(\psi_1,\psi_2,...,\psi_{2n}\right)$

Finally, we relate these two quantities for a general antisymmetric matrix $A$. We first write out the expression for the Pfaffian squared:
\begin{equation}
{\rm Pf}^2\left(A\right) = \int \left[d\xi\right] \left[d\chi\right] e^{-\frac{1}{2} \xi^T A \xi} e^{-\frac{1}{2} \chi^T A \chi}.
\label{squaresub}
\end{equation}

Next we make a transformation by introducing the Grassmann numbers $\bar{\psi}_i$ and $\psi_i$, which we relate to $\chi$ and $\xi$ as:
\begin{equation}
\begin{array}{cc}
\psi_i = \frac{1}{\sqrt{2}} \left(\chi_i + i \xi_i \right) \qquad & \bar{\psi}_i = \frac{1}{\sqrt{2}} \left(\chi_i - i \xi_i\right).
\end{array}
\end{equation}
Upon inversion, we get
\begin{equation}
\begin{array}{cc}
\chi_i = \frac{1}{\sqrt{2}} \left(\psi_i + \bar{\psi}_i \right) \qquad & \xi_i = \frac{1}{i\sqrt{2}} \left(\psi_i - \bar{\psi}_i\right).
\end{array}
\end{equation}
Substituting these into the integrand of (\ref{squaresub}) and using the antisymmetry of $A$, we get
\begin{equation}
{\rm Pf}^2\left(A\right) = \int \left[d\xi\right] \left[d\chi\right] e^{- \bar{\psi}^T A \psi}.
\end{equation}

Now we have to transform the measure. First we seek to obtain
\begin{equation}
\left[d\chi d\xi\right],
\end{equation}
where $\left[d\chi d\xi\right] = d\chi_1 d\xi_1 d\chi_2 d\xi_2 ...d\chi_{2n} d\xi_{2n}$.

To get here from $\left[d\xi\right] \left[d\chi\right]$, we must move $d\chi_1$ past $2n$ variables, then $d\chi_2$ past $2n-1$ variables, etc. In the end, we need
\begin{equation}
1 + 2+ ... + 2n = \frac{2n\left(2n+1\right)}{2} = n\left(2n+1\right)
\end{equation}
moves total, and so we have
\begin{equation}
\left[d\xi\right]\left[d\chi\right] = \left(-1\right)^{n\left(2n+1\right)}\left[d\chi d\xi\right].
\label{meas}
\end{equation}

To fully transform the measure, we must calculate the Jacobian:
\begin{equation}
J_i = \frac{\partial\left(\chi_i, \xi_i\right)}{\partial\left(\bar{\psi}_i,\psi_i\right)} 
= \det\left(
\begin{array}{cc}\frac{\partial\chi_i}{\partial\bar{\psi}_i} & \frac{\partial\chi_i}{\partial\psi_i} \\
\frac{\partial\xi_i}{\partial\bar{\psi}_i} & \frac{\partial\xi_i}{\partial\psi_i}
\end{array} \right)
= \det\left(
\begin{array}{cc}\frac{1}{\sqrt{2}} & \frac{1}{\sqrt{2}} \\
-\frac{1}{i\sqrt{2}} & \frac{1}{i\sqrt{2}}
\end{array} \right)
=-i .
\end{equation}

Thus we see that
\begin{equation}
d\chi_i d\xi_i = J_i^{-1} d\bar{\psi}_i d\psi_i = i d\bar{\psi}_i d\psi_i,
\label{jac}
\end{equation}
and so for the full measure we combine (\ref{meas}) and (\ref{jac}) to get
\begin{equation}
\begin{aligned}
\left[d\xi\right] \left[d\chi\right] &= \left(-1\right)^{n\left(2n+1\right)}\left[d\chi d\xi\right]
= \left(-1\right)^{n\left(2n+1\right)}\left(i\right)^{2n}\left[d\bar{\psi} d\psi\right] \\
&= \left(-1\right)^{n\left(2n+1\right)}\left(-1\right)^{n}\left[d\bar{\psi} d\psi\right] = \left[d\bar{\psi} d\psi\right].
\end{aligned}
\end{equation}

Finally, we obtain
\begin{equation}
{\rm Pf}^2\left(A\right) = \int \left[d\bar{\psi}d\psi\right] e^{- \bar{\psi}^T M \psi} = \det A,
\end{equation}
and so we have proven that for a general antisymmetric matrix $A$,
\begin{equation}
{\rm Pf}^2\left(A\right) = \det A.
\end{equation}

Now we generalize the $A$ antisymmetric matrix to any matrix $M$. A general matrix $M$ may be decomposed as,
\begin{equation}
M = S+A,
\end{equation}
where $S$ is a symmetric matrix, given by $S=\left(M+M^T\right)/2$, and $A$ is an antisymmetric matrix, given by $A = \left(M-M^T\right)/2$ \cite{wald1984}. The pfaffian of the matrix $M$ is then given by
\begin{equation}
\begin{aligned}
{\rm Pf}\left(M\right) &= \int \left[d\xi\right] e^{-\frac{1}{2}\left(\xi_i S_{ij} \xi_j +\xi_i A_{ij} \xi_j\right)}\\
&= \int \left[d\xi\right]  e^{-\frac{1}{2}\left(\frac{1}{2}\left(\xi_i S_{ij} \xi_j - \xi_j S_{ji} \xi_i\right) +\xi_i A_{ij} \xi_j\right)}.
\end{aligned}
\end{equation}
The second step consists of splitting $\xi_i S_{ij} \xi_j$ into two identical terms, and then using the anticommutation property of Grassman numbers and the symmetry of $S$ on the second term. But the indices $i$ and $j$ are dummy indices that are summed over, so the symmetric terms cancel each other, and the remainder is
\begin{equation}
{\rm Pf}\left(M\right) = \int \left[d\xi\right] e^{-\frac{1}{2}\xi_i A_{ij} \xi_j} = {\rm Pf}\left(A\right).
\end{equation}

Thus for a general matrix $M$, we have shown that
\begin{equation}
    {\rm Pf}^2\left(M\right) = \det A,
\end{equation}
where $A$ is the antisymmetric part of $M$.
\chapter{Comparison with Auxiliary Field Method}

The Hamiltonian fermion bag algorithm that we have developed is a \textit{continuous time} algorithm. The largest lattice size that has been studied earlier in continuous time for the $t$-$V$ model involves $N=648$ spatial sites on a honeycomb lattice \cite{Wang:2015rga}. The lattice sizes we have studied in this work greatly exceed this. In fact we have even demonstrated the viability of performing calculations up to lattice sizes involving $N=10,000$ sites using supercomputers.

The largest lattices reached until now in Hamiltonian lattice field theories involve about $N=2500$ sites, but use a \textit{discrete} time approach, where the imaginary time extent $\beta$ is divided into $N_t$ time slices  \cite{sorella}. The algorithms use the traditional auxiliary field methods so that the time for a single sweep scales as $N_t N^3$ instead of $\beta N^3$. So it is interesting to see how our algorithm stacks up against these traditional algorithms. However, it is difficult to perform a direct comparison since in continuous time we work with $N_t=\infty$. Naively, we could argue that we are infinitely better. But such a comparisom may not be fair.

\begin{table}[t]

  \subfloat{\includegraphics[width=.44\linewidth]{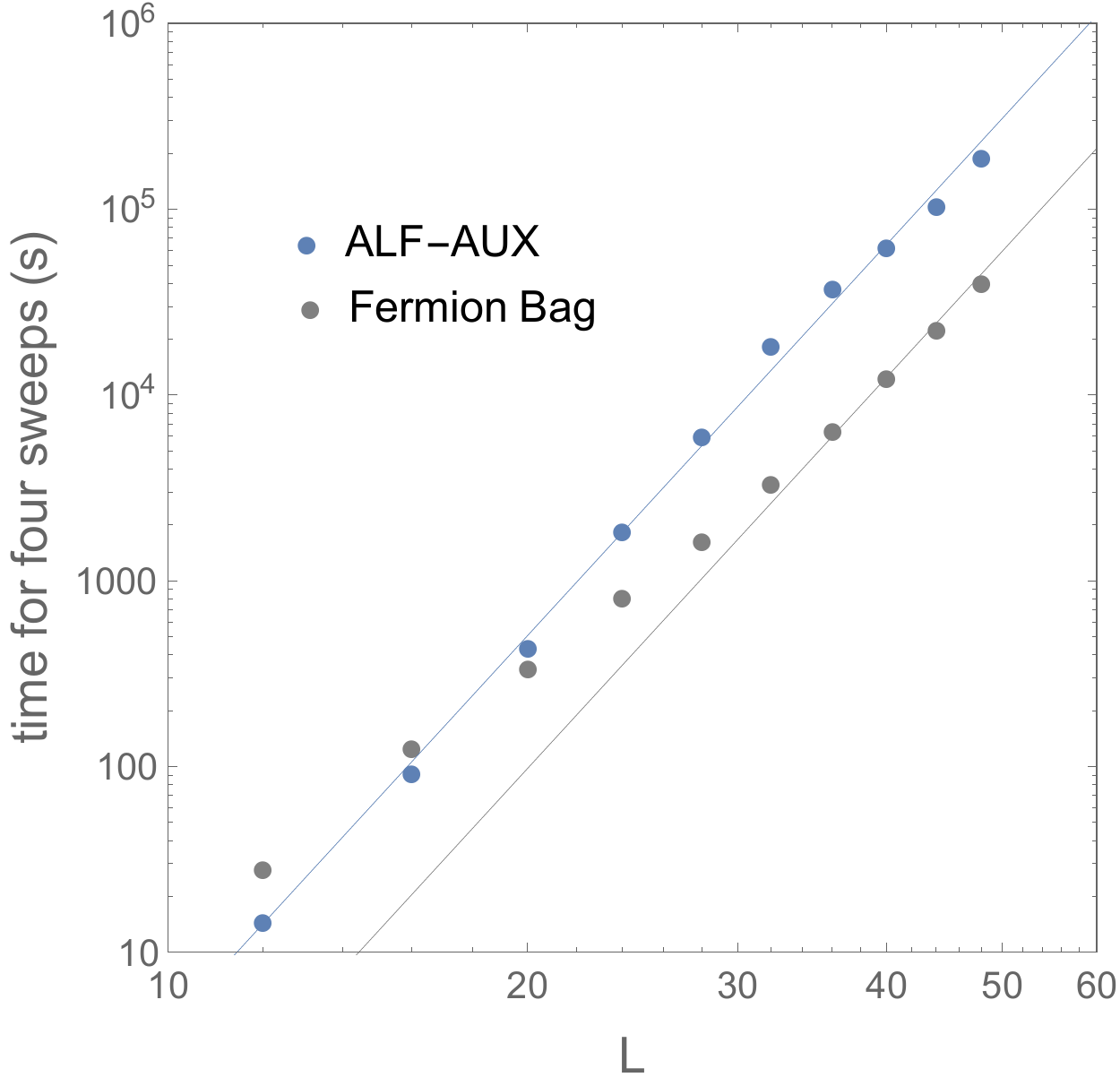}} \quad \quad
  \subfloat{\adjustbox{valign=B,raise=0\baselineskip}{\begin{tabular}{|c|c|c|}
\hline
\hline
\multicolumn{3}{|c|}{\textbf{Timing Comparison}} \\
\hline
\hline
L & \small Fermion Bag (s) & \small ALF-AUX (s) \\
\hline
$\quad$12$\quad$ &  27.82 & 14.49\\
\hline
16 & 124.80 & 91.94 \\
\hline
20 & 335.67 & 433.30\\
\hline
24 & 810.39 & 1834.40\\
\hline
28 & 1624.74 & 5968.59\\
\hline
32 & 3315.44 & 18319.72\\
\hline
36 & 6377.32 & 37486.95\\
\hline
40 & 12301.09 & 62507.31\\
\hline
44 & 22466.78 & 103771.97\\
\hline
48 & 39759.52 & 188659.16\\
\hline
\end{tabular}}}
\\
  \caption[Timing comparison for the Hamiltonian fermion bag algorithm (Fermion Bag), and the auxiliary field algorithm (ALF-AUX).]{Timing comparison for the Hamiltonian fermion bag algorithm (Fermion Bag), and the auxiliary field algorithm (ALF-AUX). The solid lines are $\tau_1 = 7.57467\times 10^{-8} L^7$ for the fermion bag algorithm and $\tau_2=3.93116\times 10^{-7} L^7$ for the auxiliary field algorithm. The exact timing numbers are given by the table to the right.}
  \label{auxfb}

\end{table}

In order to make a reasonable comparison, we developed a discrete time version of our method. Fakher Assaad was willing to share with us the results for the $t$-$V$ using the Algorithms for Lattice Fermions (ALF) project \cite{SciPostPhys.3.2.013}, which employs the finite temperature auxiliary field quantum Monte Carlo algorithm. We compared the run time for a single sweep for the two algorithms using $\beta/N_t=0.1$. This comparison is found in Figure~\ref{auxfb}. Both times are for a single core of an Intel(R) Xeon(R) CPU X5650 @ 2.67GHz model with x86-64 architecture. For both algorithms the expected scaling of $O(L^7)$ can be seen for larger lattices. Note that for the Fermion Bag algorithm the scaling is better than $O\left(L^7\right)$ at smaller lattices. This is because the fermion bag technique reduces the prefactor of the $L^7$ scaling term. As seen in the table, for the larger lattices our results are roughly 5 times faster than those for the auxiliary field algorithm (the continuous time results for our algorithm are only slightly slower than these discrete results). Thus, if NT is increased our times would not scale linearly with NT as Dr. Assaad's code would.

On the other hand it is important to emphasize that a comparison of the time for one sweep has limited meaning since the auto correlation times of different methods may be quite different. Since in our approach we do not have a background auxiliary field at every space-time point, we can also perform other fast updates like the move-update, described in Chapter 5. This update does not scale as $\beta N^3$ and could help us significantly reduce auto-correlation times in our method. A more careful study of the autocorrelation times in our algorithm would be useful.% Start with'\chapter{Title}'
%You can always add more appendices here if you want

%-----------------------------------------------------------------------------%
% BIBLIOGRAPHY -- uncomment \nocite{*} to include items in 'mybib.bib' file
% that aren't cited in the text.  Change the style to match your
% discipline's standards.  Of course, if your bibliography file isn't called
% 'mybib.bib' you might want to change that here too :)
%-----------------------------------------------------------------------------%
%\nocite{*} - if you use this it will put EVERYTHING in your .bib file into the references even if you don't cite it in the text
\bibliographystyle{./Bibliography/jasa} %Formats bibliography
\bibliographystyle{unsrt}
\cleardoublepage
\normalbaselines %Fixes spacing of bibliography
\addcontentsline{toc}{chapter}{Bibliography} %adds Bibliography to your table of contents
\bibliography{./Bibliography/References} %your bibliography file - change the path if needed

\begin{thebibliography}{}
\newcommand{\enquote}[1]{``#1''}

\bibitem[Agranovich and Maradudin(1992)Agranovich and Maradudin]{inbook}
Agranovich, V. and Maradudin, A.~A. (1992), \emph{Electronic Phase
  Transitions}, vol.~32, chap.~4, pp. 177--235, Elsevier Science Publishers B.
  V.

\bibitem[Armour et~al.(2010)Armour, Hands, and Strouthos]{Armour:2009vj}
Armour, W., Hands, S., and Strouthos, C. (2010), \enquote{{Monte Carlo
  Simulation of the Semimetal-Insulator Phase Transition in Monolayer
  Graphene},} \emph{Phys. Rev.}, B81, 125105.

\bibitem[Assaad(1999)Assaad]{PhysRevLett.83.796}
Assaad, F.~F. (1999), \enquote{Quantum Monte Carlo Simulations of the
  Half-Filled Two-Dimensional Kondo Lattice Model,} \emph{Phys. Rev. Lett.},
  83, 796--799.

\bibitem[Assaad and Grover(2016)Assaad and Grover]{PhysRevX.6.041049}
Assaad, F.~F. and Grover, T. (2016), \enquote{Simple Fermionic Model of
  Deconfined Phases and Phase Transitions,} \emph{Phys. Rev. X}, 6, 041049.

\bibitem[Auerbach and Levin(1986)Auerbach and Levin]{PhysRevLett.57.877}
Auerbach, A. and Levin, K. (1986), \enquote{Kondo Bosons and the Kondo Lattice:
  Microscopic Basis for the Heavy Fermi Liquid,} \emph{Phys. Rev. Lett.}, 57,
  877--880.

\bibitem[Ayyar and Chandrasekharan(2015)Ayyar and
  Chandrasekharan]{PhysRevD.91.065035}
Ayyar, V. and Chandrasekharan, S. (2015), \enquote{Massive fermions without
  fermion bilinear condensates,} \emph{Phys. Rev. D}, 91, 065035.

\bibitem[Ayyar and Chandrasekharan(2016)Ayyar and
  Chandrasekharan]{PhysRevD.93.081701}
Ayyar, V. and Chandrasekharan, S. (2016), \enquote{Origin of fermion masses
  without spontaneous symmetry breaking,} \emph{Phys. Rev. D}, 93, 081701.

\bibitem[Ayyar and Chandrasekharan(2017)Ayyar and
  Chandrasekharan]{PhysRevD.96.114506}
Ayyar, V. and Chandrasekharan, S. (2017), \enquote{Generating a nonperturbative
  mass gap using Feynman diagrams in an asymptotically free theory,}
  \emph{Phys. Rev. D}, 96, 114506.

\bibitem[Ayyar et~al.(2017)Ayyar, Chandrasekharan, and
  Rantaharju]{Ayyar:2017xmi}
Ayyar, V., Chandrasekharan, S., and Rantaharju, J. (2017), \enquote{{Benchmark
  results in the 2D lattice Thirring model with a chemical potential},} .

\bibitem[Beard and Wiese(1996)Beard and Wiese]{Bea96}
Beard, B.~B. and Wiese, U.-J. (1996), \enquote{Simulations of Discrete Quantum
  Systems in Continuous Euclidean Time,} \emph{Phys. Rev. Lett.}, 77,
  5130--5133.

\bibitem[Bercx et~al.(2017)Bercx, Goth, Hofmann, and
  Assaad]{SciPostPhys.3.2.013}
Bercx, M., Goth, F., Hofmann, J.~S., and Assaad, F.~F. (2017), \enquote{{The
  ALF (Algorithms for Lattice Fermions) project release 1.0. Documentation for
  the auxiliary field quantum Monte Carlo code},} \emph{SciPost Phys.}, 3, 013.

\bibitem[Blankenbecler et~al.(1981)Blankenbecler, Scalapino, and
  Sugar]{PhysRevD.24.2278}
Blankenbecler, R., Scalapino, D.~J., and Sugar, R.~L. (1981), \enquote{Monte
  Carlo calculations of coupled boson-fermion systems. I,} \emph{Phys. Rev. D},
  24, 2278--2286.

\bibitem[Boninsegni et~al.(2006)Boninsegni, Prokof'ev, and
  Svistunov]{PhysRevE.74.036701}
Boninsegni, M., Prokof'ev, N.~V., and Svistunov, B.~V. (2006), \enquote{Worm
  algorithm and diagrammatic Monte Carlo: A new approach to continuous-space
  path integral Monte Carlo simulations,} \emph{Phys. Rev. E}, 74, 036701.

\bibitem[Burovski et~al.(2008)Burovski, Kozik, Prokof'ev, Svistunov, and
  Troyer]{PhysRevLett.101.090402}
Burovski, E., Kozik, E., Prokof'ev, N., Svistunov, B., and Troyer, M. (2008),
  \enquote{Critical Temperature Curve in BEC-BCS Crossover,} \emph{Phys. Rev.
  Lett.}, 101, 090402.

\bibitem[Campostrini et~al.(2014)Campostrini, Pelissetto, and
  Vicari]{PhysRevB.89.094516}
Campostrini, M., Pelissetto, A., and Vicari, E. (2014), \enquote{Finite-size
  scaling at quantum transitions,} \emph{Phys. Rev. B}, 89, 094516.

\bibitem[Cardy(1988)Cardy]{Cardy:1988ag}
Cardy, J.~L. (ed.) (1988), \emph{{FINITE SIZE SCALING}}.

\bibitem[Chandrasekharan(2010)Chandrasekharan]{Chandrasekharan:2009wc}
Chandrasekharan, S. (2010), \enquote{{The Fermion bag approach to lattice field
  theories},} \emph{Phys. Rev.}, D82, 025007.

\bibitem[Chandrasekharan(2012)Chandrasekharan]{PhysRevD.86.021701}
Chandrasekharan, S. (2012), \enquote{Solutions to sign problems in lattice
  Yukawa models,} \emph{Phys. Rev. D}, 86, 021701.

\bibitem[Chandrasekharan(2013)Chandrasekharan]{Chandrasekharan:2013rpa}
Chandrasekharan, S. (2013), \enquote{Fermion bag approach to fermion sign
  problems,} \emph{The European Physical Journal A}, 49, 90.

\bibitem[Chandrasekharan and Li(2012a)Chandrasekharan and
  Li]{PhysRevD.85.091502}
Chandrasekharan, S. and Li, A. (2012a), \enquote{Fermion bag solutions to some
  sign problems in four-fermion field theories,} \emph{Phys. Rev. D}, 85,
  091502.

\bibitem[Chandrasekharan and Li(2012b)Chandrasekharan and
  Li]{PhysRevLett.108.140404}
Chandrasekharan, S. and Li, A. (2012b), \enquote{Fermion Bags, Duality, and the
  Three Dimensional Massless Lattice Thirring Model,} \emph{Phys. Rev. Lett.},
  108, 140404.

\bibitem[Chandrasekharan and Wiese(1999)Chandrasekharan and
  Wiese]{Chandrasekharan:1999ys}
Chandrasekharan, S. and Wiese, U.-J. (1999), \enquote{Meron-Cluster Solution of
  Fermion Sign Problems,} \emph{Phys. Rev. Lett.}, 83, 3116--3119.

\bibitem[Chandrasekharan et~al.(2002)Chandrasekharan, Scarlet, and
  Wiese]{Chandrasekharan:2001ya}
Chandrasekharan, S., Scarlet, B., and Wiese, U.~J. (2002), \enquote{{From spin
  ladders to the 2-d O(3) model at nonzero density},} \emph{Comput. Phys.
  Commun.}, 147, 388--393.

\bibitem[Continentino(2017)Continentino]{continentino_2017}
Continentino, M. (2017), \emph{Quantum Scaling in Many-Body Systems: An
  Approach to Quantum Phase Transitions}, Cambridge University Press, 2 edn.

\bibitem[Creutz(1977)Creutz]{PhysRevD.15.1128}
Creutz, M. (1977), \enquote{Gauge fixing, the transfer matrix, and confinement
  on a lattice,} \emph{Phys. Rev. D}, 15, 1128--1136.

\bibitem[Drut and L\"ahde(2009)Drut and L\"ahde]{PhysRevB.79.241405}
Drut, J.~E. and L\"ahde, T.~A. (2009), \enquote{Critical exponents of the
  semimetal-insulator transition in graphene: A Monte Carlo study,} \emph{Phys.
  Rev. B}, 79, 241405.

\bibitem[Duane et~al.(1987)Duane, Kennedy, Pendleton, and Roweth]{DUANE1987216}
Duane, S., Kennedy, A., Pendleton, B.~J., and Roweth, D. (1987),
  \enquote{Hybrid Monte Carlo,} \emph{Physics Letters B}, 195, 216 -- 222.

\bibitem[Fowler and Milne(1925)Fowler and Milne]{Fowler400}
Fowler, R.~H. and Milne, E.~A. (1925), \enquote{A Note on the Principle of
  Detailed Balancing,} \emph{Proceedings of the National Academy of Sciences},
  11, 400--402.

\bibitem[Gazit et~al.(2017)Gazit, Randeria, and Vishwanath]{Gazit2017}
Gazit, S., Randeria, M., and Vishwanath, A. (2017), \enquote{Emergent Dirac
  fermions and broken symmetries in confined and deconfined phases of Z2 gauge
  theories,} \emph{Nat Phys}, 13, 484--490.

\bibitem[Goldenfeld(1992)Goldenfeld]{Goldenfield}
Goldenfeld, N. (1992), \emph{Lectures on Phase Transitions and the
  Renormalization Group}, Addison-Wesley.

\bibitem[Goulko and Wingate(2010)Goulko and Wingate]{PhysRevA.82.053621}
Goulko, O. and Wingate, M. (2010), \enquote{Thermodynamics of balanced and
  slightly spin-imbalanced Fermi gases at unitarity,} \emph{Phys. Rev. A}, 82,
  053621.

\bibitem[Gubernatis et~al.(2016)Gubernatis, Kawashima, and Werner]{Gubernatis}
Gubernatis, J., Kawashima, N., and Werner, P. (2016), \emph{Quantum Monte Carlo
  Methods: Algorithms for Lattice Models}, Cambridge University Press,
  Cambridge CB2 8BS, United Kingdom.

\bibitem[Gubernatis et~al.(1985)Gubernatis, Scalapino, Sugar, and
  Toussaint]{PhysRevB.32.103}
Gubernatis, J.~E., Scalapino, D.~J., Sugar, R.~L., and Toussaint, W.~D. (1985),
  \enquote{Two-dimensional spin-polarized fermion lattice gases,} \emph{Phys.
  Rev. B}, 32, 103--116.

\bibitem[Gull et~al.(2011)Gull, Millis, Lichtenstein, Rubtsov, Troyer, and
  Werner]{RevModPhys.83.349}
Gull, E., Millis, A.~J., Lichtenstein, A.~I., Rubtsov, A.~N., Troyer, M., and
  Werner, P. (2011), \enquote{Continuous-time Monte Carlo methods for quantum
  impurity models,} \emph{Rev. Mod. Phys.}, 83, 349--404.

\bibitem[Haldane(1983)Haldane]{PhysRevLett.50.1153}
Haldane, F. D.~M. (1983), \enquote{Nonlinear Field Theory of Large-Spin
  Heisenberg Antiferromagnets: Semiclassically Quantized Solitons of the
  One-Dimensional Easy-Axis N\'eel State,} \emph{Phys. Rev. Lett.}, 50,
  1153--1156.

\bibitem[Hands et~al.(1993)Hands, Kocic, and Kogut]{HANDS199329}
Hands, S., Kocic, A., and Kogut, J. (1993), \enquote{Four-Fermi Theories in
  Fewer Than Four Dimensions,} \emph{Annals of Physics}, 224, 29 -- 89.

\bibitem[Hann et~al.(2017)Hann, Huffman, and Chandrasekharan]{Hann:2016xsw}
Hann, C.~T., Huffman, E., and Chandrasekharan, S. (2017), \enquote{{Solution to
  the sign problem in a frustrated quantum impurity model},} \emph{Annals
  Phys.}, 376, 63--75.

\bibitem[Hesselmann and Wessel(2016)Hesselmann and Wessel]{Hesselmann:2016tvh}
Hesselmann, S. and Wessel, S. (2016), \enquote{Thermal Ising transitions in the
  vicinity of two-dimensional quantum critical points,} \emph{Phys. Rev.}, B93,
  155157.

\bibitem[Hirsch(1983)Hirsch]{PhysRevB.28.4059}
Hirsch, J.~E. (1983), \enquote{Discrete Hubbard-Stratonovich transformation for
  fermion lattice models,} \emph{Phys. Rev. B}, 28, 4059--4061.

\bibitem[Hirsch and Fye(1986)Hirsch and Fye]{PhysRevLett.56.2521}
Hirsch, J.~E. and Fye, R.~M. (1986), \enquote{Monte Carlo Method for Magnetic
  Impurities in Metals,} \emph{Phys. Rev. Lett.}, 56, 2521--2524.

\bibitem[H\"ofling et~al.(2002)H\"ofling, Nowak, and
  Wetterich]{PhysRevB.66.205111}
H\"ofling, F., Nowak, C., and Wetterich, C. (2002), \enquote{Phase transition
  and critical behavior of the d=3 Gross-Neveu model,} \emph{Phys. Rev. B}, 66,
  205111.

\bibitem[Huffman(2016)Huffman]{Huffman:2016cgh}
Huffman, E. (2016), \enquote{{Monte Carlo methods in continuous time for
  lattice Hamiltonians},} [PoSLATTICE2016,258(2016)].

\bibitem[Huffman and Chandrasekharan(2016)Huffman and
  Chandrasekharan]{PhysRevE.94.043311}
Huffman, E. and Chandrasekharan, S. (2016), \enquote{Solution to sign problems
  in models of interacting fermions and quantum spins,} \emph{Phys. Rev. E},
  94, 043311.

\bibitem[Huffman and Chandrasekharan(2017)Huffman and
  Chandrasekharan]{Huffman:2017swn}
Huffman, E. and Chandrasekharan, S. (2017), \enquote{{Fermion bag approach to
  Hamiltonian lattice field theories in continuous time},} \emph{Phys. Rev.},
  D96, 114502.

\bibitem[Huffman and Mizel(2017)Huffman and Mizel]{PhysRevA.95.032131}
Huffman, E. and Mizel, A. (2017), \enquote{Violation of noninvasive
  macrorealism by a superconducting qubit: Implementation of a Leggett-Garg
  test that addresses the clumsiness loophole,} \emph{Phys. Rev. A}, 95,
  032131.

\bibitem[Huffman et~al.(2016)Huffman, Banerjee, Chandrasekharan, and
  Wiese]{Huffman:2015szi}
Huffman, E., Banerjee, D., Chandrasekharan, S., and Wiese, U.-J. (2016),
  \enquote{{Real-Time Evolution of Strongly Coupled Fermions driven by
  Dissipation},} \emph{Annals Phys.}, 372, 309--319.

\bibitem[Huffman and Chandrasekharan(2014)Huffman and
  Chandrasekharan]{PhysRevB.89.111101}
Huffman, E.~F. and Chandrasekharan, S. (2014), \enquote{Solution to sign
  problems in half-filled spin-polarized electronic systems,} \emph{Phys. Rev.
  B}, 89, 111101.

\bibitem[Iliesiu et~al.(2018)Iliesiu, Kos, Poland, Pufu, and
  Simmons-Duffin]{Iliesiu2018}
Iliesiu, L., Kos, F., Poland, D., Pufu, S.~S., and Simmons-Duffin, D. (2018),
  \enquote{Bootstrapping 3D fermions with global symmetries,} \emph{Journal of
  High Energy Physics}, 2018, 36.

\bibitem[Kadanoff(1966)Kadanoff]{Kadanoff:1966wm}
Kadanoff, L.~P. (1966), \enquote{{Scaling laws for Ising models near T(c)},}
  \emph{Physics}, 2, 263--272.

\bibitem[Kaul et~al.(2013)Kaul, Melko, and Sandvik]{RevCondMat}
Kaul, R.~K., Melko, R.~G., and Sandvik, A.~W. (2013), \enquote{Bridging
  Lattice-Scale Physics and Continuum Field Theory with Quantum Monte Carlo
  Simulations,} \emph{Annual Review of Condensed Matter Physics}, 4, 179--215.

\bibitem[Kawashima and Harada(2004)Kawashima and
  Harada]{doi:10.1143/JPSJ.73.1379}
Kawashima, N. and Harada, K. (2004), \enquote{Recent Developments of World-Line
  Monte Carlo Methods,} \emph{Journal of the Physical Society of Japan}, 73,
  1379--1414.

\bibitem[Li et~al.(2015a)Li, Jiang, and Yao]{1367-2630-17-8-085003}
Li, Z.-X., Jiang, Y.-F., and Yao, H. (2015a), \enquote{Fermion-sign-free
  Majarana-quantum-Monte-Carlo studies of quantum critical phenomena of Dirac
  fermions in two dimensions,} \emph{New Journal of Physics}, 17, 085003.

\bibitem[Li et~al.(2015b)Li, Jiang, and Yao]{Li:2014tla}
Li, Z.-X., Jiang, Y.-F., and Yao, H. (2015b), \enquote{{Solving the fermion
  sign problem in quantum Monte Carlo simulations by Majorana representation},}
  \emph{Phys. Rev.}, B91, 241117.

\bibitem[Li et~al.(2016)Li, Jiang, and Yao]{Li:2016gte}
Li, Z.-X., Jiang, Y.-F., and Yao, H. (2016), \enquote{{Majorana-time-reversal
  symmetries: a fundamental principle for sign-problem-free quantum Monte Carlo
  simulations},} \emph{Phys. Rev. Lett.}, 117, 267002.

\bibitem[Li et~al.(2017)Li, Jiang, Jian, and Yao]{Li2017}
Li, Z.-X., Jiang, Y.-F., Jian, S.-K., and Yao, H. (2017),
  \enquote{Fermion-induced quantum critical points,} \emph{Nature
  Communications}, 8, 314.

\bibitem[Lin et~al.(1993)Lin, Gubernatis, Gould, and
  Tobochnik]{doi:10.1063/1.4823192}
Lin, H., Gubernatis, J., Gould, H., and Tobochnik, J. (1993), \enquote{Exact
  Diagonalization Methods for Quantum Systems,} \emph{Computers in Physics}, 7,
  400--407.

\bibitem[Lüscher(1977)Lüscher]{lüscher1977}
Lüscher, M. (1977), \enquote{Construction of a selfadjoint, strictly positive
  transfer matrix for euclidean lattice gauge theories,} \emph{Comm. Math.
  Phys.}, 54, 283--292.

\bibitem[Metropolis et~al.(1953)Metropolis, Rosenbluth, Rosenbluth, Teller, and
  Teller]{met}
Metropolis, N., Rosenbluth, A., Rosenbluth, M., Teller, A., and Teller, E.
  (1953), \enquote{{Equation of State Calculations by Fast Computing
  Machines},} \emph{J Chem. Phys.}, 21.

\bibitem[Mihaila et~al.(2017)Mihaila, Zerf, Ihrig, Herbut, and
  Scherer]{Mihaila:2017ble}
Mihaila, L.~N., Zerf, N., Ihrig, B., Herbut, I.~F., and Scherer, M.~M. (2017),
  \enquote{{Gross-Neveu-Yukawa model at three loops and Ising critical behavior
  of Dirac systems},} .

\bibitem[Montvay and Munster(1997)Montvay and Munster]{Montvay:1994cy}
Montvay, I. and Munster, G. (1997), \emph{{Quantum fields on a lattice}},
  Cambridge University Press.

\bibitem[Negele and Orland(1988a)Negele and Orland]{negele1988quantum}
Negele, J. and Orland, H. (1988a), \emph{Quantum many-particle systems},
  Frontiers in physics, Addison-Wesley Pub. Co.

\bibitem[Negele and Orland(1988b)Negele and Orland]{Negele:1988vy}
Negele, J.~W. and Orland, H. (1988b), \emph{{Quantum Many Particle Systems}},
  Addison-Wesley.

\bibitem[Orús(2014)Orús]{ORUS2014117}
Orús, R. (2014), \enquote{A practical introduction to tensor networks: Matrix
  product states and projected entangled pair states,} \emph{Annals of
  Physics}, 349, 117 -- 158.

\bibitem[Otsuka et~al.(2016)Otsuka, Yunoki, and Sorella]{sorella}
Otsuka, Y., Yunoki, S., and Sorella, S. (2016), \enquote{{Universal quantum
  criticality in the metal-insulator transition of two-dimensional interacting
  Dirac electrons},} \emph{Phys. Rev.}, X6, 011029.

\bibitem[Prokof'ev et~al.(1998)Prokof'ev, Svistunov, and
  Tupitsyn]{PROKOFEV1998253}
Prokof'ev, N., Svistunov, B., and Tupitsyn, I. (1998), \enquote{“Worm”
  algorithm in quantum Monte Carlo simulations,} \emph{Physics Letters A}, 238,
  253 -- 257.

\bibitem[Prokof'ev and Svistunov(1998)Prokof'ev and
  Svistunov]{PhysRevLett.81.2514}
Prokof'ev, N.~V. and Svistunov, B.~V. (1998), \enquote{Polaron Problem by
  Diagrammatic Quantum Monte Carlo,} \emph{Phys. Rev. Lett.}, 81, 2514--2517.

\bibitem[R.~K.~Pathria(2011)R.~K.~Pathria]{pathria}
R.~K.~Pathria, P. D.~B. (2011), \emph{Statistical Mechanics, Third Edition},
  Academic Press, Burlington, MA.

\bibitem[Rosa et~al.(2001)Rosa, Vitale, and Wetterich]{PhysRevLett.86.958}
Rosa, L., Vitale, P., and Wetterich, C. (2001), \enquote{Critical Exponents of
  the Gross-Neveu Model from the Effective Average Action,} \emph{Phys. Rev.
  Lett.}, 86, 958--961.

\bibitem[Rubtsov et~al.(2005)Rubtsov, Savkin, and
  Lichtenstein]{PhysRevB.72.035122}
Rubtsov, A.~N., Savkin, V.~V., and Lichtenstein, A.~I. (2005),
  \enquote{Continuous-time quantum Monte Carlo method for fermions,}
  \emph{Phys. Rev. B}, 72, 035122.

\bibitem[Sandvik and Kurkij\"arvi(1991)Sandvik and
  Kurkij\"arvi]{PhysRevB.43.5950}
Sandvik, A.~W. and Kurkij\"arvi, J. (1991), \enquote{Quantum Monte Carlo
  simulation method for spin systems,} \emph{Phys. Rev. B}, 43, 5950--5961.

\bibitem[Scalapino et~al.(1984)Scalapino, Sugar, and
  Toussaint]{PhysRevB.29.5253}
Scalapino, D.~J., Sugar, R.~L., and Toussaint, W.~D. (1984), \enquote{Monte
  Carlo study of a two-dimensional spin-polarized fermion lattice gas,}
  \emph{Phys. Rev. B}, 29, 5253--5255.

\bibitem[Schollw\"ock(2005)Schollw\"ock]{RevModPhys.77.259}
Schollw\"ock, U. (2005), \enquote{The density-matrix renormalization group,}
  \emph{Rev. Mod. Phys.}, 77, 259--315.

\bibitem[Schollwöck(2011)Schollwöck]{SCHOLLWOCK201196}
Schollwöck, U. (2011), \enquote{The density-matrix renormalization group in
  the age of matrix product states,} \emph{Annals of Physics}, 326, 96 -- 192,
  January 2011 Special Issue.

\bibitem[Shankar(2011)Shankar]{Shankar}
Shankar, R. (2011), \emph{Principles of Quantum Mechanics, 2nd Edition}, Plenum
  Press.

\bibitem[Swendsen and Wang(1987)Swendsen and Wang]{PhysRevLett.58.86}
Swendsen, R.~H. and Wang, J.-S. (1987), \enquote{Nonuniversal critical dynamics
  in Monte Carlo simulations,} \emph{Phys. Rev. Lett.}, 58, 86--88.

\bibitem[Troyer and Wiese(2005)Troyer and Wiese]{PhysRevLett.94.170201}
Troyer, M. and Wiese, U.-J. (2005), \enquote{Computational Complexity and
  Fundamental Limitations to Fermionic Quantum Monte Carlo Simulations,}
  \emph{Phys. Rev. Lett.}, 94, 170201.

\bibitem[Wald(1984)Wald]{wald1984}
Wald, R.~M. (1984), \emph{General Relativity}, University of Chicago Press,
  Chicago, IL.

\bibitem[Wang et~al.(2014)Wang, Corboz, and Troyer]{1367-2630-16-10-103008}
Wang, L., Corboz, P., and Troyer, M. (2014), \enquote{Fermionic quantum
  critical point of spinless fermions on a honeycomb lattice,} \emph{New
  Journal of Physics}, 16, 103008.

\bibitem[Wang et~al.(2015a)Wang, Iazzi, Corboz, and Troyer]{Wang:2015rga}
Wang, L., Iazzi, M., Corboz, P., and Troyer, M. (2015a), \enquote{Efficient
  continuous-time quantum Monte Carlo method for the ground state of correlated
  fermions,} \emph{Phys. Rev.}, B91, 235151.

\bibitem[Wang et~al.(2015b)Wang, Liu, Iazzi, Troyer, and Harcos]{Wang:2015vha}
Wang, L., Liu, Y.-H., Iazzi, M., Troyer, M., and Harcos, G. (2015b),
  \enquote{{Split orthogonal group: A guiding principle for sign-problem-free
  fermionic simulations},} \emph{Phys. Rev. Lett.}, 115, 250601.

\bibitem[Wang et~al.(2016)Wang, Liu, and Troyer]{PhysRevB.93.155117}
Wang, L., Liu, Y.-H., and Troyer, M. (2016), \enquote{Stochastic series
  expansion simulation of the $t\text{\ensuremath{-}}V$ model,} \emph{Phys.
  Rev. B}, 93, 155117.

\bibitem[Wei et~al.(2016)Wei, Wu, Li, Zhang, and Xiang]{Wei:2016sgb}
Wei, Z.~C., Wu, C., Li, Y., Zhang, S., and Xiang, T. (2016), \enquote{{Majorana
  Positivity and the Fermion Sign Problem of Quantum Monte Carlo Simulations},}
  \emph{Phys. Rev. Lett.}, 116, 250601.

\bibitem[Wiese(1993)Wiese]{WIESE1993235}
Wiese, U.-J. (1993), \enquote{Bosonization and cluster updating of lattice
  fermions,} \emph{Physics Letters B}, 311, 235 -- 240.

\bibitem[Wilson(1971)Wilson]{PhysRevB.4.3174}
Wilson, K.~G. (1971), \enquote{Renormalization Group and Critical Phenomena. I.
  Renormalization Group and the Kadanoff Scaling Picture,} \emph{Phys. Rev. B},
  4, 3174--3183.

\bibitem[Wilson and Kogut(1974)Wilson and Kogut]{WILSON197475}
Wilson, K.~G. and Kogut, J. (1974), \enquote{The renormalization group and the
  ϵ expansion,} \emph{Physics Reports}, 12, 75 -- 199.

\bibitem[Zerf et~al.(2017)Zerf, Mihaila, Marquard, Herbut, and
  Scherer]{Zerf:2017zqi}
Zerf, N., Mihaila, L.~N., Marquard, P., Herbut, I.~F., and Scherer, M.~M.
  (2017), \enquote{{Four-loop critical exponents for the Gross-Neveu-Yukawa
  models},} .

\bibitem[Zinn-Justin(1996)Zinn-Justin]{Zinn-Justin:2280881}
Zinn-Justin, J. (1996), \emph{{Quantum field theory and critical phenomena; 3rd
  ed.}}, International series of monographs on physics, Clarendon Press,
  Oxford.

\bibitem[Zohar et~al.(2016)Zohar, Cirac, and Reznik]{0034-4885-79-1-014401}
Zohar, E., Cirac, J.~I., and Reznik, B. (2016), \enquote{Quantum simulations of
  lattice gauge theories using ultracold atoms in optical lattices,}
  \emph{Reports on Progress in Physics}, 79, 014401.

\end{thebibliography}
%-----------------------------------------------------------------------------%

%-----------------------------------------------------------------------------%
% BIOGRAPHY -- Start file with '\biography'.  Mandatory for Ph.D.
%-----------------------------------------------------------------------------%
\biography
\normalsize
Emilie Huffman was born on March 23, 1990 in Charlotte, North Carolina, USA. She received a Bachelor of Science in Physics and a Bachelor of Arts in Mathematics from Union University in 2012, and a Master of Arts degree in Physics from Duke University in 2015. She graduated from Duke University in 2018 with a PhD in Physics.

She was supported in graduate school by a National Physical Science Consortium fellowship. From June-December 2018 she will be a visiting researcher at the University of W\"{u}rzburg in Germany, and in 2019 she will begin a postdoctoral fellowship at the Perimeter Institute in Waterloo, Canada.

\section*{Publications}
\paragraph{}
1. E.~F.~Huffman and S.~Chandrasekharan, Phys.\ Rev.\ B {\bf 89}, no. 11, 111101 

  (2014) doi:10.1103/PhysRevB.89.111101
  [arXiv:1311.0034 [cond-mat.str-el]]. \cite{PhysRevB.89.111101}

\paragraph{}
2. E.~Huffman, D.~Banerjee, S.~Chandrasekharan and U.~J.~Wiese, Annals Phys.\  

{\bf 372}, 309 (2016) doi:10.1016/j.aop.2016.05.019
  [arXiv:1512.00412 [quant-ph]]. \cite{Huffman:2015szi}
  
\paragraph{}
3. E.~Huffman and S.~Chandrasekharan, Phys.\ Rev.\ E {\bf 94}, no. 4, 043311 (2016)

  doi:10.1103/PhysRevE.94.043311
  [arXiv:1605.07420 [cond-mat.str-el]]. \cite{PhysRevE.94.043311}
  
\paragraph{}
4. C.~T.~Hann, E.~Huffman and S.~Chandrasekharan, Annals Phys.\  {\bf 376}, 63 (2017)

  doi:10.1016/j.aop.2016.11.006
  [arXiv:1608.05144 [cond-mat.str-el]]. \cite{Hann:2016xsw}
  
\paragraph{}
5. E.~Huffman and A.~Mizel, Phys.\ Rev.\ A {\bf 95}, no. 3, 032131 (2016)

doi:10.1103/PhysRevA.95.032131 [arXiv:1609.05957 [quant-ph]] \cite{PhysRevA.95.032131}

\paragraph{}
6. E.~Huffman and S.~Chandrasekharan, Phys.\ Rev.\ D {\bf 96}, no. 11, 114502 (2017)

  doi:10.1103/PhysRevD.96.114502
  [arXiv:1709.03578 [hep-lat]]. \cite{Huffman:2017swn}
%Your biography is limited to one page and must contain
%\begin{enumerate}
%\normalbaselines
%\item Full name
%\item Date and place of birth
%\item Every degree you've earned, including this one, and where you earned it
 %     from.
%\end{enumerate}
%Mostly, that information is to narrow down which John Smith wrote that
%dissertation on the mating habits of sea cucumbers.  Sexy!

%You may also include
%\begin{enumerate}
%\item Any awards you've won related to your discipline since your
%undergraduate degree.
%\item Any fellowships you've held
%\item Anything you've published (papers, books, book chapters).  Don't be
%afraid to cite it here, so that the full bibliographic record of your
%article appears in the bibliography!
%\item Where your next job will be, if you know
%\end{enumerate}

%-----------------------------------------------------------------------------
% You're done :)
\end{document}